\documentclass[11pt,a4paper]{article}
\usepackage{hyperref}

\usepackage{color}
\usepackage{graphicx}
\usepackage{amssymb,amsfonts}
\usepackage{amsmath,amssymb}
\usepackage{float}
\usepackage{amsmath,amsfonts}
\usepackage{comment}
\usepackage{fancyvrb}
\usepackage{verbatimbox}
\usepackage{fullpage}
\usepackage{appendix,stfloats}
\usepackage{bbm}
\usepackage{subfigure}
\usepackage{blkarray}
\usepackage{multirow}

\usepackage{wrapfig}

\usepackage{amsmath,amssymb,amsthm,graphicx,mathtools}
\usepackage{algorithmicx}
\usepackage[ruled]{algorithm}
\usepackage{algpseudocode}
\usepackage{enumerate}
\usepackage{subfigure,color}
\usepackage{array,dsfont}

\usepackage{wrapfig}

\usepackage{tikz,amsmath}
\usepackage{tikz-qtree}
\usetikzlibrary{calc}
\usetikzlibrary{arrows}
\usepackage{verbatim}
\usetikzlibrary{shapes.geometric}

\def \bw{\mathbf w}

\usepackage{breqn,dsfont}

\usepackage{amsmath,amssymb,mathtools,bm,etoolbox}
\usepackage{tikz}
\usepackage{rotating}

\newcommand*{\factor}{0.23}
\newcommand*{\factorzero}{0.25}
\newcommand*{\factorone}{0.23}
\newcommand*{\factorbig}{0.34}

\usepackage[hyperref]{acl2019}
\usepackage{times}
\usepackage{latexsym}

\usepackage{url}
\usepackage{arydshln}
\usepackage{fancyvrb}
\usepackage{fvextra}

\aclfinalcopy 

\title{Topic Modeling with Wasserstein Autoencoders}

\author{Feng Nan$^{\dag}$, Ran Ding \thanks{\enskip This work was done when the author was with Amazon.}\enskip $^{\ddag}$, Ramesh Nallapati$^{\dag}$, Bing Xiang$^{\dag}$ \\ Amazon Web  Services$^{\dag}$, Compass Inc.$^{\ddag}$\\ 
	{\tt \{nanfen, rnallapa, bxiang\}@amazon.com$^{\dag}$, ran.ding@compass.com$^{\ddag}$}}
\date{}

\begin{document}
\maketitle
\begin{abstract}
	We propose a novel neural topic model in the Wasserstein autoencoders (WAE) framework. Unlike existing variational autoencoder based models, we directly enforce Dirichlet prior on the latent document-topic vectors. We exploit the structure of the latent space and apply a suitable kernel in minimizing the Maximum Mean Discrepancy (MMD) to perform distribution matching. We discover that  MMD performs much better than the Generative Adversarial Network (GAN) in matching high dimensional Dirichlet distribution. We further discover that incorporating randomness in the encoder output during training leads to significantly more coherent topics. To measure the diversity of the produced topics, we propose a simple topic uniqueness metric. Together with the widely used coherence measure NPMI, we offer a more wholistic evaluation of topic quality. Experiments on several real datasets show that our model produces significantly better topics than existing topic models. 
\end{abstract}

\section{Introduction}

Probabilistic topic models \cite{NIPS2010_3902} have been widely used to explore large collections of documents in an unsupervised manner. They can discover the underlying themes and organize the documents accordingly. 
The most popular probabilistic topic model is the Latent Dirichlet Allocation (LDA) \cite{Blei03latentdirichlet}, where the authors developed a variational Bayesian (VB) algorithm to perform approximate inference; subsequently \cite{2004PNAS..101.5228G} proposed an alternative inference method using collapsed Gibbs sampling. 

More recently, deep neural networks have been successfully used for such probabilistic models with the 
emergence of variational autoencoders (VAE) \cite{kingma2013auto}. The key advantage of such neural network based models is that inference can be carried out easily via a forward pass of the recognition network, without the need for expensive iterative inference scheme per example as in VB and collapsed Gibbs sampling. 
Topic models that fall in this framework include NVDM \cite{miao2016neural}, ProdLDA \cite{srivastava2017autoencoding} and NTM-R \cite{DBLP:conf/emnlp/DingNX18}. 
At a high level, these models consist of an encoder network that maps the Bag-of-Words (BoW) input to a latent document-topic vector and a decoder network that maps the document-topic vector to a discrete distribution over the words in the vocabulary. They are autoencoders in the sense that the output of the decoder aims to reconstruct the word distribution of the input BoW representation. Besides the reconstruction loss, VAE-based methods also minimize a KL-divergence term between the prior and posterior of the latent vector distributions. 
Despite their popularity, these VAE-based topic models suffer from several conceptual and practical challenges. 
First, the Auto-Encoding Variational Bayes \cite{kingma2013auto} framework of VAE relies on a reparameterization trick that only works with the ``location-scale" family of distributions. Unfortunately, the Dirichlet distribution, which largely accounted for the modeling success of LDA, does not belong to this family. 
The Dirichlet prior on the latent document-topic vector nicely captures the intuition that a document typically belongs to a sparse subset of topics. The VAE-based topic models have to resort to various Gaussian approximations toward this effect. For example, NVDM and NTM-R simply use Gaussian instead of Dirichlet prior; ProdLDA uses Laplace approximation of the Dirichlet distribution in the softmax basis as prior. 
Second, the KL divergence term in the VAE objective forces posterior distributions for all examples to match the prior, essentially making the encoder output independent of the input. This leads to the problem commonly known as \emph{posterior collapse} \cite{he2019lagging}. Although various heuristics such as KL-annealing \cite{45404} have been proposed to address this problem, they are shown to be ineffective in more complex datasets \cite{pmlr-v80-kim18e}.

In this work we leverage the expressive power and efficiency of neural networks and propose a novel neural topic model to address the above difficulties. 
Our neural topic model belongs to a broader family of Wasserstein autoencoders (WAE) \cite{tolstikhin2017wasserstein}. We name our neural topic model W-LDA \footnote{Code available at \url{https://github.com/awslabs/w-lda}} to emphasize the connection with WAE. 
Compared to the VAE-based topic models, our model has a few advantages. 
First, we encourage the latent document-topic vectors to follow the Dirichlet prior directly via distribution matching, without any Gaussian approximation; by preserving the Dirichlet prior, our model represents a much more faithful generalization of LDA to neural network based topic models. 
Second, our model matches the \emph{aggregated} posterior to the prior. As a result, the latent codes of different examples get to stay away from each other, promoting a better reconstruction \cite{tolstikhin2017wasserstein}. We are thus able to avoid the problem of posterior collapse.

To evaluate the quality of the topics from W-LDA and other models, we measure the coherence of the representative words of the topics using the widely accepted Normalized Pointwise Mutual Information (NPMI) \cite{DBLP:conf/iwcs/AletrasS13} score, which is shown to closely match human judgments \cite{lau2014machine}. 
While NPMI captures topic coherence, it is also important that the discovered topics are diverse (not repetitive). Yet such a measure has been missing in the topic model literature.\footnote{Most papers on topic modeling only present a selected small subset of non-repetitive topics for qualitative evaluation. The diversity among the topics is not measured.} We therefore propose a simple Topic Uniqueness (TU) measure for this purpose. Given a set of representative words from all the topics, the TU score is inversely proportional to the number of times each word is repeated in the set. High TU score means the representative words are rarely repeated and the topics are unique to each other. 
Using both TU and NPMI, we are able to provide a more wholistic measure of topic quality.  
%
To summarize our main contributions:
\begin{itemize}
	\item We introduce a uniqueness measure to evaluate topic quality more wholistically. 
	\item W-LDA produces significantly better quality topics than existing topic models in terms of topic coherence and uniqueness.
	\item We experiment with both the WAE-GAN and WAE-MMD variants \cite{tolstikhin2017wasserstein} for distribution matching and demonstrate key performance advantage of the latter with a carefully chosen kernel, especially in high dimensional settings.
	\item We discover a novel technique of adding noise to W-LDA to significantly boost topic coherence. This technique can potentially be applied to WAE in general and is of independent interest.
\end{itemize}

\section{Related Work}

Adversarial Autoencoder (AAE) \cite{makhzani2015adversarial} was proposed as an alternative to VAE. The main difference is that AAE regularizes the aggregated posterior to be close to a prior distribution whereas VAE regularizes the posterior to be close to the prior. Wasserstein autoencoders (WAE) \cite{tolstikhin2017wasserstein} provides justification for AAE from the Wasserstein distance minimization point of view. In addition to adversarial training used in AAE, the authors also suggested using Maximum Mean Discrepancy (MMD) for distribution matching. Compared to VAE, AAE/WAEs are shown to produce better quality samples. 

AAE has been applied in the task of unaligned text style transfer and semi-supervised natural language inference by ARAE \cite{kim2017adversarially}.
To be best of our knowledge, W-LDA is the first topic model based on the WAE framework. 
Recently, Adversarial Topic model (ATM) \cite{wang2018atm} proposes using GAN with Dirichlet prior to learn topics. The generator takes in samples from Dirichlet distribution and maps to a document-word distribution layer to form the fake samples. The discriminator tries to distinguish between the real documents from the fake documents. It also pre-processes the BoW representation of documents using TF-IDF. The evaluation is limited to topic coherence. 
A critical difference of W-LDA and ATM is that ATM tries to perform distribution matching in the vocabulary space whereas W-LDA in the latent document-topic space. Since the vocabulary space has much higher dimension (size of the vocabulary) than the latent document-topic space (number of topics), we believe it is much more challenging for ATM to train and perform well compared to W-LDA. 

Our work is also related to the topic of learning disentangled representations. A disentangled representation can be defined as one where single latent units are sensitive to changes
in single generative factors, while being relatively invariant to changes in other factors \cite{representationLearning2013}. In topic modeling, such disentanglement means that the learned topics are coherent and distinct. \cite{rubenstein2018latent} demonstrated that WAE learns better disentangled representation than VAE. 
Interestingly, \cite{rubenstein2018latent} argue for adding randomness to the encoder output to address the dimensionality mismatch between the intrinsic data and the latent space. 
One of our contributions is to discover that by properly adding randomness, we can significantly improve disentanglement (topic coherence and uniqueness) of WAE. Therefore we offer yet another evidence to the advantage of randomized WAE.

\section{Background}
\subsection{Latent Dirichlet Allocation}
LDA is the most popular topic model. Suppose there are $V$ words in the vocabulary, each document is represented as a BoW $\bw = (w_1, \dots, w_N)$, where $w_n$ is the word at position $n$ and assume there are $N$ words in the document. The number of topics $K$ is pre-specified. Each topic $\beta_k, k=1,\dots,K$ is a probability distribution over the words in the vocabulary. Each document is assumed to have a mixed membership of the topics $\theta \in \Re^K, \sum_k \theta_k = 1, \theta_k \geq 0$. The generative process for each document starts with drawing a document-topic vector from the Dirichlet prior distribution with parameter $\alpha$. To generate the $n$th word in the document, a topic $z_n \in \{1,\dots,K\}$ is drawn according to the multinomial distribution $\theta$ and the word is then drawn according to the multinomial distribution $\beta_{z_n}$. 
Thus, the marginal likelihood of the document $p(\bw | \alpha, \beta)$ is 
\begin{equation*}
	\int_{\theta} \left( \prod_{n=1}^{N} \sum_{z_n=1}^{K} p(w_n| z_n, \beta) p(z_n|\theta) \right) p(\theta|\alpha) d\theta
\end{equation*}
Given a document $\bw$, the inference task is to determine the conditional distribution $p(\theta|\bw)$.

\subsection{Wasserstein Auto-encoder}
The latent variable generative model posits that a target domain example (eg. document $\bw$) is generated by first sampling a latent code $\theta$ from a prior distribution $P_{\Theta}$ and then passed through a decoder network. The resulting distribution in the target domain is $P_{\text{dec}}$ with density:
\begin{equation}
	p_{\text{dec}}(\bw) = \int_{\theta} p_{\text{dec}}(\bw|\theta) p(\theta) d \theta.
\end{equation}
The key result of \cite{tolstikhin2017wasserstein} is that in order to minimize the optimal transport distance between $P_{\text{dec}}$ and the target distribution $P_{\bw}$, it is equivalent to minimizing the following objective for some scalar value of $\lambda$:
\begin{equation}
\inf_{Q(\theta| \bw)} \mathbb{E}_{P_{\bw}} \mathbb{E}_{Q(\theta| \bw)} [c(\bw, \text{dec}(\theta))] + \lambda \cdot \mathcal{D}_{\Theta}(Q_{\Theta}, P_{\Theta}),
\end{equation}
where $c$ is a cost function and $Q_{\Theta} := \mathbb{E}_{P_{\bw}} Q(\theta| \bw)$ is the aggregated posterior or the encoded distribution of the examples; $\mathcal{D}_{\Theta}(Q_{\Theta}, P_{\Theta})$ is an arbitrary divergence between $Q_{\Theta}$ and $P_{\Theta}$. 
Similar to VAE, the WAE objective consists of a reconstruction term and a regularization term. Note the key difference is that the regularization term for WAE is on the aggregated posterior whereas the term for VAE is on the posterior distribution.

Two different divergences were proposed for $\mathcal{D}_{\Theta}(Q_{\Theta}, P_{\Theta})$.
The first is GAN-based, setting $\mathcal{D}_{\Theta}(Q_{\theta}, P_{\Theta}) = D_{JS}(Q_{\Theta}, P_{\Theta})$ \cite{NIPS2014_5423}. A discriminator (an adversary) is introduced trying to separate ``true'' points sampled from $P_{\Theta}$ and ``fake'' ones sampled from $Q_{\Theta}$. The second is Maximum Mean Discrepancy (MMD)-based \cite{gretton2012kernel}, setting $\mathcal{D}_{\theta}(Q_{\theta}, P_{\theta}) = \text{MMD}_{\mathbf{k}} (Q_{\Theta}, P_{\Theta})$. For a kernel function $\mathbf{k}: \Theta \times \Theta \to \Re$, the MMD is defined as 
\begin{align}
& \text{MMD}_{\mathbf{k}} (Q_{\Theta}, P_{\Theta}) \notag \\ & = \| \int_{\Theta} \mathbf{k}(\theta, \cdot) d P_{\Theta}(\theta) - \int_{\Theta} \mathbf{k}(\theta, \cdot) d Q_{\Theta}(\theta) \|_{\mathcal{H}_{\mathbf{k}}}, \label{eq:mmd}
\end{align}
where $\mathcal{H}$ is the Reproducing Kernel Hilbert Space (RKHS) of real-valued functions mapping $\Theta$ to $\Re$ and $\mathbf{k}$ is the kernel function; $\mathbf{k}(\theta, \cdot)$ can be considered as the feature mapping of $\theta$ to a higher dimensional space. 

\section{W-LDA}
We now introduce our W-LDA model. 
We consider the BoW representation of documents. With a slight abuse of notation, a document is a BoW $\bw$, where $w_i$ is the number of occurrences of the $i$th vocabulary word in the document. 
\subsection{Encoder-decoder}
The encoder of W-LDA consists of an Multi-Layer Perceptron (MLP) mapping $\bw$ to an output layer of $K$ units before applying softmax to obtain the document-topic vector $\theta \in \mathbb{S}^{K-1}$. The encoder acts as the recognition network to perform efficient inference: $Q(\theta|\bw) \approx p(\theta|\bw)$. Unlike VAE-based method, we have the option to use deterministic encoder $\theta = \text{enc}(\bw)$, which is conceptually and computationally simpler. In this case $Q(\theta|\bw)$ is a Dirac Delta distribution. 
Given $\theta$, the decoder consists of a single layer neural network mapping $\theta$ to an output layer of $V$ units before applying softmax to obtain $\hat{\bw}\in \mathbb{S}^{V-1}$. $\hat{\bw}$ is a probability distribution over the words in the vocabulary. Mathematically, we have
\begin{equation}
\hat{w}_i = \frac{\exp{h_i}}{\sum_{j=1}^{V} \exp{h_j}},  \mathbf{h} = \beta \theta + b,
\end{equation}
where $\beta = [\beta_1, \dots, \beta_K]$ is the matrix of topic-word vectors as in LDA and $b$ is an offset vector. 
The reconstruction loss for the autoencoder is simply the negative cross-entropy loss between the BoW $\bw$ and the $\hat{\bw}$ from the decoder:
\begin{equation}
c(\bw, \hat{\bw}) = - \sum_{i=1}^{V} w_i \log \hat{w}_i. \label{eq:recon}
\end{equation} 

\subsection{Distribution matching}
We explored both GAN and MMD-based options for $\mathcal{D}_{\Theta}(Q_{\Theta}, P_{\Theta})$.
For GAN, we additionally introduce an MLP as a discriminator network. We alternate between minimization and maximization as done in \cite{tolstikhin2017wasserstein, makhzani2015adversarial}. 
Unfortunately, we are unable to train the GAN-based W-LDA as we face a vanishing gradient problem and the encoder fails to update for distribution matching. We investigate this issue further in Section \ref{sec:results} and demonstrate through a toy example that MMD is better suited than GAN for matching high dimensional Dirichlet distributions.
We therefore focus on the MMD-based method. 
The immediate question is which kernel function to use for MMD. 
Since our task is to match the Dirichlet distribution, it is natural to seek kernel functions that are based on meaningful distance metrics on the simplex. We therefore choose to use the information diffusion kernel \cite{DBLP:conf/nips/LaffertyL02}, which uses the geodesic distance:
\begin{equation*}
	d(\theta, \theta')= 2 \arccos \left( \sum_{k=1}^{K} \sqrt{\theta_k \theta'_k} \right).
\end{equation*}
Intuitively, it first maps points on the simplex to a sphere via $\theta_k \to \sqrt{\theta_k}$ and then measures the distance between points on the curved surface. 
Compared to the more common L-2 distance, the geodesic distance is much more sensitive to points near the boundary of the simplex, which is especially important for sparse data \cite{DBLP:conf/nips/LaffertyL02}.
The information diffusion kernel we use is 
\begin{equation}
\mathbf{k}(\theta, \theta')= \exp \left( -\arccos^2 \left( \sum_{k=1}^{K} \sqrt{\theta_k \theta'_k} \right) \right). \label{eq:kernel}
\end{equation}
The MMD in \eqref{eq:mmd} can be unbiasedly estimated using $m$ samples via \eqref{eq:mmd_estimate}, where $\{\theta_1, \dots, \theta_m \}$ are sampled from $Q_{\Theta}$ and $\{\theta'_1, \dots, \theta'_m \}$ are sampled from $P_{\Theta}$. This form can be more easily understood by writing the norm in \eqref{eq:mmd} in terms of an inner product and expand the product of sums. 
\begin{figure*}[!t]
	\begin{equation}
\widehat{\text{MMD}}_{\mathbf{k}} (Q_{\Theta}, P_{\Theta}) = \frac{1}{m(m-1)} \sum_{i\neq j} \mathbf{k}(\theta_i, \theta_j) + \frac{1}{m(m-1)} \sum_{i\neq j} \mathbf{k}(\theta'_i, \theta'_j) - \frac{2}{m^2} \sum_{i, j} \mathbf{k}(\theta_i, \theta'_j).
\label{eq:mmd_estimate}
\end{equation} \vspace{-5mm}
\end{figure*}

In practice, the reconstruction loss \eqref{eq:recon} can be orders of magnitude larger than the regularization term $\mathcal{D}_{\Theta}(Q_{\Theta}, P_{\Theta})$. We therefore need to multiply a scaling factor to the reconstruction loss in order to balance the two terms. Yet, we would like to avoid introducing an additional hyperparameter. Consider a baseline case where the document length is $s$ and contains only one unique word; further assume the output of the decoder is completely uninformative, i.e. $\hat{w}_i = 1 / V, i =1, \dots, V$; then $s \log V$ is the reconstruction loss. 
By setting the scaling factor to $1/(s \log V)$, we can normalize the reconstruction loss to 1 with respect to this baseline case. Empirical study suggests that such a choice works well across multiple datasets. 

\subsection{Adding noise}
One of the key discoveries of this paper is that adding noise to the document-topic vectors during training leads to substantially better topics. Specifically, for each training example we sample a random Dirichlet vector from the prior $\theta_{\text{noise}} \sim P_{\Theta}$ and mix with the encoder output $\theta=\text{enc}(\bw)$:
\begin{equation}
\theta_+ = (1-\alpha) \theta + \alpha \theta_{\text{noise}}, \label{eq:noise}
\end{equation}
where $\alpha \in [0,1]$ is the mixing proportion. $\alpha=0$ is equivalent to not adding any noise; $\alpha=1$ is equivalent to using purely noise and ignore the encoder output altogether. 
We use $\theta_+$ as input to the decoder and compute the reconstruction loss for stochastic gradient optimization. 
Note that although adding noise appears similar to the reparameterization trick in VAEs, it is much more flexible and not restricted to the ``location-scale'' family of distributions as in VAEs. 

\section{Topic extraction and TU measure}
We can extract the top words based on the decoder matrix weights. Specifically, The representative words of the $k$th topic are those corresponding to the top entries of $\beta_k$ sorted in descending order. As explained in the introduction, we evaluate the quality of the topics in terms of both topic uniqueness (TU) and coherence (NPMI). We propose a simple measure of TU defined as follows. 
Given the top $L$ words from each of the $K$ topics, the TU for topic $k$ is
$
\text{TU}(k) = \frac{1}{L} \sum_{l=1}^{L} \frac{1}{\text{cnt}(l,k)}, k = 1, \dots, K,
$
where $\text{cnt}(l,k)$ is the total number of times the $l^{th}$ top word in topic $k$ appears in the top words across all topics. For example, if the $l^{th}$ top word in topic $k$ appears only in topic $k$, then $\text{cnt}(l,k)=1$; on the other hand, if the word appears in all the topics then $\text{cnt}(l,k)=K$. Finally, the average TU is computed as 
$\text{TU} = \frac{1}{K} \sum_{k=1}^{K} \text{TU}(k)$.
The range of the TU value is between $1/K$ and 1. A higher TU value means the produced topics are more diverse.

\section{Experiments and Results} \label{sec:experiments}
We conduct experiments on a synthetic corpus generated according to the LDA model and six widely used real world benchmark datasets: 20NG (the same version as \cite{srivastava2017autoencoding}), AGNews, \footnote{\scriptsize \url{http://www.di.unipi.it/~gulli/AG_corpus_of_news_articles.html}} DBpedia \cite{dbpedia} , Yelp review polarity from the Yelp Dataset Challenge in 2015, NYTimes \cite{Dua:2017} and Wikitext-103 \cite{DBLP:journals/corr/MerityXBS16}. 
We use the same version of AGNews, DBpedia and Yelp review polarity as \cite{Zhang:2015:CCN:2969239.2969312}. 
These datasets have very different characteristics in terms of vocabulary size, document length and the number of samples. Four of them have class labels associated with the documents. Table \ref{table:datasets} summarizes the basic statistics.

\begin{table}[]
	\centering
	\resizebox{0.49\textwidth}{!}{ 
		\begin{tabular}{|c|c|c|c|c|c|}
			\hline
			dataset       & \#train & \#test & vocab & avg.doc.len & \#class\\ \hline
			Synthetic LDA & 10000 & - & 100 & 30 & - \\ \hline
			20NG & 10926   & 7266   & 1995       &       52.5       & 20  \\ \hline
			AGNews & 96000 & 7600 & 31827 & 17.6  & 4 \\ \hline
			DBPedia & 448000 & 70000 & 10248 & 21.3 & 14 \\ \hline
			Yelp P. & 448000 & 38000 & 20000 & 57.5 & 2 \\ \hline
			NYTimes       &   242798      &   29977     &    102660        &    330.6  &     -     \\ \hline
			Wikitext-103  &  28472       &   60     &    20000        &      1392.2      &  -  \\ \hline
		\end{tabular}
	}
	\caption{Dataset summary} \label{table:datasets} \vspace{-3mm}
\end{table}

\subsection{Baselines}
We evaluate W-LDA against existing topic model methods:
{\bf 1.} Collapsed Gibbs Sampling LDA as implemented in the Mallet package \cite{McCallumMALLET}; {\bf 2.} Online LDA as implemented in the Gensim package \cite{rehurek_lrec};  {\bf 3.} ProdLDA \cite{tolstikhin2017wasserstein}: VAE-based, uses Gaussian approximation of the Dirichlet prior in the softmax space;  {\bf 4.} NTM-R \cite{DBLP:conf/emnlp/DingNX18}: VAE-based, improvement of NVDM  \cite{miao2016neural}, uses pretrained word embeddings for coherence regularization.

\subsection{Synthetic topic recovery}
We first verify the ability of W-LDA in recovering topics via a synthetic experiment. We construct a corpus of 10000 documents following the LDA generative process. The vocabulary size is 100 and there are 5 topics and Dirichlet parameters are 0.1. We run all methods with 5 latent topics and compare the recovered top 10 words for each topic against the ground truth. We compute the maximum precision among all permutations to align the topics and report the result in Table \ref{table:lda_recovery}. Note a top-10 word in a predicted topic is a false positive if it is not among the top-10 words in the ground truth topic. 
We also compare the topic words produced by W-LDA against the ground truth in Table \ref{table:lda_recovery_topics}. W-LDA clearly recovers the ground truth very well, even the relative importance of most top words. Details of the experiments can be found in the Appendix.
\begin{table}[]
	\centering
	\resizebox{0.5\textwidth}{!}{ 
		\begin{tabular}{|c|c|c|c|c|}
			\hline
			LDA (C.G.) & Online LDA & ProdLDA & NTM-R  & W-LDA  \\ \hline
			0.88 & 0.98    & 0.76 & 0.52 & 0.94\\ \hline
		\end{tabular}
	}
	\caption{Precision in topic recovery: W-LDA is competitive with the best models.} \label{table:lda_recovery} \vspace{-5mm}
\end{table}

\begin{table}[]
	\centering
	\resizebox{0.3\textwidth}{!}{ 
		\begin{tabular}{|c|}
			\hline
			\begin{tabular}[c]{@{}c@{}}46,  4, 44, 30, 81, 40, 87, 13, 58, 62 \\ 46, 4, 44, 30, 81, 40, 13, 87, 62, 58\end{tabular}                                                                                                                                                                              \\ \hline
			\begin{tabular}[c]{@{}c@{}}13, 81, 29, 33, 27,  1,  7, 83,  2, 39 \\ 13, 81, 29, 27, 33, 1, 7, 83, 39, 2 \end{tabular}                                               \\ \hline
			\begin{tabular}[c]{@{}c@{}}88, 67, 16, 13, 14,  3, 75,  8, 61, 71 \\88, 67, 16, 13, 14, 3, 75, 8,  \textbf{44},  \textbf{32} \end{tabular}                                                         \\ \hline
			\begin{tabular}[c]{@{}c@{}}38, 17, 57, 48, 23, 56, 50, 83, 16, 82 \\ 38, 17, 57, 48, 23, 50, 56, 83, 16, 82\end{tabular}                                      \\ \hline
			\begin{tabular}[c]{@{}c@{}}44, 86, 32, 62, 20, 99, 83, 88, 51, 31 \\ 44, 86, 32, 62, 20, 88, 99, 83,  \textbf{16}, 31\end{tabular} \\ \hline
		\end{tabular}
	}\caption{Top 10 word indices ordered in decreasing importance. Each cell corresponds to a topic, in which the first row is the ground truth and the second row is W-LDA output. The false positives are in bold. W-LDA recovers the ground truth topics very well.}
	\label{table:lda_recovery_topics} \vspace{-5mm}
\end{table}

\begin{figure}%
	\centering
	\subfigure{%
		\includegraphics[scale=\factorbig]{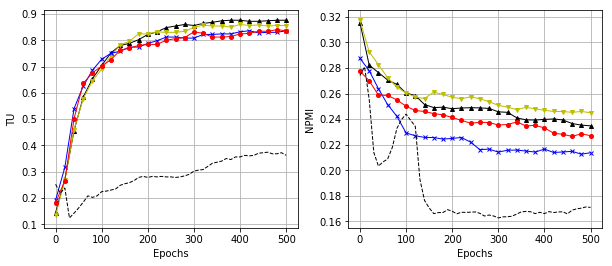}}\\
	\subfigure{%
		\includegraphics[scale=\factorbig]{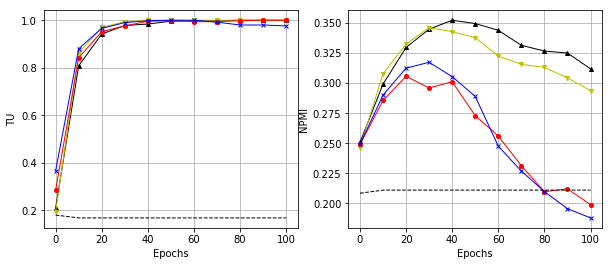}}\\
	\subfigure{%
		\includegraphics[scale=\factorbig]{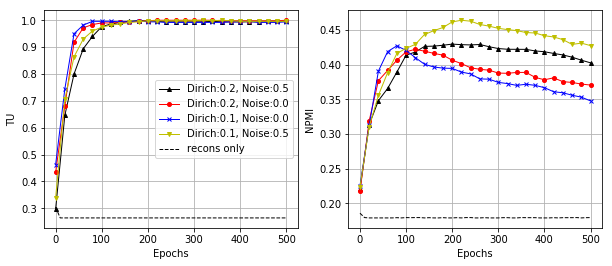}}
	\caption[TU and NPMI for various Dirichlet parameters and noise $\alpha$.]{W-LDA: TU and NPMI for various Dirichlet parameters and noise $\alpha$ for 20NG (top row); NYTimes (2nd row) and Wikitext-103 (bottom row). Adding Dirichlet noise generally improves topic NPMI. Minimizing reconstruction loss only (without distribution matching in latent space) generally leads to mode collapse of latent space where only one dimension is non-zero and the failure to learn the topics.}
	\label{fig:ablation_dirichlet}%
	\vspace{-5mm}
\end{figure}

\subsection{Parameter settings for benchmarking}
The parameter settings to run the real world datasets are as follows.
For LDA with collapsed Gibbs sampling we use the default Mallet parameter settings and run 2000 iterations. For Online LDA we use the default Gensim parameter settings and run 100 passes. 
For ProdLDA, we use the original implementation provided by the authors.\footnote{\scriptsize \url{https://github.com/akashgit/autoencoding_vi_for_topic_models}} We tune the dropout probability on the latent vector (the $\texttt{keep_prob}$ parameter in the original implementation) as we find it has significant impact on topic quality. We vary it from 0.4 (recommended value in the original paper) to 1. We find that setting it to 0.4 gives the highest NPMI; setting it to 1 gives better TU but much lower NPMI.
For NTM-R, we vary the Word Embedding Topic Coherence (WETC) coefficient in $[0, 1, 2, 5, 10, 50]$ and observe that setting it to 10 usually gives the best results in terms of NPMI and TU; setting it to 50 indeed raises the NPMI but the TU becomes very low and the topics consist of repetitive and generic words. 
For W-LDA, we set the Dirichlet parameter to 0.1 and 0.2 and use MMD with the information diffusion kernel \eqref{eq:kernel}; we set the noise coefficient $\alpha=\{0, 0.1, 0.2, 0.3, 0.4, 0.5, 0.6\}$. Similar to ProdLDA, we use ADAM optimizer with high momentum $\beta_1=0.99$ and learning rate of $0.002$ as they can overcome initial local minima. To be consistent, we set the encoder layers of W-LDA and ProdLDA the same as NTM-R, with two hidden layers and 100 neurons in each layer. 

For the evaluation of topic quality, we monitor the NPMI and TU for all algorithms over a reasonable number of iterations (either when the topic quality begins to deteriorate or stops improving over a number of iterations) and report the best results from the different parameter settings. 

\begin{table*}[t]
	\centering
	\resizebox{0.7\textwidth}{!}{ 
	\begin{tabular}{|c|c|c|c|c|c|}
		\hline
		&  LDA (C.G.) & Online LDA         & ProdLDA   & NTM-R        & \textbf{W-LDA}        \\ \hline
		20NG     &     \textbf{0.264}/0.85         & 0.252/0.79 &   0.267/0.58     & 0.240/0.62   & 0.252/\textbf{0.86}    \\ \hline
		AGNews & 0.239/0.76 & 0.213/0.80 & 0.245/0.68 & 0.220/0.69 &  \textbf{0.270/0.89} \\ \hline
DBpedia & 0.257/0.81 & 0.230/0.81 & \textbf{0.334}/0.49 & 0.222/0.71 & 0.295/ \textbf{1.00} \\ \hline
Yelp.P. &\textbf{0.238}/0.68 & 0.233/0.74 & 0.215/0.63 &0.224/0.40 & 0.235/\textbf{0.82} \\ \hline
		\multicolumn{1}{|c|}{NYTimes} & 0.300/0.81 & 0.291/0.80 &   0.319/0.67      & 0.218/0.88 &  \textbf{0.356/1.00} \\ \hline
		Wikitext-103               & 0.289/0.75   & 0.282/0.78 &    0.400/0.62   &    0.215/0.91        &  \textbf{0.464/1.00}  \\ \hline		
	\end{tabular}
	}
	\caption{Benchmark results for 50 topics. The numbers in each cell are NPMI/TU. Overall our method (W-LDA) achieves much higher NPMI as well as TU than existing methods.} \label{table:benchmark}
	\vspace{-5mm}
\end{table*}

\subsection{Benchmark results and ablation study} \label{sec:results}
The benchmark results are summarized in Table \ref{table:benchmark}.
We observe that LDA with collapsed Gibbs sampling produces similar topics as Online LDA. 
Although the NPMI of the topics produced by ProdLDA is high, the TU score is low, which means the topics are repetitive. For a qualitative inspection, we identify several repetitive topics that ProdLDA produces on the Wikitext-103 in Table \ref{table:sample_topics_prodlda} together with the best aligned topics from W-LDA. The topics from W-LDA are much more unique. A complete comparison of the topics from all of the methods can be found in the Appendix. 
NTM-R generally achieves a higher TU than LDAs and ProdLDA but has lower NPMI. 
Overall W-LDA achieves much higher NPMI as well as TU than existing methods, especially on NYTimes and Wikitext-103. 

\paragraph{Document classification:} Since W-LDA is not based on variational inference, we cannot compute the ELBO based perplexity as a performance metric as in \cite{miao2016neural,srivastava2017autoencoding,DBLP:conf/emnlp/DingNX18}. 
To compare the predictive performance of the latent document-topic vectors across all models, we use document classification accuracy instead. Detailed setup can be found in the Appendix to save space. The accuracies on the test set are summarized in Table \ref{table:doc_classification}. 
\begin{table}[]
	\centering
	\resizebox{0.5\textwidth}{!}{ 
		\begin{tabular}{|c|c|c|c|c|c|}
			\hline
			& LDA (C.G.) & Online LDA & ProdLDA & NTM-R  & W-LDA  \\ \hline
			20NG & 0.513 & 0.473    & 0.213 & 0.433 & 0.431 \\ \hline
			AGNews      & 0.848     & 0.825     & 0.827  & 0.857 & 0.853 \\ \hline
			DBpedia     & 0.906     & 0.890     & 0.112  & 0.916 & 0.938 \\ \hline
			Yelp P. & 0.869     & 0.865     & 0.777  & 0.862 & 0.856 \\ \hline
		\end{tabular}
	}
	\caption{Test accuracies for the document classification task. W-LDA is competitive with the best models.} \label{table:doc_classification} \vspace{-0mm}
\end{table}
We observe that the latent vectors from W-LDA have competitive classification accuracy with LDAs and NTM-R. ProdLDA performs significantly poorly on DBpedia dataset; further inspection shows that the distribution of the document-topic vectors produced by ProdLDA on test and training data are quite different. Next, we carry out ablation study on W-LDA.

\begin{table}[]
	\centering
	\resizebox{0.49\textwidth}{!}{ 
		\begin{tabular}{|c|}
			\hline
			\begin{tabular}[c]{@{}c@{}}season, playoff, league, nhl, game, rookie, touchdown, player, coach, goaltender\\ season, nhl, playoff, game, rookie, shutout, player, league, roster,  goaltender \\ \hdashline
				touchdown, fumble, quarterback, kickoff, punt, yardage, cornerback, linebacker, rushing, preseason
			\end{tabular}                                                                                                                                                                              \\ \hline
			\begin{tabular}[c]{@{}c@{}}infantry, casualty, troop, battalion, artillery, reinforcement, brigade, flank, division, army\\ brigade, casualty, troop, infantry, artillery, flank, battalion, commanded, division, regiment\\ artillery, casualty, destroyer, battalion, squadron, reinforcement, troop, regiment, guadalcanal, convoy \\ \hdashline 
				battalion, brigade, infantry, platoon, bridgehead, regiment, panzer, rok, pusan, counterattack
			\end{tabular}                                      \\ \hline
			\begin{tabular}[c]{@{}c@{}}mph, km, tropical, westward, landfall, flooding, northwestward, rainfall, northeastward, extratropical \\ mph, km, landfall, tropical, storm, hurricane, rainfall, flooding, extratropical, saffir\\ km, mph, tropical, westward, rainfall, flooding, convection, landfall, extratropical, storm \\ \hdashline
				dissipating, tropical, dissipated, extratropical, cyclone, shear, northwestward, southwestward, saffir, convection \end{tabular}                                                \\ \hline
		\end{tabular}
	}\caption{Comparison of select ProdLDA and W-LDA topics on Wikitext-103. ProdLDA topics are repetitive (above the dashed line in each cell); W-LDA topics are unique (below the dashed line in each cell). }
	\label{table:sample_topics_prodlda}
	\vspace{-3mm}
\end{table}

\paragraph{Distribution matching:} 
What if we only minimize the reconstruction loss of the auto-encoder, without the loss term associated with the distribution matching? We found that across all datasets in general, the learning tends to get stuck in bad local minima where only one dimension in the latent space is non-zero. The decoder weights also fail to produce meaningful topics at all. The NPMI and TU values are plotted in dashed lines in Figure \ref{fig:ablation_dirichlet}. This confirms the importance of  distribution matching in our topic model.  

\paragraph{Dirichlet parameter and noise effects:}
We study the effect of the Dirichlet parameter that controls the sparsity and the amount of noise added to the latent vector during training. 
Due to space limit, we only plot the TU and NPMI curves for 3 datasets in Figure \ref{fig:ablation_dirichlet}. The full set of plots on all datasets can be found in the Appendix. We observe that NPMI can be significantly improved by setting the noise coefficient $\alpha$ to $0.5$ compared to 0 (no added noise). It may appear surprising that such a high level of noise is beneficial; however, we note that due to the sparsity of the Dirichlet noise, the significant elements of the  encoder output $\theta$ would remain significant in $\theta_+$ in Eq. \eqref{eq:noise}. In other words, the variance from the noise does not wash out the signal; it helps spread out the latent space to benefit the training of the decoder network. This highlights the importance of randomness in the WAE framework on the one hand and the importance of Dirichlet assumption in the topic model on the other hand. 
The effect of setting the Dirichlet parameter to 0.1 or 0.2 is more mixed, signaling that the inherent topic sparsity in these datasets can be different. 

\paragraph{MMD vs GAN:}
We encountered vanishing gradient problem for the GAN-based W-LDA. The encoder was not able to learn to match the prior distribution. To investigate further we compare MMD and GAN in distribution matching via a toy experiment. 
Our setup is as follows. $100000$ input vectors are drawn from a 2D spherical Gaussian distribution. 
The encoder network consists of two hidden layers with 2 neurons in each layer and a 2D output layer with softmax. There is no decoder and no reconstruction loss. The goal is to train the encoder network so that the output appears to come from a 2D Dirichlet prior distribution of parameter 0.1. Due to space limit, Figure 1 in the Appendix shows that both GAN and MMD training successfully match the Dirichlet prior.
Next, we increase the number of neurons in each hidden and output layer to 50 and set the prior to a 50D Dirichlet distribution of parameter 0.1. Since there is no easy way to visualize the 50D distribution, we use t-SNE \cite{t-sne} to reduce the vectors to 2D and scatter plot the encoder output vectors (red) together with samples from the true Dirichlet prior (green) in Figure \ref{fig:adv_mmd_50d}.
Since the samples from the 50D Dirichlet prior tends to be sparse, there are roughly 50 green clusters corresponding to the 50 modes. We see that GAN (first row) fails to match the Dirichlet prior. On the other hand, MMD (second row) is able to gradually match the Dirichlet prior by capturing more and more clusters (modes). 

Given recent report that GAN learns challenging distributions much better than MMD \cite{li2017mmd}, our model offers an alternative view in support of the latter.
The success of using MMD in W-LDA is perhaps not surprising; the Dirichlet distribution is supported in the space of simplex, which behaves much more regularly than the space of pixels in images. Furthermore, the information diffusion  kernel that we choose is able to exploit such regularity in the geometry.

\begin{figure*}[t]%
	\centering
	\subfigure{%
		\includegraphics[scale=\factorzero]{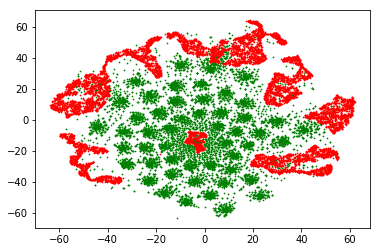}}%
	\subfigure{%
		\includegraphics[scale=\factorzero]{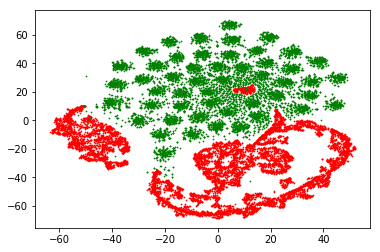}}
	\subfigure{%
		\includegraphics[scale=\factorzero]{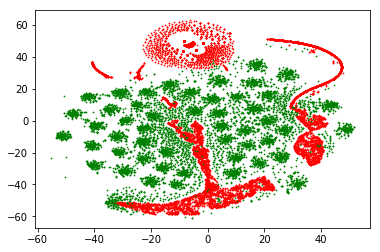}}%
	\subfigure{%
		\includegraphics[scale=\factorzero]{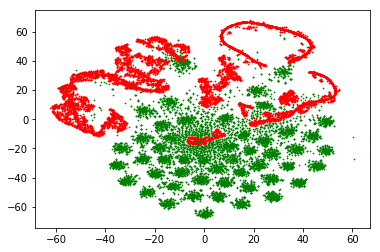}} \\ %
	\subfigure{%
		\includegraphics[scale=\factorone]{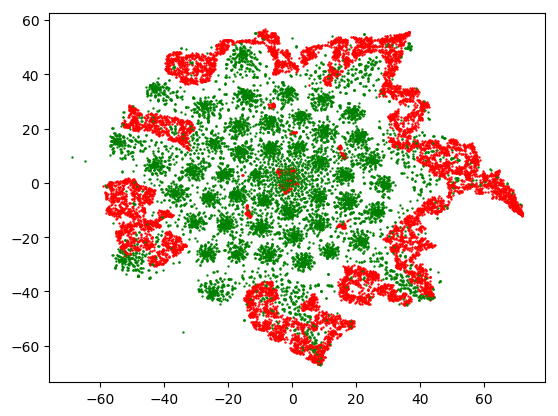}}%
	\subfigure{%
		\includegraphics[scale=\factorone]{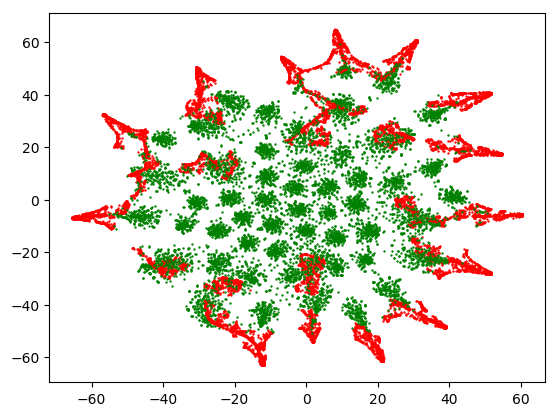}}
	\subfigure{%
		\includegraphics[scale=\factorone]{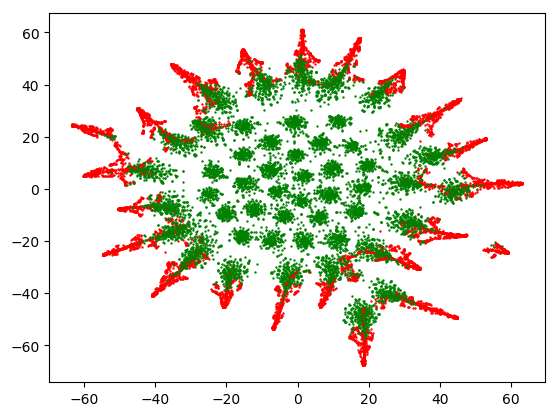}}%
	\subfigure{%
		\includegraphics[scale=\factorone]{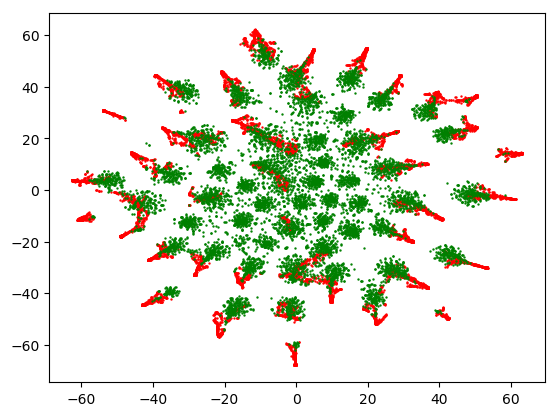}}%
	\caption[t-SNE plot of 50 dimensional encoder output vectors (red) and samples from the Dirichlet prior (green) over epochs. ]{t-SNE plot of encoder output vectors (red) and samples from the Dirichlet prior (green) over epochs. First row corresponds to epochs 0,10,30,99 of GAN training; second row corresponds to those of MMD training.}%
	\label{fig:adv_mmd_50d} \vspace{-3mm}
\end{figure*}

\section{Conclusion and Future Work}
We have proposed W-LDA, a neural network based topic model. Unlike existing neural network based models, W-LDA can directly enforce Dirichlet prior, which plays a central role in the sparse mixed membership model of LDA. To measure topic diversity, we have proposed a topic uniqueness measure in addition to the widely used NPMI for coherence. We report significant improvement of topic quality in both coherence and diversity over existing topic models. 
We further make two novel discoveries: first, MMD out-performs GAN in matching high dimensional Dirichlet distributions; second, carefully adding noise to the encoder output can significantly boost topic coherence without harming diversity. We believe these discoveries are of independent interest to the broader research on MMD, GAN and WAE. 

While we were not successful in training W-LDA using the GAN-based method, we acknowledge that many new formulations of GAN have been proposed to overcome mode collapse and vanishing gradient such as \cite{arjovsky2017wasserstein,gulrajani2017improved}. A future direction is to improve the GAN-based training of W-LDA.

Another future direction is to experiment with more complex priors than the Dirichlet prior. The W-LDA framework that we have proposed offers the flexibility of matching more sophisticated prior distributions via MMD or GAN. For example, the nested Chinese restaurant process can be used as a nonparametric prior to induce hierarchical topic models \cite{NIPS2003_2466}.




\bibliographystyle{acl_natbib}
\bibliography{generative}

\clearpage
\section{Appendix: synthetic topic recovery experiment details}
We construct a synthetic corpus of 10000 documents following the LDA generative process. The vocabulary size is 100 and there are 5 topics and Dirichlet parameters is 0.1. 
For all models we set the number of topics to be 5.
For LDA with collapsed Gibbs sampling, we use the default parameters of Mallet and run 2000 iterations. For Online LDA we run 200 iterations using the default parameters. We set the encoder network to have two hidden layers with 10 units each for the NTM-R, ProdLDA and W-LDA. For these 3 methods, we run 50 epochs and evaluate the topics every 10 epochs to choose the best epoch. We disable the WETC parameter for NTM-R because there is no word embedding. We set the Dirichlet parameter to 0.1 for W-LDA without adding noise. For ProdLDA we set the $\texttt{keep_prob}$ parameter to 1. 

\section{Appendix: additional TU and NPMI plots for W-LDA}
Due to space limit, we only provided TU and NPMI plots for 3 datasets in Figure 1 in the main paper. Here we provide the complete plots for all datasets in Figure \ref{fig:ablation_dirichlet}.
\begin{figure}%
	\centering
	\subfigure{%
		\includegraphics[scale=\factorbig]{Figs/20news_paper.png}}\\
	\subfigure{%
		\includegraphics[scale=\factorbig]{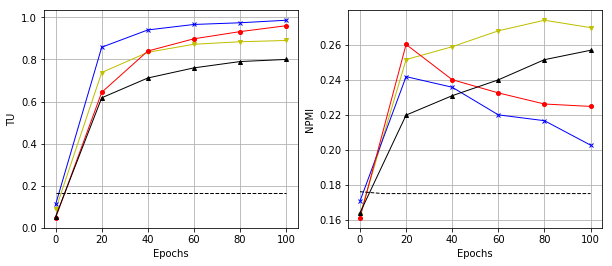}}\\
	\subfigure{%
		\includegraphics[scale=\factorbig]{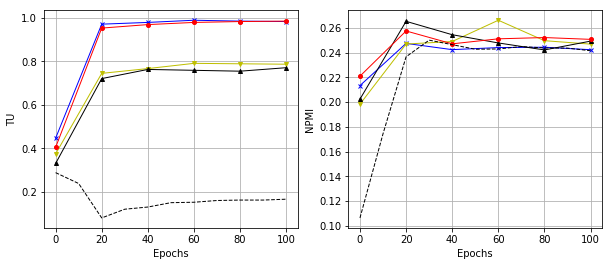}}\\
	\subfigure{%
		\includegraphics[scale=\factorbig]{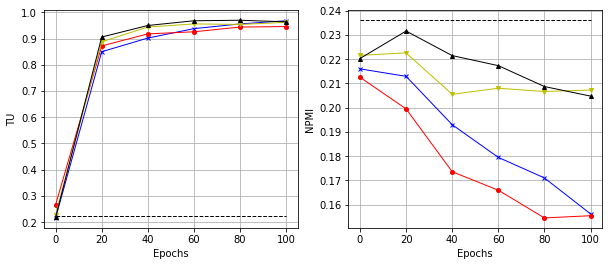}}\\
	\subfigure{%
		\includegraphics[scale=\factorbig]{Figs/nytimes_paper.png}}\\
	\subfigure{%
		\includegraphics[scale=\factorbig]{Figs/wiki_paper.png}}
	\caption[TU and NPMI for various Dirichlet parameters and noise $\alpha$.]{W-LDA: TU and NPMI for various Dirichlet parameters and noise $\alpha$ for 20NG (top row); NYTimes (2nd row) and Wikitext-103 (bottom row). Adding Dirichlet noise generally improves topic NPMI. Minimizing reconstruction loss only (without distribution matching in latent space) generally leads to mode collapse of latent space where only one dimension is non-zero and the failure to learn the topics.}
	\label{fig:ablation_dirichlet}%
\end{figure}
Note that even though the NPMI in Yelp P. without distribution matching is high, the TU is very low. The topics turn out to consist of highly repetitive words such as ``good'', ``nice'', ``love''.

\section{Appendix: document classification}
Besides exploring the corpus using interpretable topics, another usage for topic model is to act as a feature transformation of documents for downstream task such as document classification.
We compare the predictive performance of the latent document-topic vectors across all models. We set the number of topics for all models to be 50. For the neural network based models, we extract the output of the encoder as the features for document classification. For LDA, we extract the inferred document-topic vectors. A linear multiclass classifier with cross entropy loss is minimized using Adam optimizer with learning rate of 0.01 for 100 iterations for all models. Finally we choose the best parameter setting for each model based on the accuracy on a separate validation set.
For NTM, we vary the topic coherence parameter between 0 and 50; for ProdLDA we vary the $\texttt{keep_prob}$ parameter between 0.4 and 1. For W-LDA, we set the Dirichlet parameter to 0.1 and vary the Dirichlet prior parameter between 0.1 and 0.7. The accuracies on the test set are summarized in Table \ref{table:doc_classification}. 
\begin{table}[]
	\centering
	\resizebox{0.5\textwidth}{!}{ 
		\begin{tabular}{|l|l|l|l|l|l|}
			\hline
			& LDA (C.G.) & Online LDA & ProdLDA & NTM-R  & W-LDA  \\ \hline
			20NG & 0.5129 & 0.4725    & 0.2133 & 0.4334 & 0.4308 \\ \hline
			AGNews      & 0.8478     & 0.8253     & 0.8265  & 0.8567 & 0.8529 \\ \hline
			DBpedia     & 0.9059     & 0.8902     & 0.1124  & 0.9159 & 0.9382 \\ \hline
			Yelp P. & 0.8685     & 0.8652     & 0.7773  & 0.8616 & 0.8563 \\ \hline
		\end{tabular}
	}
	\caption{Test accuracies for the document classification task. W-LDA is competitive with the best models.} \label{table:doc_classification}
\end{table}
We observe that the latent vectors from W-LDA have competitive classification accuracy with LDAs and NTM-R. ProdLDA performs significantly poorly on DBpedia dataset; further inspection shows that the distribution of the document-topic vectors produced by ProdLDA on test and training data are quite different. 

\section{Appendix: MMD vs GAN in distribution matching}
In our experiments we encountered vanishing gradient problem for the GAN-based W-LDA. The encoder was not able to learn to match the prior distribution. To investigate further we compare MMD and GAN in distribution matching via a synthetic experiment. We show that both approaches perform well for low dimensional Dirichlet distribution yet MMD performs much better than GAN in higher dimensional setting. 
Our setup is as follows. $100,000$ input vectors are drawn from a 2D spherical Gaussian distribution. 
The encoder network consists of two hidden layers with 2 neurons in each layer and a 2D output layer with softmax. The goal is to train the encoder network so that the output appears to come from a 2D Dirichlet prior distribution of parameter 0.1.

\begin{figure*}
	\subfigure{%
		\includegraphics[scale=\factor]{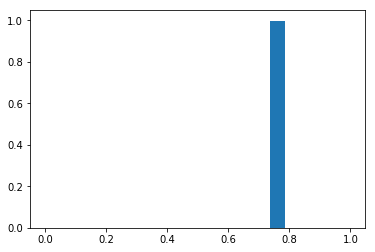}}%
	\subfigure{%
		\includegraphics[scale=\factor]{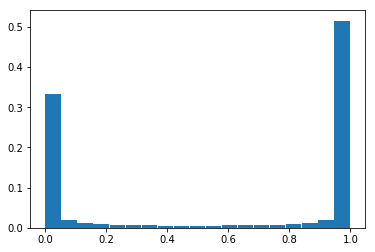}}
	\subfigure{%
		\includegraphics[scale=\factor]{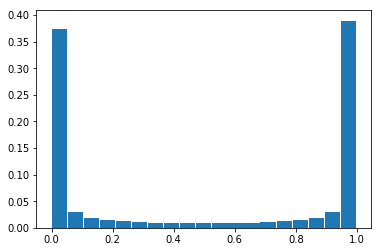}} 
	\subfigure{%
		\includegraphics[scale=\factor]{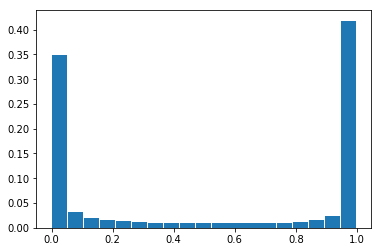}} \\ \vspace{-2mm}
	\subfigure{%
		\includegraphics[scale=\factor]{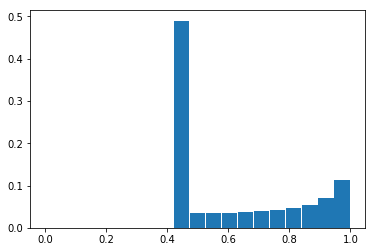}}%
	\subfigure{%
		\includegraphics[scale=\factor]{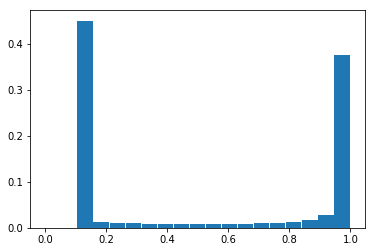}}
	\subfigure{%
		\includegraphics[scale=\factor]{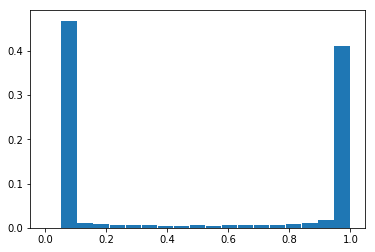}}%
	\subfigure{%
		\includegraphics[scale=\factor]{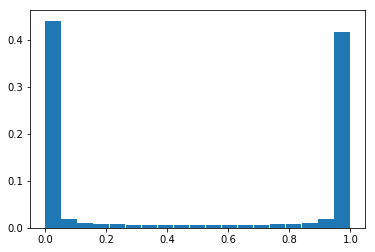}}
	\includegraphics[scale=\factor]{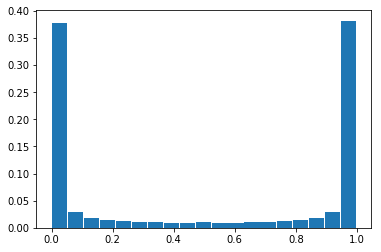}
	\caption[Encoded latent distribution over epochs.]{Histogram for the encoded latent distribution over epochs. First row corresponds to epochs 0, 10, 20 and 50 of GAN training; second row corresponds to epochs 0, 10, 20 and 50 of MMD training; the right most figure on the second row corresponds to the histogram of the prior distribution: 2D Dirichlet of parameter 0.1}
	\label{fig:adv_mmd_2d}
\end{figure*}

Since the 2 dimensions of the output vector sum to 1, we can visualize the resulting distribution via the histogram of the first dimension. The histogram from the true 2D Dirichlet prior of parameter 0.1 is shown in the right most sub-figure on the second row of Figure \ref{fig:adv_mmd_2d}. 
After 20 epochs of GAN training, the encoder output distribution is able to match that of the prior as shown in the first row of Figure \ref{fig:adv_mmd_2d}.
Similarly, MMD training is able to match that of the prior as shown in the second row of Figure \ref{fig:adv_mmd_2d}.
Next, we increase the number of neurons in each hidden and output layer to 50 and set the prior to a Dirichlet distribution of parameter 0.1. Since there is no easy way to visualize the 50 dimensional distribution, we use t-SNE \cite{t-sne} to reduce the vectors to 2D and scatter plot the encoder output vectors (red) together with samples from the true Dirichlet prior (green).
Figure \ref{fig:adv_mmd_50d} shows such a plot. Since the samples from the 50 dimensional Dirichlet prior tends to be sparse, there are roughly 50 green clusters corresponding to the 50 modes. We see that GAN (first row) fails to match the Dirichlet prior. On the other hand, MMD (second row) is able to gradually match the Dirichlet prior by capturing more and more clusters (modes). 
\begin{figure*}
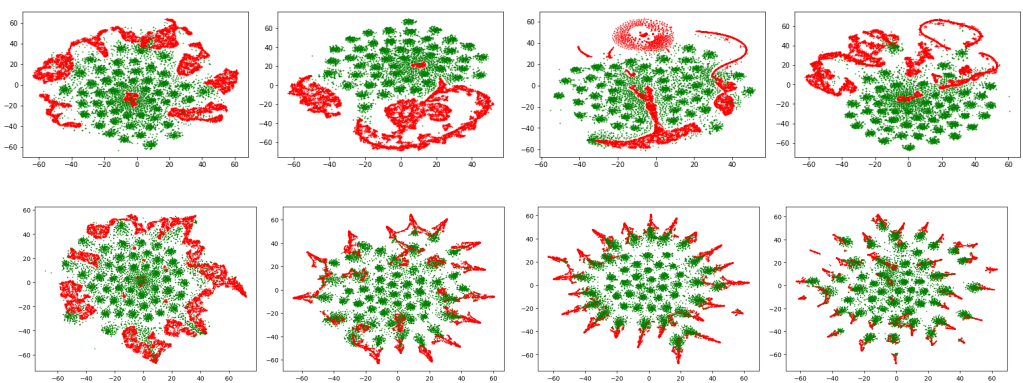
%
	\centering
	\subfigure{%
		\includegraphics[scale=\factorzero]{Figs/enc_out_epoch0_adv50_plot.png}}%
	\subfigure{%
		\includegraphics[scale=\factorzero]{Figs/enc_out_epoch10_adv50_plot.png}}
	\subfigure{%
		\includegraphics[scale=\factorzero]{Figs/enc_out_epoch30_adv50_plot.png}}%
	\subfigure{%
		\includegraphics[scale=\factorzero]{Figs/enc_out_epoch99_adv50_plot.png}} \\ %
	\subfigure{%
		\includegraphics[scale=\factorone]{Figs/enc_out_epoch0_mmd50_plot.png}}%
	\subfigure{%
		\includegraphics[scale=\factorone]{Figs/enc_out_epoch10_mmd50_plot.png}}
	\subfigure{%
		\includegraphics[scale=\factorone]{Figs/enc_out_epoch30_mmd50_plot.png}}%
	\subfigure{%
		\includegraphics[scale=\factorone]{Figs/enc_out_epoch99_mmd50_plot.png}}%
	\caption[t-SNE plot of 50 dimensional encoder output vectors (red) and samples from the Dirichlet prior (green) over epochs. ]{t-SNE plot of encoder output vectors (red) and samples from the Dirichlet prior (green) over epochs. First row corresponds to epochs 0,10,30,99 of GAN training; second row corresponds to those of MMD training}%
	\label{fig:adv_mmd_50d}%
\end{figure*}

\onecolumn
\section{Appendix: topic words}
The numbers at the beginning of each row are topic ID, TU and NPMI for each topic. 
\subsection{Topic words on 20NG}
LDA Collapsed Gibbs sampling:
NPMI=0.264, TU=0.854
\begin{Verbatim}[breaklines=true, fontsize=\tiny]
[ 0 - 0.85 - 0.30594]:  ['question', 'answer', 'correct', 'order', 'wrong', 'claim', 'knowledge', 'doubt', 'original', 'reason']
[ 1 - 0.49762 - 0.28311]:  ['thing', 'find', 'idea', 'couple', 'make', 'ago', 'put', 'guess', 'read', 'happy']
[ 2 - 1 - 0.41173]:  ['god', 'jesus', 'bible', 'christian', 'church', 'christ', 'faith', 'christianity', 'lord', 'sin']
[ 3 - 0.88333 - 0.27473]:  ['life', 'hell', 'man', 'death', 'love', 'body', 'dead', 'world', 'die', 'point']
[ 4 - 0.79762 - 0.32737]:  ['case', 'point', 'fact', 'make', 'situation', 'clear', 'avoid', 'idea', 'simply', 'position']
[ 5 - 0.95 - 0.35637]:  ['drive', 'scsi', 'disk', 'hard', 'controller', 'floppy', 'ide', 'rom', 'tape', 'card']
[ 6 - 0.95 - 0.19716]:  ['mr', 'president', 'stephanopoulos', 'package', 'today', 'house', 'press', 'myers', 'george', 'continue']
[ 7 - 0.61429 - 0.26655]:  ['work', 'job', 'lot', 'school', 'year', 'business', 'experience', 'make', 'learn', 'time']
[ 8 - 0.73333 - 0.29981]:  ['time', 'day', 'long', 'times', 'week', 'end', 'give', 'night', 'stop', 'rest']
[ 9 - 0.95 - 0.28707]:  ['gun', 'control', 'police', 'crime', 'carry', 'rate', 'weapon', 'defense', 'times', 'firearm']
[ 10 - 0.9 - 0.32923]:  ['windows', 'dos', 'os', 'screen', 'software', 'driver', 'mode', 'pc', 'ibm', 'memory']
[ 11 - 0.93333 - 0.27132]:  ['price', 'offer', 'sale', 'interested', 'buy', 'sell', 'mail', 'shipping', 'company', 'condition']
[ 12 - 0.88333 - 0.29119]:  ['team', 'hockey', 'season', 'league', 'nhl', 'year', 'game', 'division', 'city', 'pick']
[ 13 - 0.95 - 0.30172]:  ['evidence', 'argument', 'true', 'exist', 'truth', 'science', 'existence', 'theory', 'atheism', 'statement']
[ 14 - 0.9 - 0.30554]:  ['israel', 'jews', 'jewish', 'israeli', 'peace', 'arab', 'land', 'state', 'islam', 'human']
[ 15 - 0.75833 - 0.32471]:  ['state', 'government', 'law', 'rights', 'bill', 'states', 'federal', 'public', 'court', 'united']
[ 16 - 0.81667 - 0.24906]:  ['small', 'large', 'size', 'type', 'area', 'difference', 'free', 'order', 'work', 'set']
[ 17 - 0.93333 - 0.26958]:  ['health', 'medical', 'number', 'food', 'disease', 'care', 'pain', 'blood', 'study', 'msg']
[ 18 - 0.85833 - 0.23656]:  ['chip', 'encryption', 'clipper', 'government', 'law', 'technology', 'enforcement', 'escrow', 'privacy', 'phone']
[ 19 - 0.9 - 0.14479]:  ['period', 'la', 'power', 'pp', 'win', 'van', 'play', 'ny', 'cal', 'de']
[ 20 - 0.83333 - 0.24515]:  ['good', 'pretty', 'nice', 'worth', 'bad', 'level', 'class', 'quality', 'luck', 'thing']
[ 21 - 1 - 0.22251]:  ['car', 'bike', 'engine', 'speed', 'dod', 'road', 'ride', 'front', 'oil', 'dealer']
[ 22 - 0.80833 - 0.19851]:  ['file', 'output', 'entry', 'program', 'build', 'section', 'info', 'read', 'int', 'number']
[ 23 - 0.93333 - 0.21397]:  ['window', 'server', 'motif', 'application', 'widget', 'display', 'subject', 'mit', 'sun', 'set']
[ 24 - 0.85833 - 0.23623]:  ['post', 'article', 'group', 'posting', 'news', 'newsgroup', 'reply', 'read', 'response', 'mail']
[ 25 - 0.9 - 0.26876]:  ['image', 'graphics', 'version', 'ftp', 'color', 'format', 'package', 'jpeg', 'gif', 'contact']
[ 26 - 0.76429 - 0.31668]:  ['sense', 'make', 'moral', 'choice', 'person', 'personal', 'human', 'means', 'objective', 'understand']
[ 27 - 0.8 - 0.27119]:  ['back', 'side', 'left', 'put', 'head', 'end', 'turn', 'top', 'hand', 'picture']
[ 28 - 0.59762 - 0.28526]:  ['people', 'person', 'make', 'live', 'thing', 'talk', 'give', 'stop', 'realize', 'means']
[ 29 - 0.73333 - 0.25073]:  ['book', 'word', 'read', 'law', 'reference', 'find', 'matthew', 'text', 'context', 'david']
[ 30 - 0.88333 - 0.2469]:  ['water', 'war', 'military', 'time', 'air', 'south', 'plan', 'nuclear', 'force', 'ago']
[ 31 - 0.9 - 0.18688]:  ['cs', 'uk', 'ed', 'ac', 'john', 'david', 'ca', 'mark', 'jim', 'tom']
[ 32 - 0.78333 - 0.20401]:  ['key', 'bit', 'number', 'public', 'des', 'message', 'algorithm', 'security', 'part', 'block']
[ 33 - 0.66429 - 0.26271]:  ['big', 'bad', 'make', 'lot', 'stuff', 'remember', 'back', 'gm', 'guy', 'guess']
[ 34 - 0.9 - 0.25369]:  ['home', 'woman', 'wife', 'building', 'left', 'mother', 'door', 'remember', 'family', 'leave']
[ 35 - 0.95 - 0.21317]:  ['power', 'ground', 'current', 'wire', 'cable', 'supply', 'circuit', 'hot', 'box', 'run']
[ 36 - 0.9 - 0.30954]:  ['system', 'data', 'systems', 'software', 'computer', 'design', 'analysis', 'level', 'digital', 'high']
[ 37 - 0.95 - 0.22316]:  ['university', 'research', 'national', 'information', 'center', 'april', 'california', 'office', 'washington', 'conference']
[ 38 - 0.875 - 0.33608]:  ['armenian', 'turkish', 'armenians', 'people', 'turkey', 'armenia', 'turks', 'greek', 'genocide', 'government']
[ 39 - 0.83333 - 0.24725]:  ['game', 'year', 'play', 'hit', 'baseball', 'goal', 'player', 'average', 'flyers', 'shot']
[ 40 - 1 - 0.22238]:  ['black', 'fire', 'light', 'white', 'face', 'fbi', 'red', 'local', 'thought', 'koresh']
[ 41 - 0.93333 - 0.22892]:  ['code', 'line', 'source', 'set', 'include', 'simple', 'library', 'language', 'write', 'object']
[ 42 - 0.85 - 0.23873]:  ['card', 'video', 'mac', 'bit', 'apple', 'monitor', 'board', 'ram', 'memory', 'modem']
[ 43 - 0.83333 - 0.26739]:  ['mail', 'list', 'send', 'information', 'internet', 'email', 'anonymous', 'request', 'ftp', 'address']
[ 44 - 0.65833 - 0.30522]:  ['reason', 'wrong', 'agree', 'point', 'true', 'feel', 'find', 'opinion', 'reading', 'experience']
[ 45 - 0.9 - 0.13391]:  ['db', 'call', 'copy', 'al', 'section', 'mov', 'cs', 'place', 'bh', 'dangerous']
[ 46 - 0.9 - 0.21481]:  ['world', 'history', 'media', 'germany', 'german', 'europe', 'usa', 'american', 'great', 'part']
[ 47 - 0.85833 - 0.21134]:  ['problem', 'work', 'advance', 'fine', 'friend', 'find', 'recently', 'error', 'machine', 'cross']
[ 48 - 1 - 0.37787]:  ['space', 'nasa', 'earth', 'launch', 'satellite', 'shuttle', 'orbit', 'moon', 'mission', 'lunar']
[ 49 - 0.83929 - 0.30421]:  ['money', 'cost', 'pay', 'support', 'insurance', 'make', 'private', 'million', 'administration', 'government']
\end{Verbatim}
Online LDA:
NPMI=0.252, TU=0.788
\begin{Verbatim}[breaklines=true, fontsize=\tiny]
[ 0 - 0.78333 - 0.33403]:  ['drive', 'disk', 'scsi', 'hard', 'controller', 'ide', 'floppy', 'tape', 'system', 'bus']
[ 1 - 0.9 - 0.25403]:  ['jews', 'greek', 'jewish', 'turkish', 'turkey', 'greece', 'turks', 'adam', 'western', 'movement']
[ 2 - 0.86667 - 0.17326]:  ['new', 'period', 'york', 'chicago', 'st', 'pp', 'second', 'pittsburgh', 'los', 'power']
[ 3 - 0.68333 - 0.24503]:  ['encryption', 'government', 'law', 'technology', 'enforcement', 'privacy', 'security', 'new', 'clipper', 'escrow']
[ 4 - 0.76429 - 0.23766]:  ['widget', 'application', 'window', 'use', 'display', 'set', 'server', 'xt', 'motif', 'resource']
[ 5 - 0.95 - 0.23512]:  ['good', 'article', 'book', 'read', 'ago', 'paper', 'reading', 'reference', 'excellent', 'bob']
[ 6 - 0.825 - 0.24322]:  ['card', 'video', 'monitor', 'bit', 'screen', 'port', 'mode', 'vga', 'color', 'bus']
[ 7 - 0.9 - 0.29799]:  ['available', 'ftp', 'graphics', 'software', 'data', 'information', 'also', 'version', 'contact', 'package']
[ 8 - 0.62708 - 0.31925]:  ['one', 'people', 'think', 'true', 'may', 'question', 'say', 'point', 'evidence', 'even']
[ 9 - 0.82292 - 0.18638]:  ['bike', 'dod', 'pain', 'day', 'one', 'side', 'back', 'ride', 'like', 'first']
[ 10 - 1 - 0.37787]:  ['space', 'nasa', 'launch', 'earth', 'satellite', 'orbit', 'shuttle', 'moon', 'lunar', 'mission']
[ 11 - 1 - 0.1984]:  ['line', 'radio', 'tv', 'mark', 'audio', 'try', 'end', 'two', 'edge', 'center']
[ 12 - 0.80625 - 0.24895]:  ['power', 'board', 'memory', 'supply', 'ram', 'case', 'battery', 'motherboard', 'one', 'pin']
[ 13 - 0.5625 - 0.26666]:  ['people', 'right', 'government', 'rights', 'state', 'well', 'society', 'system', 'law', 'militia']
[ 14 - 0.36042 - 0.3857]:  ['like', 'people', 'think', 'get', 'know', 'one', 'really', 'want', 'say', 'something']
[ 15 - 0.9 - 0.35912]:  ['god', 'religion', 'believe', 'atheism', 'christian', 'religious', 'exist', 'belief', 'islam', 'existence']
[ 16 - 0.71429 - 0.26711]:  ['image', 'color', 'jpeg', 'gif', 'file', 'format', 'quality', 'use', 'bit', 'convert']
[ 17 - 0.91667 - 0.22672]:  ['black', 'man', 'cover', 'white', 'art', 'frank', 'red', 'jim', 'new', 'green']
[ 18 - 0.8 - 0.2531]:  ['thanks', 'please', 'anyone', 'know', 'help', 'mail', 'like', 'advance', 'post', 'need']
[ 19 - 0.57054 - 0.20668]:  ['chip', 'number', 'phone', 'clipper', 'use', 'serial', 'company', 'one', 'get', 'want']
[ 20 - 0.8 - 0.24168]:  ['university', 'program', 'research', 'national', 'conference', 'science', 'new', 'april', 'organization', 'billion']
[ 21 - 0.85 - 0.11704]:  ['year', 'last', 'win', 'la', 'cal', 'min', 'det', 'van', 'mon', 'tor']
[ 22 - 0.775 - 0.22759]:  ['game', 'goal', 'scsi', 'play', 'shot', 'puck', 'flyers', 'net', 'penalty', 'bit']
[ 23 - 0.65208 - 0.39624]:  ['god', 'jesus', 'one', 'church', 'bible', 'christ', 'christian', 'us', 'faith', 'people']
[ 24 - 0.60625 - 0.30005]:  ['money', 'buy', 'one', 'price', 'pay', 'insurance', 'cost', 'get', 'like', 'new']
[ 25 - 0.95 - 0.1792]:  ['ca', 'uk', 'cs', 'david', 'de', 'michael', 'ac', 'tom', 'john', 'andrew']
[ 26 - 0.81667 - 0.19665]:  ['sale', 'price', 'offer', 'new', 'shipping', 'condition', 'dos', 'cd', 'sell', 'interested']
[ 27 - 0.9 - 0.24906]:  ['sound', 'mike', 'record', 'oh', 'night', 'okay', 're', 'last', 'eric', 'sorry']
[ 28 - 0.51458 - 0.34604]:  ['much', 'time', 'one', 'like', 'good', 'better', 'think', 'get', 'well', 'really']
[ 29 - 0.85625 - 0.13969]:  ['db', 'al', 'cs', 'mov', 'bh', 'channel', 'byte', 'pop', 'push', 'one']
[ 30 - 0.76875 - 0.34827]:  ['armenian', 'armenians', 'turkish', 'people', 'genocide', 'armenia', 'one', 'russian', 'soviet', 'azerbaijan']
[ 31 - 0.9 - 0.25769]:  ['list', 'internet', 'mail', 'address', 'news', 'email', 'send', 'posting', 'anonymous', 'information']
[ 32 - 0.67262 - 0.28665]:  ['windows', 'dos', 'software', 'use', 'system', 'mac', 'problem', 'pc', 'file', 'driver']
[ 33 - 0.85833 - 0.27275]:  ['gun', 'file', 'crime', 'bill', 'law', 'control', 'police', 'weapon', 'states', 'firearm']
[ 34 - 0.74762 - 0.24866]:  ['study', 'health', 'number', 'rate', 'use', 'april', 'among', 'report', 'page', 'risk']
[ 35 - 0.76667 - 0.21434]:  ['window', 'sun', 'keyboard', 'server', 'mouse', 'motif', 'xterm', 'font', 'mit', 'get']
[ 36 - 0.62292 - 0.26581]:  ['car', 'engine', 'speed', 'front', 'oil', 'one', 'may', 'get', 'like', 'right']
[ 37 - 0.95 - 0.1132]:  ['vs', 'gm', 'la', 'pt', 'pm', 'ma', 'mg', 'md', 'tm', 'mi']
[ 38 - 0.9 - 0.28248]:  ['israel', 'israeli', 'arab', 'san', 'land', 'arabs', 'francisco', 'palestinian', 'state', 'jews']
[ 39 - 0.9 - 0.27416]:  ['medical', 'disease', 'public', 'soon', 'cancer', 'trial', 'treatment', 'health', 'gordon', 'medicine']
[ 40 - 0.81875 - 0.2017]:  ['fire', 'fbi', 'koresh', 'gas', 'dog', 'batf', 'compound', 'one', 'people', 'story']
[ 41 - 0.73125 - 0.2485]:  ['key', 'des', 'public', 'algorithm', 'bit', 'nsa', 'encryption', 'one', 'rsa', 'ripem']
[ 42 - 0.85 - 0.36136]:  ['team', 'game', 'season', 'hockey', 'league', 'year', 'play', 'nhl', 'player', 'baseball']
[ 43 - 0.66429 - 0.19931]:  ['entry', 'section', 'must', 'use', 'cross', 'program', 'info', 'number', 'source', 'may']
[ 44 - 0.77917 - 0.24336]:  ['us', 'war', 'country', 'government', 'military', 'american', 'people', 'world', 'nuclear', 'america']
[ 45 - 0.9 - 0.16519]:  ['master', 'feature', 'slave', 'pin', 'systems', 'tank', 'model', 'jumper', 'drive', 'japanese']
[ 46 - 0.56042 - 0.22238]:  ['mr', 'people', 'know', 'president', 're', 'us', 'one', 'stephanopoulos', 'think', 'go']
[ 47 - 0.82054 - 0.21982]:  ['ground', 'wire', 'hot', 'circuit', 'use', 'one', 'wiring', 'neutral', 'cable', 'current']
[ 48 - 0.80833 - 0.24895]:  ['output', 'file', 'program', 'int', 'printf', 'char', 'entry', 'input', 'oname', 'stream']
[ 49 - 0.90625 - 0.19526]:  ['code', 'media', 'call', 'one', 'object', 'stuff', 'date', 'btw', 'way', 'deal']
\end{Verbatim}

ProdLDA :
NPMI=0.268, TU=0.59
\begin{Verbatim}[breaklines=true, fontsize=\tiny]
[ 0 - 0.58333 - 0.21393]:  ['int', 'char', 'oname', 'buf', 'printf', 'output', 'null', 'entry', 'file', 'stream']
[ 1 - 0.7 - 0.19171]:  ['stephanopoulos', 'administration', 'president', 'senior', 'sector', 'congress', 'mr', 'russian', 'package', 'russia']
[ 2 - 0.43333 - 0.095146]:  ['tor', 'det', 'que', 'pit', 'nj', 'min', 'la', 'buf', 'van', 'cal']
[ 3 - 0.65 - 0.18382]:  ['bike', 'brake', 'gear', 'gateway', 'rider', 'manual', 'quadra', 'filter', 'mhz', 'motherboard']
[ 4 - 0.345 - 0.46605]:  ['interface', 'rom', 'controller', 'disk', 'ram', 'floppy', 'motherboard', 'mb', 'slot', 'scsi']
[ 5 - 0.70833 - 0.40336]:  ['israel', 'israeli', 'arab', 'arabs', 'islamic', 'lebanon', 'lebanese', 'palestinian', 'jew', 'murder']
[ 6 - 0.56667 - 0.32953]:  ['privacy', 'escrow', 'encryption', 'security', 'wiretap', 'enforcement', 'secure', 'encrypt', 'anonymous', 'ripem']
[ 7 - 0.43333 - 0.3356]:  ['jesus', 'passage', 'matthew', 'doctrine', 'scripture', 'holy', 'prophet', 'church', 'prophecy', 'pope']
[ 8 - 0.55 - 0.273]:  ['export', 'ftp', 'mit', 'xt', 'widget', 'server', 'unix', 'directory', 'vendor', 'font']
[ 9 - 0.425 - 0.36579]:  ['jesus', 'faith', 'passage', 'god', 'doctrine', 'belief', 'christ', 'existence', 'church', 'biblical']
[ 10 - 0.60833 - 0.12043]:  ['app', 'professor', 'rider', 'annual', 'league', 'genocide', 'francisco', 'armenian', 'art', 'arab']
[ 11 - 0.65 - 0.20985]:  ['stephanopoulos', 'mr', 'president', 'senate', 'consideration', 'meeting', 'myers', 'promise', 'decision', 'package']
[ 12 - 0.71667 - 0.29307]:  ['xt', 'image', 'xlib', 'amiga', 'toolkit', 'processing', 'resource', 'jpeg', 'workstation', 'server']
[ 13 - 0.8 - 0.31247]:  ['anonymous', 'privacy', 'cryptography', 'rsa', 'cipher', 'electronic', 'ftp', 'ripem', 'internet', 'pgp']
[ 14 - 0.56667 - 0.16196]:  ['stephanopoulos', 'president', 'clipper', 'scheme', 'mr', 'escrow', 'myers', 'restriction', 'nsa', 'wiretap']
[ 15 - 0.395 - 0.40117]:  ['armenians', 'turkish', 'armenian', 'turks', 'armenia', 'genocide', 'massacre', 'muslim', 'turkey', 'jews']
[ 16 - 0.5 - 0.29259]:  ['holy', 'jesus', 'son', 'father', 'lord', 'spirit', 'matthew', 'prophecy', 'satan', 'prophet']
[ 17 - 0.95 - 0.16966]:  ['health', 'hus', 'among', 'child', 'culture', 'md', 'volume', 'laboratory', 'age', 'safety']
[ 18 - 0.31667 - 0.34482]:  ['jesus', 'god', 'matthew', 'passage', 'prophecy', 'christ', 'holy', 'faith', 'lord', 'prophet']
[ 19 - 0.85 - 0.11794]:  ['db', 'byte', 'mov', 'bh', 'cs', 'ax', 'pop', 'push', 'west', 'ah']
[ 20 - 0.45 - 0.092708]:  ['tor', 'det', 'que', 'pit', 'van', 'nj', 'cal', 'la', 'gm', 'min']
[ 21 - 0.83333 - 0.3036]:  ['conclude', 'universe', 'existence', 'atheism', 'atheist', 'religious', 'belief', 'conclusion', 'evidence', 'truth']
[ 22 - 0.4 - 0.35819]:  ['hitter', 'season', 'defensive', 'puck', 'braves', 'baseball', 'playoff', 'league', 'coach', 'team']
[ 23 - 0.63333 - 0.32329]:  ['windows', 'colormap', 'window', 'microsoft', 'application', 'menu', 'dos', 'screen', 'widget', 'default']
[ 24 - 0.37833 - 0.34674]:  ['scsi', 'motherboard', 'ide', 'quadra', 'ram', 'vga', 'meg', 'mhz', 'adapter', 'isa']
[ 25 - 0.53333 - 0.30136]:  ['hitter', 'coach', 'offense', 'career', 'team', 'season', 'baseball', 'pitcher', 'dog', 'defensive']
[ 26 - 0.56667 - 0.16056]:  ['detroit', 'winnipeg', 'det', 'playoff', 'calgary', 'tor', 'vancouver', 'pp', 'rangers', 'gm']
[ 27 - 0.69167 - 0.37335]:  ['god', 'belief', 'faith', 'truth', 'reject', 'absolute', 'bible', 'christianity', 'christian', 'revelation']
[ 28 - 0.44167 - 0.24637]:  ['turkish', 'jews', 'greece', 'greek', 'muslims', 'jewish', 'matthew', 'lebanese', 'pope', 'christ']
[ 29 - 0.63333 - 0.16721]:  ['wiring', 'wire', 'oname', 'buf', 'entry', 'char', 'outlet', 'int', 'output', 'printf']
[ 30 - 0.575 - 0.33351]:  ['rom', 'disk', 'controller', 'floppy', 'feature', 'interface', 'connector', 'slot', 'mb', 'jumper']
[ 31 - 0.67 - 0.21175]:  ['armenians', 'apartment', 'woman', 'neighbor', 'troops', 'secretary', 'armenian', 'girl', 'armenia', 'afraid']
[ 32 - 0.42 - 0.35202]:  ['greek', 'turks', 'armenian', 'greece', 'minority', 'armenians', 'muslim', 'muslims', 'genocide', 'lebanese']
[ 33 - 0.5 - 0.30162]:  ['puck', 'flyers', 'season', 'score', 'hitter', 'braves', 'coach', 'team', 'nhl', 'career']
[ 34 - 0.545 - 0.21961]:  ['ide', 'scsi', 'meg', 'bus', 'isa', 'dos', 'hd', 'controller', 'adapter', 'slave']
[ 35 - 0.68333 - 0.29779]:  ['os', 'server', 'pixel', 'vendor', 'image', 'processing', 'documentation', 'xterm', 'unix', 'mit']
[ 36 - 0.8 - 0.2106]:  ['file', 'gun', 'united', 'congress', 'handgun', 'journal', 'prohibit', 'february', 'firearm', 'senate']
[ 37 - 0.68333 - 0.3552]:  ['winnipeg', 'calgary', 'montreal', 'detroit', 'rangers', 'nhl', 'hockey', 'leafs', 'louis', 'minnesota']
[ 38 - 0.63333 - 0.27956]:  ['heaven', 'god', 'eternal', 'braves', 'christ', 'christianity', 'pray', 'sin', 'dog', 'satan']
[ 39 - 0.7 - 0.3484]:  ['satellite', 'mission', 'space', 'nasa', 'shuttle', 'lunar', 'spacecraft', 'launch', 'international', 'earth']
[ 40 - 0.395 - 0.22319]:  ['hockey', 'nhl', 'league', 'armenian', 'massacre', 'turkish', 'draft', 'armenians', 'genocide', 'turks']
[ 41 - 0.7 - 0.1999]:  ['motherboard', 'amp', 'hd', 'brake', 'mhz', 'monitor', 'tire', 'upgrade', 'bike', 'compatible']
[ 42 - 0.56667 - 0.25204]:  ['widget', 'visual', 'resource', 'xt', 'application', 'colormap', 'app', 'export', 'default', 'converter']
[ 43 - 0.7 - 0.28056]:  ['earth', 'space', 'shuttle', 'mission', 'orbit', 'km', 'nasa', 'sky', 'lunar', 'foundation']
[ 44 - 0.60333 - 0.34564]:  ['mhz', 'scsi', 'modem', 'ram', 'vga', 'processor', 'cache', 'port', 'screen', 'printer']
[ 45 - 0.66667 - 0.19954]:  ['encryption', 'key', 'escrow', 'clipper', 'algorithm', 'enforcement', 'des', 'secure', 'wiretap', 'session']
[ 46 - 0.73333 - 0.086758]:  ['mw', 'db', 'wm', 'na', 'rg', 'van', 'md', 'mov', 'sl', 'bh']
[ 47 - 0.40333 - 0.40371]:  ['scsi', 'controller', 'mb', 'cache', 'disk', 'card', 'windows', 'floppy', 'vga', 'ram']
[ 48 - 0.57 - 0.21702]:  ['armenians', 'father', 'armenian', 'apartment', 'armenia', 'february', 'azerbaijan', 'woman', 'soviet', 'investigation']
[ 49 - 0.64167 - 0.31175]:  ['militia', 'sentence', 'jews', 'constitution', 'arab', 'israeli', 'lebanese', 'arabs', 'israel', 'nazi']
%\end{verbnobox}
\end{Verbatim}

NTM-R:
NPMI=0.24, TU=0.624
\begin{Verbatim}[breaklines=true, fontsize=\tiny]
[0-0.78333-0.22157]: ['marriage', 'exist', 'evidence', 'surely', 'sick', 'perhaps', 'appear', 'air', 'serious', 'raise']
[1-0.465-0.16851]: ['monitor', 'jesus', 'surrender', 'lot', 'dave', 'drive', 'put', 'disk', 'love', 'soon']
[2-0.33167-0.37993]: ['ide', 'controller', 'vga', 'card', 'floppy', 'adapter', 'hd', 'scsi', 'mb', 'video']
[3-0.73667-0.23189]: ['lebanon', 'surrender', 'evidence', 'reaction', 'islamic', 'death', 'soon', 'government', 'happen', 'effect']
[4-0.83333-0.39082]: ['armenian', 'armenians', 'turks', 'armenia', 'turkish', 'genocide', 'turkey', 'israel', 'arab', 'israeli']
[5-0.79-0.20377]: ['mask', 'punishment', 'surrender', 'try', 'religious', 'guess', 'patient', 'always', 'islam', 'bible']
[6-0.64167-0.24654]: ['year', 'consider', 'certain', 'besides', 'day', 'blame', 'pretty', 'evidence', 'damage', 'go']
[7-0.28667-0.33766]: ['ide', 'scsi', 'drive', 'disk', 'controller', 'floppy', 'isa', 'card', 'bus', 'ram']
[8-0.83333-0.3028]: ['hockey', 'toronto', 'cal', 'coach', 'game', 'league', 'winnipeg', 'rangers', 'detroit', 'playoff']
[9-0.54167-0.25723]: ['fan', 'season', 'team', 'toronto', 'game', 'last', 'year', 'braves', 'hit', 'miss']
[10-0.80833-0.18641]: ['insurance', 'false', 'difficult', 'find', 'clipper', 'relatively', 'regard', 'chip', 'etc', 'damn']
[11-0.395-0.24484]: ['please', 'sale', 'email', 'version', 'mail', 'modem', 'thanks', 'mailing', 'macintosh', 'ftp']
[12-0.88333-0.19338]: ['weapon', 'federal', 'military', 'warrant', 'population', 'government', 'judge', 'worry', 'attitude', 'ago']
[13-0.55278-0.22393]: ['interested', 'advance', 'os', 'dos', 'thanks', 'box', 'apple', 'windows', 'monitor', 'file']
[14-0.75833-0.18371]: ['round', 'year', 'go', 'else', 'money', 'digital', 'air', 'lot', 'wait', 'clinton']
[15-0.62833-0.18964]: ['mail', 'ftp', 'sale', 'workstation', 'email', 'eric', 'via', 'project', 'thanks', 'test']
[16-0.775-0.17498]: ['san', 'nasa', 'clipper', 'administration', 'americans', 'houston', 'gun', 'gm', 'closer', 'president']
[17-0.60333-0.24235]: ['realize', 'arab', 'israeli', 'jews', 'religious', 'surrender', 'shall', 'raise', 'atheism', 'carry']
[18-0.68333-0.29585]: ['hitter', 'hit', 'baseball', 'coach', 'team', 'flyers', 'staff', 'braves', 'season', 'player']
[19-0.55333-0.24486]: ['motif', 'image', 'mode', 'thanks', 'appreciate', 'pc', 'widget', 'vga', 'available', 'graphics']
[20-0.41667-0.24645]: ['cable', 'disk', 'ram', 'thanks', 'board', 'mb', 'modem', 'video', 'sale', 'adapter']
[21-0.7-0.26178]: ['widget', 'input', 'key', 'toolkit', 'chip', 'window', 'menu', 'error', 'default', 'int']
[22-0.565-0.38053]: ['god', 'christian', 'heaven', 'faith', 'christianity', 'jesus', 'hell', 'sin', 'interpretation', 'bible']
[23-0.44278-0.18352]: ['appreciate', 'thanks', 'card', 'windows', 'post', 'luck', 'vga', 'anybody', 'advance', 'thank']
[24-0.71667-0.19024]: ['window', 'toolkit', 'server', 'key', 'motif', 'pgp', 'mit', 'session', 'utility', 'stream']
[25-0.76167-0.23294]: ['properly', 'catholic', 'thanks', 'bible', 'sex', 'easy', 'moral', 'religion', 'mine', 'appropriate']
[26-0.83333-0.18606]: ['design', 'doctor', 'car', 'alive', 'imagine', 'brain', 'go', 'suppose', 'something', 'student']
[27-0.78333-0.22233]: ['israel', 'kill', 'jews', 'arab', 'woman', 'americans', 'responsible', 'nothing', 'civil', 'gordon']
[28-0.44778-0.27568]: ['windows', 'modem', 'server', 'version', 'vga', 'appreciate', 'client', 'binary', 'file', 'mouse']
[29-0.69167-0.27641]: ['key', 'escrow', 'encryption', 'clipper', 'chip', 'secure', 'enforcement', 'privacy', 'crypto', 'algorithm']
[30-0.5-0.18861]: ['hit', 'year', 'last', 'baseball', 'pick', 'love', 'address', 'ago', 'thanks', 'anyone']
[31-0.38167-0.4242]: ['jesus', 'god', 'christ', 'belief', 'faith', 'christian', 'bible', 'scripture', 'sin', 'church']
[32-0.50278-0.19305]: ['windows', 'client', 'font', 'advance', 'info', 'thanks', 'graphics', 'color', 'appreciate', 'anybody']
[33-0.57333-0.17343]: ['driver', 'file', 'help', 'anybody', 'anyone', 'hello', 'ftp', 'cool', 'jesus', 'set']
[34-0.88333-0.18867]: ['win', 'chicago', 'game', 'average', 'tie', 'car', 'bike', 'yeah', 'nice', 'hot']
[35-0.575-0.32561]: ['serious', 'christ', 'mary', 'eternal', 'god', 'faith', 'truth', 'freedom', 'scripture', 'man']
[36-0.57778-0.22004]: ['reply', 'windows', 'driver', 'version', 'file', 'thanks', 'find', 'ask', 'legal', 'switch']
[37-0.34778-0.32627]: ['controller', 'scsi', 'ide', 'bus', 'motherboard', 'port', 'mb', 'windows', 'isa', 'card']
[38-0.45333-0.24598]: ['cable', 'drive', 'rom', 'ftp', 'printer', 'pc', 'scsi', 'cd', 'disk', 'thanks']
[39-0.75833-0.22125]: ['proposal', 'encryption', 'clipper', 'secure', 'fairly', 'expensive', 'far', 'government', 'enough', 'traffic']
[40-0.95-0.10837]: ['det', 'van', 'pit', 'tor', 'period', 'min', 'pp', 'gm', 'que', 'ny']
[41-0.50333-0.33905]: ['satan', 'christian', 'jesus', 'scripture', 'moral', 'eternal', 'objective', 'truth', 'christ', 'belief']
[42-0.54167-0.15676]: ['bh', 'hd', 'rg', 'bus', 'ide', 'isa', 'db', 'md', 'floppy', 'drive']
[43-0.26778-0.21658]: ['printer', 'vga', 'card', 'anybody', 'windows', 'monitor', 'sale', 'controller', 'isa', 'port']
[44-0.75-0.1717]: ['motorola', 'db', 'ac', 'contact', 'toolkit', 'sale', 'xt', 'clock', 'macintosh', 'hr']
[45-0.61167-0.2934]: ['morality', 'moral', 'atheism', 'cause', 'bible', 'person', 'god', 'accurate', 'sin', 'disease']
[46-0.88333-0.25513]: ['stuff', 'ahead', 'fall', 'disease', 'food', 'thing', 'know', 'actually', 'anyone', 'expect']
[47-0.8-0.24088]: ['gun', 'trust', 'gang', 'something', 'blame', 'child', 'reading', 'avoid', 'abuse', 'pretty']
[48-0.68111-0.15852]: ['dos', 'hear', 'bob', 'package', 'anyway', 'windows', 'david', 'consider', 'surrender', 'site']
[49-0.41444-0.21188]: ['ftp', 'site', 'sale', 'monitor', 'windows', 'thanks', 'email', 'please', 'gif', 'newsgroup']	
\end{Verbatim}

W-LDA:
NPMI=0.252, TU=0.856
\begin{Verbatim}[breaklines=true, fontsize=\tiny]
[0-0.9-0.31117]: ['leafs', 'stanley', 'coach', 'nhl', 'hockey', 'team', 'wings', 'roger', 'cup', 'rangers']
[1-0.9-0.21338]: ['char', 'entry', 'widget', 'toolkit', 'int', 'oname', 'printf', 'contest', 'xlib', 'mit']
[2-0.9-0.2541]: ['xterm', 'window', 'colormap', 'expose', 'widget', 'client', 'xlib', 'null', 'button', 'server']
[3-0.9-0.21862]: ['amp', 'wave', 'voltage', 'audio', 'electronics', 'circuit', 'heat', 'cycle', 'bell', 'noise']
[4-0.9-0.19192]: ['plane', 'voltage', 'motif', 'edge', 'instruction', 'tube', 'algorithm', 'input', 'draw', 'surface']
[5-0.69-0.30885]: ['sorry', 'guess', 'like', 'get', 'anyone', 'think', 'know', 'someone', 'thanks', 'one']
[6-0.65-0.2681]: ['dos', 'driver', 'printer', 'card', 'windows', 'video', 'microsoft', 'isa', 'mode', 'pc']
[7-0.81667-0.20558]: ['helmet', 'bike', 'ride', 'detector', 'rider', 'motorcycle', 'road', 'radar', 'eye', 'cop']
[8-0.85-0.18271]: ['apartment', 'armenians', 'azerbaijan', 'neighbor', 'armenian', 'floor', 'afraid', 'secretary', 'building', 'woman']
[9-0.86667-0.28224]: ['orbit', 'earth', 'theory', 'mass', 'star', 'universe', 'space', 'moon', 'physical', 'material']
[10-0.9-0.30376]: ['scsi', 'ide', 'controller', 'bus', 'isa', 'jumper', 'drive', 'mhz', 'mb', 'disk']
[11-0.83333-0.20827]: ['neutral', 'outlet', 'wire', 'wiring', 'ground', 'electrical', 'panel', 'circuit', 'lunar', 'orbit']
[12-0.9-0.21013]: ['drive', 'floppy', 'meg', 'cd', 'motherboard', 'hd', 'external', 'boot', 'supply', 'brand']
[13-0.8-0.22575]: ['oh', 'yeah', 'guess', 'sick', 'hey', 'employer', 'sorry', 'disclaimer', 'wonder', 'excuse']
[14-0.82-0.181]: ['advance', 'gif', 'convert', 'format', 'graphic', 'graphics', 'thanks', 'ftp', 'site', 'anybody']
[15-0.70333-0.17932]: ['ford', 'curious', 'anyone', 'manual', 'recall', 'band', 'ago', 'paint', 'car', 'stuff']
[16-0.85-0.33088]: ['jesus', 'god', 'christ', 'matthew', 'spirit', 'lord', 'holy', 'passage', 'heaven', 'eternal']
[17-0.64-0.20301]: ['thanks', 'anybody', 'hello', 'appreciate', 'excuse', 'thread', 'friend', 'anyone', 'adams', 'mirror']
[18-0.9-0.30013]: ['homosexual', 'sexual', 'punishment', 'gay', 'sex', 'murder', 'commit', 'islamic', 'male', 'penalty']
[19-0.85-0.23501]: ['resurrection', 'hell', 'kent', 'eternal', 'evidence', 'body', 'heaven', 'koresh', 'claim', 'death']
[20-0.9-0.26142]: ['dog', 'ball', 'hitter', 'hr', 'pitcher', 'hit', 'braves', 'hall', 'ryan', 'ab']
[21-0.75333-0.13538]: ['uucp', 'curious', 'anyone', 'al', 'dave', 'compare', 'hear', 'someone', 'office', 'mine']
[22-0.85-0.21743]: ['doctor', 'pain', 'koresh', 'compound', 'fbi', 'tear', 'batf', 'fire', 'gas', 'treatment']
[23-0.95-0.22785]: ['sale', 'shipping', 'condition', 'offer', 'excellent', 'pair', 'sell', 'manual', 'inch', 'price']
[24-0.85-0.30941]: ['hitter', 'puck', 'defensive', 'season', 'offense', 'score', 'braves', 'game', 'team', 'career']
[25-0.72333-0.14718]: ['connector', 'curious', 'newsgroup', 'help', 'pin', 'anyone', 'soul', 'greatly', 'hello', 'thanks']
[26-1-0.38384]: ['israel', 'israeli', 'arabs', 'arab', 'lebanon', 'lebanese', 'civilian', 'peace', 'palestinian', 'war']
[27-0.76667-0.41534]: ['mission', 'satellite', 'shuttle', 'lunar', 'nasa', 'space', 'spacecraft', 'launch', 'orbit', 'solar']
[28-0.95-0.23404]: ['msg', 'morality', 'objective', 'moral', 'food', 'science', 'absolute', 'existence', 'scientific', 'definition']
[29-0.85-0.29938]: ['monitor', 'apple', 'vga', 'quadra', 'video', 'card', 'motherboard', 'mac', 'simm', 'cache']
[30-0.95-0.31918]: ['church', 'catholic', 'pope', 'doctrine', 'worship', 'authority', 'scripture', 'christ', 'lewis', 'tradition']
[31-1-0.087238]: ['mw', 'tor', 'det', 'que', 'ax', 'pit', 'rg', 'van', 'min', 'wm']
[32-0.85-0.18213]: ['tony', 'yeah', 'honda', 'student', 'watch', 'hear', 'listen', 'david', 'liberal', 'ticket']
[33-0.9-0.35688]: ['christianity', 'christian', 'bible', 'religion', 'faith', 'gay', 'belief', 'homosexual', 'islam', 'truth']
[34-0.95-0.23899]: ['keyboard', 'anonymous', 'usenet', 'privacy', 'internet', 'mailing', 'request', 'injury', 'posting', 'user']
[35-0.7-0.36042]: ['medicine', 'disease', 'drug', 'patient', 'medical', 'treatment', 'study', 'health', 'doctor', 'scientific']
[36-0.85-0.39602]: ['turkish', 'turks', 'armenian', 'genocide', 'armenians', 'armenia', 'greece', 'turkey', 'azerbaijan', 'greek']
[37-0.95-0.20527]: ['mouse', 'modem', 'printer', 'port', 'serial', 'print', 'hp', 'postscript', 'connect', 'resolution']
[38-0.85-0.22659]: ['surrender', 'gordon', 'soon', 'patient', 'eat', 'brain', 'girl', 'medicine', 'disease', 'treat']
[39-0.77-0.2276]: ['mail', 'please', 'address', 'mailing', 'advance', 'thanks', 'email', 'interested', 'appreciate', 'thank']
[40-1-0.24035]: ['stephanopoulos', 'president', 'mr', 'george', 'senate', 'myers', 'bush', 'meeting', 'consideration', 'clinton']
[41-0.85-0.25791]: ['swap', 'windows', 'nt', 'gateway', 'dos', 'memory', 'screen', 'menu', 'ram', 'microsoft']
[42-0.88333-0.15345]: ['apr', 'tom', 'frank', 'nasa', 'article', 'gmt', 'trial', 'space', 'university', 'id']
[43-0.95-0.28355]: ['escrow', 'encryption', 'clipper', 'key', 'wiretap', 'encrypt', 'des', 'nsa', 'rsa', 'algorithm']
[44-0.8-0.30457]: ['car', 'brake', 'tire', 'ford', 'engine', 'oil', 'saturn', 'dealer', 'transmission', 'fuel']
[45-0.71667-0.27262]: ['bike', 'bmw', 'battery', 'honda', 'rear', 'tank', 'ride', 'seat', 'sport', 'engine']
[46-1-0.23824]: ['handgun', 'homicide', 'gun', 'firearm', 'insurance', 'crime', 'ban', 'billion', 'seattle', 'fund']
[47-0.95-0.28678]: ['winnipeg', 'calgary', 'montreal', 'louis', 'philadelphia', 'rangers', 'minnesota', 'pittsburgh', 'ottawa', 'detroit']
[48-1-0.29992]: ['militia', 'amendment', 'constitution', 'bear', 'court', 'libertarian', 'federal', 'violate', 'rights', 'shall']
[49-0.71667-0.20957]: ['motorcycle', 'dod', 'bmw', 'ride', 'bike', 'truck', 'tire', 'lock', 'shop', 'module']
\end{Verbatim}

\subsection{Topic words on NYTimes:}
LDA Collapsed Gibbs sampling:
NPMI=0.30, TU=0.808
\begin{Verbatim}[breaklines=true, fontsize=\tiny]
[ 0 - 0.78333 - 0.20576]:  ['cup', 'food', 'minutes', 'add', 'oil', 'tablespoon', 'wine', 'sugar', 'water', 'fat']
[ 1 - 0.80333 - 0.2849]:  ['race', 'won', 'team', 'zzz_olympic', 'sport', 'track', 'gold', 'win', 'racing', 'medal']
[ 2 - 0.55667 - 0.37877]:  ['team', 'yard', 'game', 'season', 'play', 'player', 'quarterback', 'football', 'zzz_nfl', 'coach']
[ 3 - 1 - 0.34611]:  ['car', 'driver', 'truck', 'road', 'drive', 'seat', 'driving', 'vehicle', 'vehicles', 'wheel']
[ 4 - 0.67833 - 0.36089]:  ['company', 'business', 'sales', 'product', 'customer', 'million', 'market', 'companies', 'consumer', 'industry']
[ 5 - 0.80833 - 0.31807]:  ['meeting', 'question', 'asked', 'told', 'official', 'decision', 'interview', 'talk', 'reporter', 'comment']
[ 6 - 0.875 - 0.27026]:  ['art', 'century', 'history', 'french', 'artist', 'painting', 'museum', 'show', 'collection', 'zzz_london']
[ 7 - 0.87 - 0.21794]:  ['zzz_new_york', 'building', 'resident', 'area', 'million', 'mayor', 'project', 'zzz_los_angeles', 'local', 'center']
[ 8 - 0.78333 - 0.19082]:  ['daily', 'question', 'american', 'newspaper', 'beach', 'palm', 'statesman', 'information', 'today', 'zzz_washington']
[ 9 - 0.9 - 0.35134]:  ['family', 'father', 'home', 'son', 'friend', 'wife', 'mother', 'daughter', 'brother', 'husband']
[ 10 - 0.875 - 0.29023]:  ['hair', 'fashion', 'wear', 'designer', 'shirt', 'show', 'wearing', 'black', 'red', 'suit']
[ 11 - 0.7 - 0.27419]:  ['government', 'zzz_china', 'zzz_united_states', 'country', 'countries', 'foreign', 'political', 'european', 'leader', 'chinese']
[ 12 - 0.83333 - 0.29267]:  ['sense', 'fact', 'zzz_america', 'power', 'perhap', 'history', 'question', 'view', 'moment', 'real']
[ 13 - 0.88333 - 0.25602]:  ['water', 'fish', 'weather', 'boat', 'bird', 'wind', 'miles', 'storm', 'air', 'light']
[ 14 - 0.85833 - 0.30065]:  ['show', 'television', 'network', 'series', 'zzz_nbc', 'viewer', 'media', 'broadcast', 'station', 'night']
[ 15 - 0.9 - 0.41115]:  ['palestinian', 'zzz_israel', 'peace', 'zzz_israeli', 'israeli', 'zzz_yasser_arafat', 'leader', 'israelis', 'violence', 'attack']
[ 16 - 0.75333 - 0.27807]:  ['power', 'energy', 'oil', 'plant', 'gas', 'zzz_california', 'prices', 'million', 'water', 'environmental']
[ 17 - 0.95 - 0.26389]:  ['fight', 'hand', 'left', 'pound', 'body', 'weight', 'head', 'arm', 'hard', 'face']
[ 18 - 1 - 0.35861]:  ['drug', 'patient', 'doctor', 'medical', 'cell', 'cancer', 'hospital', 'health', 'treatment', 'care']
[ 19 - 0.925 - 0.30406]:  ['religious', 'church', 'zzz_god', 'gay', 'group', 'jewish', 'priest', 'faith', 'religion', 'jew']
[ 20 - 0.62333 - 0.3712]:  ['run', 'season', 'hit', 'team', 'game', 'inning', 'baseball', 'yankees', 'player', 'games']
[ 21 - 0.62833 - 0.32998]:  ['company', 'million', 'companies', 'firm', 'deal', 'zzz_enron', 'stock', 'business', 'billion', 'financial']
[ 22 - 0.95 - 0.27674]:  ['guy', 'bad', 'feel', 'thought', 'big', 'kid', 'kind', 'dog', 'word', 'remember']
[ 23 - 0.85833 - 0.25315]:  ['job', 'worker', 'employees', 'contract', 'manager', 'business', 'union', 'working', 'company', 'executive']
[ 24 - 0.775 - 0.26369]:  ['percent', 'number', 'study', 'found', 'result', 'survey', 'article', 'level', 'problem', 'group']
[ 25 - 0.70833 - 0.21533]:  ['black', 'white', 'zzz_mexico', 'american', 'country', 'immigrant', 'zzz_united_states', 'mexican', 'group', 'flag']
[ 26 - 0.83333 - 0.36285]:  ['zzz_bush', 'president', 'zzz_white_house', 'bill', 'zzz_clinton', 'zzz_senate', 'zzz_congress', 'administration', 'republican', 'political']
[ 27 - 0.62 - 0.23437]:  ['round', 'won', 'shot', 'player', 'tour', 'play', 'golf', 'zzz_tiger_wood', 'win', 'set']
[ 28 - 0.86667 - 0.33338]:  ['film', 'movie', 'character', 'actor', 'movies', 'director', 'zzz_hollywood', 'play', 'minutes', 'starring']
[ 29 - 0.37333 - 0.37747]:  ['team', 'game', 'point', 'season', 'coach', 'play', 'player', 'games', 'basketball', 'win']
[ 30 - 0.85 - 0.34187]:  ['court', 'case', 'law', 'lawyer', 'decision', 'legal', 'lawsuit', 'judge', 'zzz_florida', 'ruling']
[ 31 - 0.83333 - 0.23369]:  ['room', 'hotel', 'house', 'town', 'restaurant', 'wall', 'home', 'tour', 'trip', 'night']
[ 32 - 0.95 - 0.28975]:  ['women', 'children', 'child', 'parent', 'girl', 'age', 'young', 'woman', 'mother', 'teen']
[ 33 - 0.84167 - 0.36553]:  ['music', 'song', 'play', 'band', 'musical', 'show', 'album', 'sound', 'stage', 'record']
[ 34 - 0.68333 - 0.26919]:  ['law', 'group', 'government', 'official', 'federal', 'public', 'rules', 'agency', 'states', 'issue']
[ 35 - 0.95 - 0.28208]:  ['web', 'site', 'www', 'mail', 'information', 'online', 'sites', 'zzz_internet', 'internet', 'telegram']
[ 36 - 0.78333 - 0.27546]:  ['fire', 'night', 'hour', 'dead', 'police', 'morning', 'street', 'left', 'building', 'killed']
[ 37 - 0.75833 - 0.30963]:  ['zzz_afghanistan', 'zzz_taliban', 'war', 'bin', 'laden', 'government', 'official', 'zzz_pakistan', 'forces', 'zzz_u_s']
[ 38 - 0.83333 - 0.28581]:  ['school', 'student', 'program', 'teacher', 'high', 'college', 'education', 'class', 'test', 'public']
[ 39 - 0.95 - 0.13694]:  ['fax', 'syndicate', 'con', 'article', 'purchased', 'zzz_canada', 'una', 'publish', 'zzz_paris', 'representatives']
[ 40 - 0.80333 - 0.32636]:  ['money', 'million', 'tax', 'plan', 'pay', 'billion', 'cut', 'fund', 'cost', 'program']
[ 41 - 0.88333 - 0.41706]:  ['campaign', 'zzz_al_gore', 'zzz_george_bush', 'election', 'voter', 'vote', 'political', 'presidential', 'republican', 'democratic']
[ 42 - 0.85833 - 0.32241]:  ['computer', 'system', 'technology', 'software', 'zzz_microsoft', 'window', 'digital', 'user', 'company', 'program']
[ 43 - 0.9 - 0.35597]:  ['police', 'case', 'death', 'officer', 'investigation', 'prison', 'charges', 'trial', 'prosecutor', 'zzz_fbi']
[ 44 - 0.85 - 0.31374]:  ['percent', 'market', 'stock', 'economy', 'quarter', 'growth', 'economic', 'analyst', 'rate', 'rates']
[ 45 - 0.60833 - 0.26599]:  ['attack', 'military', 'zzz_u_s', 'zzz_united_states', 'terrorist', 'zzz_bush', 'official', 'war', 'zzz_american', 'security']
[ 46 - 0.44 - 0.30756]:  ['team', 'point', 'game', 'season', 'play', 'player', 'games', 'goal', 'shot', 'zzz_laker']
[ 47 - 0.85 - 0.27431]:  ['human', 'scientist', 'anthrax', 'animal', 'disease', 'found', 'test', 'food', 'research', 'virus']
[ 48 - 0.9 - 0.3247]:  ['flight', 'plane', 'airport', 'passenger', 'pilot', 'travel', 'security', 'air', 'airline', 'crew']
[ 49 - 0.9 - 0.31896]:  ['book', 'writer', 'author', 'wrote', 'read', 'word', 'writing', 'magazine', 'newspaper', 'paper']
\end{Verbatim}

Online LDA:
NPMI=0.291, TU=0.804
\begin{Verbatim}[breaklines=true, fontsize=\tiny]
[ 0 - 0.93333 - 0.29401]:  ['women', 'gay', 'sex', 'girl', 'woman', 'look', 'fashion', 'female', 'wear', 'hair']
[ 1 - 0.95 - 0.35632]:  ['car', 'driver', 'truck', 'race', 'vehicle', 'vehicles', 'zzz_ford', 'wheel', 'driving', 'road']
[ 2 - 0.44 - 0.30627]:  ['point', 'game', 'team', 'play', 'season', 'games', 'zzz_laker', 'shot', 'player', 'basketball']
[ 3 - 0.75 - 0.31578]:  ['election', 'ballot', 'zzz_florida', 'vote', 'votes', 'recount', 'court', 'zzz_al_gore', 'voter', 'count']
[ 4 - 0.88333 - 0.34432]:  ['computer', 'web', 'zzz_internet', 'site', 'online', 'system', 'mail', 'internet', 'sites', 'software']
[ 5 - 1 - 0.27207]:  ['con', 'una', 'las', 'mas', 'por', 'dice', 'como', 'los', 'anos', 'sus']
[ 6 - 0.87 - 0.2716]:  ['study', 'test', 'found', 'data', 'percent', 'researcher', 'evidence', 'result', 'finding', 'scientist']
[ 7 - 0.93333 - 0.28208]:  ['show', 'television', 'network', 'zzz_nbc', 'series', 'viewer', 'zzz_cb', 'zzz_abc', 'broadcast', 'producer']
[ 8 - 0.8 - 0.3368]:  ['court', 'case', 'law', 'lawyer', 'police', 'trial', 'death', 'officer', 'prosecutor', 'prison']
[ 9 - 0.82 - 0.33016]:  ['percent', 'tax', 'economy', 'money', 'cut', 'fund', 'market', 'stock', 'billion', 'economic']
[ 10 - 0.64 - 0.30092]:  ['team', 'player', 'million', 'season', 'contract', 'deal', 'manager', 'agent', 'fan', 'league']
[ 11 - 0.88333 - 0.30657]:  ['need', 'feel', 'word', 'question', 'look', 'right', 'mean', 'kind', 'fact', 'course']
[ 12 - 0.81667 - 0.25913]:  ['religious', 'zzz_american', 'jewish', 'zzz_god', 'religion', 'jew', 'american', 'german', 'political', 'zzz_america']
[ 13 - 0.9 - 0.22142]:  ['cup', 'minutes', 'add', 'tablespoon', 'food', 'oil', 'pepper', 'wine', 'sugar', 'teaspoon']
[ 14 - 0.69167 - 0.25718]:  ['zzz_china', 'zzz_united_states', 'zzz_u_s', 'chinese', 'zzz_japan', 'zzz_american', 'countries', 'foreign', 'japanese', 'official']
[ 15 - 0.72333 - 0.21791]:  ['match', 'tennis', 'set', 'boat', 'won', 'point', 'zzz_pete_sampras', 'final', 'game', 'player']
[ 16 - 0.76667 - 0.16417]:  ['zzz_texas', 'telegram', 'com', 'zzz_austin', 'zzz_houston', 'visit', 'www', 'services', 'web', 'file']
[ 17 - 0.81667 - 0.27063]:  ['room', 'building', 'house', 'look', 'wall', 'floor', 'door', 'home', 'small', 'light']
[ 18 - 0.93333 - 0.42167]:  ['music', 'song', 'band', 'album', 'musical', 'sound', 'singer', 'record', 'jazz', 'show']
[ 19 - 0.95 - 0.28682]:  ['water', 'weather', 'air', 'wind', 'storm', 'feet', 'snow', 'rain', 'mountain', 'miles']
[ 20 - 0.76667 - 0.30412]:  ['military', 'attack', 'war', 'terrorist', 'zzz_u_s', 'laden', 'zzz_american', 'bin', 'zzz_pentagon', 'forces']
[ 21 - 1 - 0.37961]:  ['drug', 'patient', 'doctor', 'health', 'medical', 'disease', 'hospital', 'care', 'cancer', 'treatment']
[ 22 - 0.93333 - 0.26762]:  ['black', 'white', 'flag', 'zzz_black', 'racial', 'irish', 'protest', 'crowd', 'american', 'african']
[ 23 - 0.64 - 0.32174]:  ['company', 'companies', 'million', 'business', 'market', 'percent', 'stock', 'sales', 'analyst', 'customer']
[ 24 - 0.95 - 0.35968]:  ['book', 'magazine', 'newspaper', 'author', 'wrote', 'writer', 'writing', 'published', 'read', 'reader']
[ 25 - 0.72 - 0.2336]:  ['company', 'zzz_enron', 'firm', 'zzz_microsoft', 'million', 'lawsuit', 'companies', 'lawyer', 'case', 'settlement']
[ 26 - 0.5 - 0.25483]:  ['official', 'zzz_fbi', 'government', 'agent', 'terrorist', 'information', 'zzz_cuba', 'attack', 'security', 'zzz_united_states']
[ 27 - 0.83333 - 0.26933]:  ['art', 'zzz_new_york', 'artist', 'century', 'painting', 'show', 'museum', 'collection', 'history', 'director']
[ 28 - 0.88333 - 0.20031]:  ['priest', 'church', 'horse', 'race', 'horses', 'bishop', 'abuse', 'zzz_kentucky_derby', 'pope', 'won']
[ 29 - 0.65833 - 0.22441]:  ['government', 'zzz_mexico', 'country', 'zzz_united_states', 'mexican', 'immigrant', 'border', 'countries', 'president', 'worker']
[ 30 - 0.715 - 0.25157]:  ['goal', 'shot', 'play', 'game', 'king', 'round', 'zzz_tiger_wood', 'player', 'fight', 'win']
[ 31 - 0.88333 - 0.34572]:  ['family', 'home', 'friend', 'father', 'children', 'mother', 'son', 'wife', 'told', 'daughter']
[ 32 - 0.87 - 0.20923]:  ['land', 'town', 'animal', 'farm', 'fish', 'bird', 'local', 'farmer', 'million', 'miles']
[ 33 - 0.70333 - 0.35313]:  ['run', 'game', 'hit', 'inning', 'season', 'yankees', 'games', 'pitcher', 'home', 'zzz_dodger']
[ 34 - 0.95 - 0.20151]:  ['plant', 'mayor', 'zzz_rudolph_giuliani', 'zzz_los_angeles', 'flower', 'garden', 'tree', 'trees', 'zzz_southern_california', 'seed']
[ 35 - 0.95 - 0.28879]:  ['cell', 'scientist', 'research', 'human', 'science', 'stem', 'brain', 'space', 'technology', 'experiment']
[ 36 - 0.75 - 0.14192]:  ['com', 'daily', 'palm', 'beach', 'question', 'statesman', 'american', 'information', 'zzz_eastern', 'austin']
[ 37 - 0.75 - 0.41055]:  ['zzz_george_bush', 'zzz_al_gore', 'president', 'zzz_bush', 'campaign', 'zzz_clinton', 'zzz_white_house', 'presidential', 'zzz_bill_clinton', 'republican']
[ 38 - 0.87 - 0.28937]:  ['school', 'student', 'program', 'teacher', 'children', 'high', 'education', 'college', 'job', 'percent']
[ 39 - 0.825 - 0.29646]:  ['zzz_taliban', 'zzz_afghanistan', 'zzz_pakistan', 'zzz_russia', 'government', 'zzz_russian', 'afghan', 'country', 'zzz_vladimir_putin', 'leader']
[ 40 - 0.825 - 0.33129]:  ['film', 'movie', 'character', 'play', 'actor', 'movies', 'director', 'book', 'zzz_hollywood', 'love']
[ 41 - 0.82 - 0.29137]:  ['oil', 'power', 'plant', 'energy', 'gas', 'prices', 'zzz_california', 'fuel', 'million', 'cost']
[ 42 - 0.875 - 0.40189]:  ['palestinian', 'zzz_israel', 'zzz_israeli', 'peace', 'israeli', 'zzz_yasser_arafat', 'leader', 'israelis', 'official', 'violence']
[ 43 - 0.87 - 0.26507]:  ['food', 'product', 'drink', 'eat', 'weight', 'pound', 'smoking', 'diet', 'percent', 'tobacco']
[ 44 - 0.65 - 0.19242]:  ['com', 'www', 'information', 'site', 'fax', 'web', 'article', 'syndicate', 'visit', 'contact']
[ 45 - 0.78333 - 0.33252]:  ['zzz_olympic', 'games', 'sport', 'medal', 'team', 'gold', 'athletes', 'event', 'won', 'competition']
[ 46 - 0.65833 - 0.20844]:  ['flight', 'plane', 'airport', 'passenger', 'attack', 'zzz_new_york', 'building', 'security', 'worker', 'official']
[ 47 - 0.7 - 0.43799]:  ['campaign', 'political', 'election', 'vote', 'democratic', 'voter', 'zzz_party', 'republican', 'zzz_republican', 'governor']
[ 48 - 0.54 - 0.37981]:  ['game', 'team', 'season', 'play', 'coach', 'yard', 'player', 'football', 'games', 'quarterback']
[ 49 - 0.825 - 0.30119]:  ['bill', 'zzz_congress', 'zzz_bush', 'plan', 'federal', 'government', 'administration', 'law', 'group', 'zzz_senate']
\end{Verbatim}
ProdLDA:
NPMI=0.319, TU=0.668

\begin{Verbatim}[breaklines=true, fontsize=\tiny]
[0-1-0.19456]: ['zzz_discover', 'molecules', 'data', 'zzz_eric_haseltine', 'ion', 'gigahertz', 'computing', 'zzz_dna', 'horsepower', 'molecule']
[1-0.95-0.18699]: ['zzz_focus', 'zzz_mississippi_valley', 'zzz_national_forecast', 'zzz_ohio_valley', 'torque', 'moisture', 'gusty', 'zzz_bernard_gladstone', 'zzz_middle_atlantic', 'zzz_winston_cup']
[2-0.625-0.14594]: ['prosecutor', 'murder', 'distinguishable', 'zzz_bantam', 'zzz_how_to_and_miscellaneous', 'zzz_ray_lewis', 'zzz_my_cheese', 'zzz_bill_phillip', 'zzz_michael_d_orso', 'zzz_fiction']
[3-0.85-0.30119]: ['film', 'comedy', 'movie', 'zzz_fare', 'zzz_judi_dench', 'zzz_billy_bob_thornton', 'zzz_steve_buscemi', 'starring', 'adaptation', 'zzz_cable_cast']
[4-0.9-0.29068]: ['zzz_federal_energy_regulatory_commission', 'zzz_enron', 'administration', 'megawatt', 'zzz_congress', 'utilities', 'lawmaker', 'legislation', 'zzz_southern_california_edison', 'zzz_senate']
[5-1-0.43922]: ['constitutional', 'justices', 'zzz_supreme_court', 'zzz_ruth_bader_ginsburg', 'ruling', 'federal', 'zzz_chief_justice_william_h_rehnquist', 'zzz_justices_sandra_day_o_connor', 'zzz_u_s_circuit_court', 'zzz_florida_supreme_court']
[6-0.9-0.26193]: ['victorian', 'artist', 'sculptures', 'painting', 'garden', 'decorative', 'zzz_post_office_box', 'zzz_gothic', 'boutiques', 'galleries']
[7-0.7-0.50355]: ['zzz_afghanistan', 'qaida', 'zzz_pentagon', 'zzz_taliban', 'bin', 'laden', 'zzz_rumsfeld', 'zzz_osama', 'zzz_defense_secretary_donald_rumsfeld', 'terrorism']
[8-0.50833-0.33875]: ['zzz_federal_reserve', 'prices', 'zzz_fed', 'stock', 'companies', 'rates', 'economy', 'investor', 'billion', 'inflation']
[9-0.50833-0.096284]: ['distinguishable', 'zzz_bantam', 'zzz_how_to_and_miscellaneous', 'bookstores', 'wholesaler', 'zzz_phillip_mcgraw', 'zzz_nonfiction', 'zzz_anne_stephenson', 'zzz_berkley', 'zzz_harpercollin']
[10-0.88333-0.55375]: ['zzz_winston_cup', 'zzz_nascar', 'zzz_daytona', 'championship', 'zzz_dale_earnhardt_jr', 'zzz_nascar_winston_cup', 'zzz_tony_stewart', 'zzz_jeff_gordon', 'restrictor', 'zzz_dale_jarrett']
[11-1-0.3249]: ['chiffon', 'zzz_randolph_duke', 'strapless', 'tulle', 'zzz_valentino', 'beaded', 'dresses', 'couture', 'zzz_hal_rubenstein', 'zzz_versace']
[12-1-0.077046]: ['gutty', 'zzz_ansel_williamson', 'zzz_caracas_cannonball', 'zzz_rosa_hoot', 'zzz_osage_indian', 'zzz_black_gold', 'arrestingly', 'zzz_canonero_ii', 'zzz_david_alexander', 'zzz_aristides']
[13-0.9-0.02489]: ['oped', 'zzz_andy_alexander', 'zzz_kaplow', 'zzz_bessonette', 'andya', 'zzz_eyman', 'zzz_news_questions_q', 'zzz_lee_may_this', 'pica', 'zzz_alan_gordon']
[14-0.51667-0.47783]: ['zzz_republican', 'election', 'zzz_al_gore', 'democratic', 'republican', 'votes', 'democrat', 'voter', 'zzz_gop', 'ballot']
[15-0.43333-0.36588]: ['inning', 'season', 'scored', 'playoff', 'scoring', 'game', 'postseason', 'homer', 'goaltender', 'baseman']
[16-0.40833-0.16908]: ['distinguishable', 'zzz_how_to_and_miscellaneous', 'zzz_bantam', 'zzz_dave_pelzer', 'zzz_nonfiction', 'bookstores', 'wholesaler', 'zzz_lost_boy', 'zzz_fiction', 'zzz_berkley']
[17-0.95-0.35634]: ['user', 'software', 'zzz_microsoft', 'zzz_internet', 'zzz_aol', 'provider', 'consumer', 'download', 'zzz_microsoft_corp', 'zzz_napster']
[18-0.83333-0.3176]: ['zzz_arthur_andersen', 'zzz_justice_department', 'zzz_enron', 'prosecutor', 'auditor', 'zzz_securities', 'defendant', 'zzz_sec', 'litigation', 'plaintiff']
[19-0.56667-0.29763]: ['rebound', 'layup', 'pointer', 'halftime', 'touchdown', 'coach', 'tournament', 'zzz_laker', 'seeded', 'championship']
[20-0.56667-0.45302]: ['zzz_al_gore', 'zzz_republican', 'election', 'democratic', 'votes', 'democrat', 'voter', 'zzz_democrat', 'zzz_bush', 'ballot']
[21-0.95-0.33733]: ['nutrient', 'biotechnology', 'zzz_drug_administration', 'protein', 'pesticides', 'zzz_starlink', 'biotech', 'bacteria', 'genetically', 'species']
[22-0.51667-0.23056]: ['tiene', 'una', 'mas', 'sobre', 'anos', 'representantes', 'publicar', 'comprar', 'tienen', 'ventas']
[23-0.45833-0.36537]: ['companies', 'stock', 'investor', 'analyst', 'company', 'shareholder', 'billion', 'zzz_thomson_financial_first_call', 'zzz_exchange_commission', 'zzz_securities']
[24-0.7-0.37406]: ['zzz_taliban', 'zzz_afghanistan', 'zzz_attorney_general_john_ashcroft', 'zzz_ashcroft', 'qaida', 'zzz_pentagon', 'tribunal', 'terrorism', 'missiles', 'zzz_rumsfeld']
[25-0.36667-0.36524]: ['inning', 'season', 'scoring', 'playoff', 'scored', 'game', 'postseason', 'homer', 'defenseman', 'fielder']
[26-0.71-0.26107]: ['zzz_robert_kagan', 'unilateral', 'zzz_jeane_kirkpatrick', 'zzz_yasser_arafat', 'democracy', 'zzz_israel', 'palestinian', 'zzz_norman_levine', 'zzz_conservative', 'zzz_nlevineiip']
[27-0.65-0.35567]: ['tax', 'trillion', 'surpluses', 'zzz_fed', 'zzz_federal_reserve', 'surplus', 'zzz_social_security', 'inflation', 'economy', 'stimulus']
[28-1-0.17428]: ['zzz_technobuddy_popular', 'zzz_husted', 'zzz_cleere_rudd', 'zzz_netwatch', 'zzz_tech_savvy', 'zzz_technobuddy', 'pageex', 'zzz_bizmags_a', 'zzz_texas_consumer_q', 'zzz_tech_tools_software']
[29-0.31833-0.55049]: ['palestinian', 'zzz_israeli', 'zzz_yasser_arafat', 'zzz_west_bank', 'israelis', 'zzz_israel', 'militant', 'zzz_prime_minister_ariel_sharon', 'zzz_gaza_strip', 'zzz_palestinian']
[30-0.59167-0.3032]: ['companies', 'analyst', 'automaker', 'stock', 'consumer', 'zzz_daimlerchrysler', 'zzz_first_call_thomson_financial', 'billion', 'zzz_daimlerchrysler_ag', 'company']
[31-0.9-0.18023]: ['zzz_doubles', 'zzz_eat', 'painting', 'artist', 'decor', 'designer', 'painter', 'zzz_sightseeing', 'sculpture', 'spangly']
[32-0.46667-0.31749]: ['mas', 'sobre', 'anos', 'una', 'como', 'otros', 'ventas', 'tienen', 'sus', 'todo']
[33-0.36833-0.54809]: ['zzz_israeli', 'palestinian', 'zzz_west_bank', 'israelis', 'zzz_palestinian', 'zzz_yasser_arafat', 'militant', 'zzz_israel', 'zzz_prime_minister_ariel_sharon', 'zzz_gaza']
[34-0.5-0.28574]: ['tablespoon', 'teaspoon', 'saucepan', 'pepper', 'cholesterol', 'cup', 'chopped', 'garlic', 'sodium', 'browned']
[35-0.41667-0.30011]: ['mas', 'anos', 'sobre', 'tiene', 'sus', 'como', 'ventas', 'todo', 'representantes', 'una']
[36-0.9-0.1902]: ['toder', 'zzz_tom_oder', 'andya', 'zzz_andy_alexander', 'artd', 'tduncan', 'zzz_dalglish', 'zzz_todd_duncan', 'zzz_rick_christie', 'rickc']
[37-0.56667-0.32578]: ['pointer', 'layup', 'touchdown', 'halftime', 'tournament', 'semifinal', 'championship', 'coach', 'zzz_ncaa', 'seeded']
[38-0.43333-0.42183]: ['season', 'playoff', 'game', 'inning', 'scoring', 'scored', 'defenseman', 'scoreless', 'games', 'shutout']
[39-0.50833-0.086353]: ['distinguishable', 'zzz_how_to_and_miscellaneous', 'zzz_dave_pelzer', 'zzz_bantam', 'zzz_lost_boy', 'zzz_phillip_mcgraw', 'zzz_hyperion', 'zzz_nonfiction', 'zzz_bill_phillip', 'zzz_robert_atkin']
[40-0.635-0.37871]: ['zzz_barak', 'zzz_israel', 'zzz_yasser_arafat', 'palestinian', 'zzz_ariel_sharon', 'zzz_prime_minister_ehud_barak', 'parliamentary', 'zzz_pri', 'israelis', 'democracy']
[41-0.45-0.35843]: ['tablespoon', 'teaspoon', 'cup', 'saucepan', 'pepper', 'garlic', 'cloves', 'onion', 'chopped', 'minced']
[42-0.61667-0.40298]: ['zzz_medicare', 'tax', 'zzz_republican', 'zzz_social_security', 'prescription', 'trillion', 'zzz_senate', 'zzz_house_republican', 'republican', 'democrat']
[43-0.85-0.26198]: ['film', 'movie', 'album', 'comedy', 'actress', 'zzz_merle_ginsberg', 'genre', 'debut', 'zzz_nicole_kidman', 'musical']
[44-0.95-0.34236]: ['anthrax', 'spores', 'inhalation', 'antibiotic', 'zzz_drug_administration', 'zzz_disease_control', 'zzz_fda', 'zzz_cdc', 'zzz_ernesto_blanco', 'zzz_cipro']
[45-0.3-0.41301]: ['season', 'inning', 'playoff', 'game', 'scoring', 'scored', 'games', 'coach', 'postseason', 'defenseman']
[46-0.45833-0.3414]: ['companies', 'stock', 'shareholder', 'investor', 'analyst', 'company', 'merger', 'billion', 'zzz_at', 'zzz_securities']
[47-0.45-0.31771]: ['tablespoon', 'teaspoon', 'saucepan', 'cholesterol', 'pepper', 'parsley', 'cup', 'garlic', 'cloves', 'onion']
[48-1-0.4116]: ['bishop', 'priest', 'catholic', 'zzz_vatican', 'zzz_cardinal_bernard_f_law', 'jew', 'religious', 'zzz_christianity', 'zzz_roman_catholic', 'dioceses']
[49-0.36833-0.55235]: ['zzz_israeli', 'palestinian', 'zzz_west_bank', 'militant', 'zzz_yasser_arafat', 'zzz_gaza_strip', 'israelis', 'zzz_ramallah', 'zzz_palestinian', 'zzz_israel']
\end{Verbatim}

NTM-R:
NPMI=0.218, TU=0.874

\begin{Verbatim}[breaklines=true, fontsize=\tiny]
[0-1-0.16553]: ['zzz_dow_jones', 'zzz_first_call_thomson_financial', 'zzz_thomson_financial_first_call', 'composite', 'zzz_tom_walker', 'indexes', 'zzz_sach', 'annualized', 'zzz_fed', 'zzz_prudential_securities']
[1-0.69769-0.1441]: ['zzz_held', 'advisory', 'redevelopment', 'renovated', 'premature', 'occupancy', 'sicheianytimes', 'suites', 'una', 'zzz_atentamente']
[2-0.95-0.40759]: ['zzz_playstation', 'gameplay', 'zzz_we_want', 'zzz_dreamcast', 'gamer', 'zzz_national_geographic_today_list', 'ps2', 'zzz_publish_a_story', 'zzz_natgeo_list', 'zzz_know_about']
[3-0.95-0.20454]: ['zzz_new_hampshire', 'zzz_budget_office', 'caucuses', 'uninsured', 'zzz_sooner', 'zzz_john_mccain', 'zzz_south_carolina', 'zzz_mccain', 'milligram', 'seeded']
[4-0.95-0.20167]: ['studios', 'zzz_dvd', 'zzz_vh', 'zzz_recording_industry_association', 'soundtrack', 'zzz_paramount', 'zzz_fare', 'zzz_dreamwork', 'zzz_metallica', 'zzz_warner_brother']
[5-0.8-0.11388]: ['zzz_joseph_ellis', 'zzz_lance_armstrong', 'zzz_my_cheese', 'zzz_bill_phillip', 'zzz_crown', 'clinton', 'zzz_michael_d_orso', 'zzz_doubleday', 'zzz_mitch_albom', 'noticias']
[6-0.8-0.092516]: ['zzz_mike_scioscia', 'minced', 'zzz_secret', 'coarsely', 'zzz_scribner', 'zzz_my_cheese', 'combine', 'zzz_chronicle', 'zzz_mitch_albom', 'zzz_michael_d_orso']
[7-1-0.24547]: ['zzz_touch_tone', 'astrascope', 'zzz_news_america', 'zzz_xii', 'zzz_sagittarius', 'zzz_capricorn', 'zzz_clip_and_save', 'zzz_birthday', 'zzz_aquarius', 'zzz_pisces']
[8-0.48269-0.17408]: ['undatelined', 'zzz_held', 'misidentified', 'zzz_attn_editor', 'zzz_boston_globe', 'zzz_killed', 'herbert', 'zzz_states_news_service', 'publication', 'dowd']
[9-1-0.37552]: ['megawatt', 'zzz_opec', 'zzz_petroleum_exporting_countries', 'renewable', 'zzz_federal_communications_commission', 'refineries', 'deregulation', 'zzz_federal_energy_regulatory_commission', 'deregulated', 'pipelines']
[10-0.95-0.19122]: ['species', 'zzz_anne_stephenson', 'ecological', 'habitat', 'archaeologist', 'mammal', 'biologist', 'genes', 'zzz_duplication', 'conservationist']
[11-1-0.16745]: ['zzz_phoenix', 'zzz_rudolph_giuliani', 'zzz_army', 'zzz_brooklyn', 'station', 'stadium', 'officer', 'apartment', 'zzz_kansas_city', 'zzz_manhattan']
[12-1-0.076741]: ['zzz_technobuddy_popular', 'zzz_husted', 'zzz_netwatch', 'zzz_cleere_rudd', 'zzz_tech_savvy', 'zzz_technobuddy', 'computing', 'hacker', 'zzz_greig', 'zzz_texas_consumer_q']
[13-1-0.33571]: ['zzz_national_transportation_safety_board', 'zzz_american_airlines_flight', 'zzz_defense_secretary_donald_rumsfeld', 'zzz_federal_aviation_administration', 'zzz_joint_chief', 'zzz_david_wood', 'zzz_rumsfeld', 'zzz_u_s_central_command', 'zzz_pentagon', 'cockpit']
[14-0.95-0.18476]: ['zzz_mccain_feingold', 'zzz_common_cause', 'zzz_ir', 'zzz_recording_industry_association', 'zzz_internal_revenue_service', 'taxable', 'deduction', 'debtor', 'infringement', 'zzz_russell_feingold']
[15-0.95-0.38946]: ['zzz_troy_glaus', 'zzz_mike_scioscia', 'zzz_david_eckstein', 'zzz_edison_field', 'zzz_garret_anderson', 'zzz_angel', 'psychiatry', 'zzz_adam_kennedy', 'zzz_troy_percival', 'zzz_scott_spiezio']
[16-1-0.38596]: ['zzz_northern_alliance', 'zzz_tajik', 'zzz_pashtun', 'zzz_uzbek', 'warlord', 'zzz_kashmir', 'zzz_taliban', 'zzz_kabul', 'zzz_afghan', 'caves']
[17-1-0.47243]: ['winemaker', 'wines', 'winery', 'vineyard', 'wineries', 'zzz_publisher', 'grape', 'tannin', 'grapes', 'zzz_harry_potter_and_the_sorcerer_s_stone']
[18-0.95-0.2147]: ['zzz_o_neal', 'zzz_kobe_bryant', 'zzz_robert_horry', 'zzz_phil_jackson', 'zzz_shaquille_o_neal', 'psychiatrist', 'screenplay', 'sexuality', 'zzz_derek_fisher', 'zzz_anne_stephenson']
[19-0.90769-0.14241]: ['zzz_held', 'goalkeeper', 'midfielder', 'zzz_ml', 'midfield', 'referee', 'zzz_olympian', 'zzz_dick_ebersol', 'zzz_galaxy', 'zzz_nbc_sport']
[20-0.95-0.18766]: ['fue', 'inversiones', 'gracias', 'las', 'latinoamericanas', 'angulos', 'finanzas', 'transmitida', 'backhand', 'industrias']
[21-0.90769-0.44435]: ['zzz_gaza_strip', 'zzz_nablus', 'oslo', 'zzz_palestinian_controlled', 'zzz_hebron', 'zzz_ramallah', 'zzz_west_bank', 'fatah', 'zzz_held', 'zzz_gaza']
[22-0.75769-0.066675]: ['zzz_karl_horwitz', 'zzz_lifebeat', 'shopper', 'homeowner', 'telex', 'zzz_nonsubscriber', 'pet', 'conditioner', 'zzz_dru_sefton', 'zzz_held']
[23-0.95-0.2593]: ['filibuster', 'bipartisanship', 'zzz_lott', 'zzz_pri', 'zzz_tom_daschle', 'zzz_mccain', 'zzz_daschle', 'zzz_sen_tom_daschle', 'centrist', 'zzz_jefford']
[24-1-0.38059]: ['holes', 'fairway', 'birdies', 'birdied', 'birdie', 'bogey', 'zzz_valentino', 'putted', 'putt', 'designation']
[25-0.85-0.28633]: ['zzz_chechnya', 'zzz_chechen', 'zzz_boris_yeltsin', 'zzz_vladimir_putin', 'choreographer', 'choreography', 'dancer', 'zzz_russian', 'costumes', 'zzz_kremlin']
[26-0.85-0.2319]: ['zzz_kgb', 'zzz_kremlin', 'zzz_jiang_zemin', 'zzz_boris_yeltsin', 'zzz_hainan', 'espionage', 'zzz_alberto_fujimori', 'zzz_wen_ho_lee', 'zzz_vladimir_putin', 'zzz_taiwan']
[27-0.54936-0.11824]: ['zzz_held', 'misidentified', 'zzz_attn_editor', 'zzz_killed', 'obituary', 'misspelled', 'zzz_washington_datelined', 'slugged', 'polygraph', 'publication']
[28-0.95-0.24534]: ['segregation', 'ordination', 'zzz_lazaro_gonzalez', 'protestant', 'dioceses', 'zzz_anthony_kennedy', 'parishes', 'zzz_juan_miguel_gonzalez', 'seminaries', 'priesthood']
[29-0.8-0.23328]: ['zzz_cox_news_campaign', 'zzz_jeb_bush', 'zzz_rev_al_sharpton', 'zzz_state_katherine_harris', 'chad', 'zzz_miami_dade', 'canvassing', 'zzz_super_tuesday', 'absentee', 'zzz_pat_buchanan']
[30-0.56603-0.15392]: ['zzz_held', 'zzz_attn_editor', 'undatelined', 'zzz_washington_datelined', 'zzz_anaconda', 'zzz_boston_globe', 'zzz_taloqan', 'zzz_international_space_station', 'zzz_killed', 'crewmen']
[31-1-0.16598]: ['zzz_gibsonburg', 'eschuett', 'nwonline', 'zzz_west_madison', 'zzz_elizabeth_schuett', 'zzz_marty_kurzfeld', 'zzz_lester', 'fumble', 'zzz_lester_pozz', 'downfield']
[32-0.93333-0.18509]: ['zzz_boston_globe', 'zzz_ralph_nader', 'jobless', 'employer', 'tuition', 'productivity', 'misstated', 'advertiser', 'tonight', 'recession']
[33-0.56436-0.10917]: ['zzz_held', 'advisory', 'premature', 'publication', 'sicheianytimes', 'guard', 'internacional', 'representantes', 'zzz_cada', 'industria']
[34-0.9-0.19004]: ['manhunt', 'arraignment', 'detectives', 'released', 'zzz_karachi', 'gunshot', 'semiautomatic', 'zzz_juan_miguel_gonzalez', 'arraigned', 'slaying']
[35-0.90769-0.28004]: ['zzz_held', 'zzz_david_pelletier', 'zzz_ottavio_cinquanta', 'zzz_jamie_sale', 'zzz_jacques_rogge', 'zzz_bob_arum', 'zzz_international_skating_union', 'doping', 'zzz_anton_sikharulidze', 'zzz_u_s_olympic_committee']
[36-0.51436-0.13181]: ['publication', 'premature', 'zzz_held', 'advisory', 'guard', 'send', 'released', 'zzz_broadway', 'zzz_lance_armstrong', 'zzz_tennessee_valley']
[37-1-0.33657]: ['zzz_fda', 'zzz_d_vt', 'zzz_security_council', 'zzz_ashcroft', 'zzz_senate_judiciary_committee', 'zzz_drug_administration', 'zzz_judiciary_committee', 'statutory', 'justices', 'zzz_attorney_general_john_ashcroft']
[38-1-0.12492]: ['zzz_focus', 'zzz_lost_boy', 'zzz_diet_revolution', 'zzz_dave_pelzer', 'zzz_jared_diamond', 'zzz_don_miguel_ruiz', 'zzz_seat', 'physiologist', 'zzz_robert_kiyosaki', 'zzz_soul']
[39-1-0.26288]: ['zzz_north_american_free_trade_agreement', 'migrant', 'saharan', 'zzz_nafta', 'undocumented', 'zzz_vicente_fox', 'afghan', 'zzz_naturalization_service', 'trafficker', 'zzz_revolutionary_party']
[40-0.95-0.17834]: ['zzz_teepen_column', 'zzz_schuett', 'carbohydrates', 'natgeo', 'zzz_national_geographic_today', 'zzz_nethaway', 'additionally', 'zzz_mccarty_column', 'zzz_mccarty', 'zzz_publish_a_story']
[41-0.95-0.059983]: ['zzz_andy_alexander', 'andya', 'toder', 'zzz_tom_oder', 'zzz_dalglish', 'artd', 'zzz_rick_christie', 'zzz_carl_rauscher', 'crausher', 'eta']
[42-1-0.14993]: ['zzz_red_sox', 'unionist', 'zzz_bill_belichick', 'zzz_david_trimble', 'zzz_richard_riordan', 'zzz_southern_california_edison', 'zzz_sinn_fein', 'zzz_pacific_gas', 'zzz_carl_everett', 'walkout']
[43-0.61436-0.13404]: ['zzz_held', 'premature', 'advisory', 'publication', 'periodicos', 'llamar', 'latinoamericanas', 'cubriendo', 'semanal', 'cubrir']
[44-0.85-0.16905]: ['zzz_karl_horwitz', 'telex', 'zzz_isabel_amorim_sicherle', 'zzz_governor_bush', 'zzz_nonsubscriber', 'zzz_ariel_sharon', 'zzz_ehud_barak', 'zzz_judaism', 'zzz_ana_pena', 'zzz_camp_david']
[45-0.95-0]: ['rickc', 'zzz_paul_foutch', 'zzz_firestone', 'pfoutch', 'zzz_layout_s_done', 'zzz_news_questions_q', 'paginated', 'zzz_bessonette', 'zzz_rick_christie', 'zzz_langhenry']
[46-0.8-0.26723]: ['canvassing', 'dimpled', 'zzz_miami_dade', 'zzz_broward', 'zzz_state_katherine_harris', 'chad', 'undervotes', 'recount', 'zzz_volusia', 'layup']
[47-0.90769-0.43632]: ['zzz_wba', 'zzz_oscar_de_la_hoya', 'zzz_ioc', 'zzz_held', 'zzz_wbc', 'zzz_international_boxing_federation', 'middleweight', 'zzz_ibf', 'zzz_world_boxing_association', 'welterweight']
[48-0.38936-0.11527]: ['zzz_attn_editor', 'zzz_held', 'misidentified', 'zzz_washington_datelined', 'zzz_los_angeles_daily_new', 'undatelined', 'premature', 'advisory', 'publication', 'imprecisely']
[49-1-0.32124]: ['zzz_pete_carroll', 'zzz_cleveland_brown', 'lineman', 'zzz_bill_parcell', 'cornerback', 'zzz_bud_selig', 'zzz_offensive', 'zzz_trojan', 'zzz_sugar_bowl', 'zzz_al_groh']
\end{Verbatim}

W-LDA:
NPMI=0.356, TU=0.998

\begin{Verbatim}[breaklines=true, fontsize=\tiny]
[0-1-0.3425]: ['touchdown', 'interception', 'cornerback', 'quarterback', 'patriot', 'linebacker', 'receiver', 'yard', 'zzz_cowboy', 'zzz_ram']
[1-1-0.27811]: ['como', 'comprar', 'una', 'tiene', 'mas', 'distinguishable', 'publicar', 'sobre', 'tienen', 'prohibitivo']
[2-1-0.38656]: ['zzz_elian', 'zzz_juan_miguel_gonzalez', 'zzz_cuba', 'cuban', 'zzz_elian_gonzalez', 'zzz_fidel_castro', 'zzz_cuban_american', 'zzz_little_havana', 'zzz_lazaro_gonzalez', 'exiles']
[3-1-0.38253]: ['zzz_red_sox', 'yankees', 'zzz_world_series', 'zzz_baseball', 'baseball', 'outfielder', 'zzz_dan_duquette', 'zzz_met', 'clubhouse', 'zzz_george_steinbrenner']
[4-1-0.24506]: ['zzz_microsoft', 'antitrust', 'zzz_judge_thomas_penfield_jackson', 'monopoly', 'monopolist', 'breakup', 'remedy', 'browser', 'zzz_u_s_district_judge_thomas_penfield_jackson', 'zzz_fcc']
[5-1-0.26442]: ['zzz_security_council', 'rebel', 'colombian', 'zzz_iraq', 'zzz_colombia', 'zzz_u_n', 'zzz_congo', 'iraqi', 'zzz_andres_pastrana', 'guerrillas']
[6-1-0.4521]: ['zzz_john_mccain', 'zzz_mccain', 'zzz_bill_bradley', 'zzz_al_gore', 'primaries', 'zzz_governor_bush', 'zzz_new_hampshire', 'caucuses', 'zzz_george_bush', 'zzz_bob_jones_university']
[7-1-0.14137]: ['zzz_bernard_gladstone', 'moisture', 'astronomer', 'species', 'zzz_caption', 'zzz_focus', 'bloom', 'particles', 'shrub', 'soil']
[8-1-0.29295]: ['couture', 'dresses', 'paginated', 'skirt', 'chiffon', 'designer', 'fashion', 'beaded', 'gown', 'zzz_layout_s_done']
[9-1-0.23167]: ['zzz_falun_gong', 'unionist', 'zzz_sinn_fein', 'zzz_islamic', 'zzz_northern_ireland', 'zzz_ulster', 'zzz_islam', 'reformist', 'zzz_ira', 'iranian']
[10-1-0.56851]: ['zzz_israeli', 'zzz_yasser_arafat', 'palestinian', 'zzz_palestinian', 'zzz_west_bank', 'israelis', 'zzz_gaza', 'zzz_israel', 'zzz_barak', 'zzz_ramallah']
[11-1-0.1902]: ['zzz_andy_alexander', 'andya', 'artd', 'zzz_tom_oder', 'toder', 'zzz_dalglish', 'tduncan', 'zzz_todd_duncan', 'rickc', 'zzz_rick_christie']
[12-0.95-0.36367]: ['zzz_fbi', 'indictment', 'zzz_justice_department', 'prosecutor', 'investigation', 'pardon', 'indicted', 'investigator', 'hijacker', 'wrongdoing']
[13-1-0.37572]: ['patient', 'embryos', 'cell', 'genes', 'gene', 'embryo', 'symptom', 'zzz_national_institutes', 'disease', 'tumor']
[14-1-0.49672]: ['zzz_taliban', 'zzz_northern_alliance', 'afghan', 'zzz_kabul', 'zzz_afghanistan', 'zzz_pakistan', 'zzz_kandahar', 'zzz_pashtun', 'bin', 'laden']
[15-1-0.4287]: ['defenseman', 'puck', 'goalie', 'goaltender', 'zzz_nhl', 'zzz_stanley_cup', 'zzz_andy_murray', 'zzz_ken_hitchcock', 'zzz_ziggy_palffy', 'defensemen']
[16-1-0.30709]: ['ballot', 'recount', 'canvassing', 'zzz_florida_supreme_court', 'absentee', 'elector', 'zzz_miami_dade', 'zzz_state_katherine_harris', 'zzz_broward', 'votes']
[17-1-0.36916]: ['zzz_enron', 'zzz_securities', 'zzz_enron_corp', 'zzz_exchange_commission', 'auditor', 'accounting', 'zzz_arthur_andersen', 'zzz_sec', 'creditor', 'bankruptcy']
[18-1-0.37785]: ['missile', 'zzz_north_korea', 'zzz_anti_ballistic_missile_treaty', 'warhead', 'zzz_abm', 'zzz_vladimir_putin', 'ballistic', 'missiles', 'zzz_taiwan', 'treaty']
[19-1-0.2819]: ['zzz_ncaa', 'zzz_florida_state', 'athletic', 'zzz_bc', 'zzz_usc', 'pac', 'zzz_bowl_championship_series', 'zzz_ucla', 'zzz_big_east', 'coaches']
[20-1-0.47272]: ['album', 'guitarist', 'guitar', 'song', 'band', 'bassist', 'songwriter', 'ballad', 'zzz_grammy', 'singer']
[21-1-0.28145]: ['zzz_cb', 'zzz_nbc', 'zzz_abc', 'sitcom', 'zzz_upn', 'zzz_cable_cast', 'zzz_fare', 'episodes', 'zzz_craig_kilborn', 'zzz_fox']
[22-1-0.37844]: ['medal', 'zzz_olympic', 'medalist', 'swimmer', 'freestyle', 'athletes', 'zzz_olympian', 'zzz_sydney', 'zzz_winter_olympic', 'gold']
[23-1-0.3913]: ['film', 'movie', 'starring', 'zzz_oscar', 'screenplay', 'actor', 'filmmaking', 'comedy', 'actress', 'zzz_oscar_winning']
[24-1-0.53088]: ['zzz_tiger_wood', 'putt', 'birdie', 'bogey', 'zzz_pga', 'birdies', 'par', 'zzz_u_s_open', 'tee', 'fairway']
[25-1-0.2886]: ['composer', 'repertory', 'literary', 'musical', 'conductor', 'choreographer', 'choreography', 'playwright', 'orchestra', 'zzz_anne_stephenson']
[26-1-0.29966]: ['zzz_fed', 'zzz_dow_jones', 'zzz_nasdaq', 'index', 'zzz_federal_reserve', 'composite', 'indexes', 'zzz_dow', 'inflation', 'stock']
[27-1-0.15688]: ['zzz_doubles', 'breakfast', 'zzz_nicholas', 'lodging', 'sleigh', 'dining', 'zzz_marty_kurzfeld', 'inn', 'excursion', 'sightseeing']
[28-1-0.26239]: ['zzz_at', 'merger', 'zzz_time_warner', 'zzz_compaq', 'acquisition', 'zzz_aol_time_warner', 'zzz_aol', 'cent', 'shareholder', 'zzz_first_call_thomson_financial']
[29-1-0.5289]: ['justices', 'zzz_supreme_court', 'zzz_chief_justice_william_h_rehnquist', 'zzz_ruth_bader_ginsburg', 'unconstitutional', 'zzz_justice_antonin_scalia', 'zzz_u_s_circuit_court', 'constitutional', 'zzz_justice_sandra_day_o_connor', 'zzz_first_amendment']
[30-1-0.45378]: ['inning', 'zzz_dodger', 'homer', 'zzz_rbi', 'bullpen', 'grounder', 'fastball', 'zzz_mike_scioscia', 'zzz_anaheim_angel', 'hander']
[31-1-0.33095]: ['zzz_opec', 'electricity', 'barrel', 'zzz_petroleum_exporting_countries', 'emission', 'gasoline', 'megawatt', 'utilities', 'gas', 'deregulation']
[32-1-0.40058]: ['tax', 'zzz_medicare', 'zzz_social_security', 'surplus', 'surpluses', 'trillion', 'taxes', 'zzz_budget', 'zzz_budget_office', 'stimulus']
[33-1-0.44259]: ['priest', 'bishop', 'parish', 'zzz_cardinal_bernard_f_law', 'zzz_vatican', 'church', 'clergy', 'catholic', 'priesthood', 'parishes']
[34-1-0.43725]: ['zzz_slobodan_milosevic', 'zzz_serbian', 'zzz_serb', 'zzz_yugoslav', 'zzz_serbia', 'zzz_belgrade', 'albanian', 'zzz_kosovo', 'zzz_vojislav_kostunica', 'submarine']
[35-1-0.40499]: ['zzz_winston_cup', 'zzz_daytona', 'colt', 'lap', 'racing', 'zzz_kentucky_derby', 'zzz_nascar', 'zzz_jeff_gordon', 'zzz_dale_earnhardt', 'zzz_preakness']
[36-1-0.43098]: ['torque', 'horsepower', 'liter', 'sedan', 'zzz_suv', 'zzz_royal_ford', 'rear', 'engine', 'wheel', 'cylinder']
[37-1-0.39637]: ['airport', 'airlines', 'passenger', 'zzz_federal_aviation_administration', 'airline', 'traveler', 'flight', 'fares', 'aviation', 'baggage']
[38-1-0.45111]: ['painting', 'curator', 'exhibition', 'sculpture', 'museum', 'sculptures', 'galleries', 'zzz_modern_art', 'painter', 'gallery']
[39-1-0.49464]: ['zzz_laker', 'zzz_phil_jackson', 'zzz_nba', 'zzz_o_neal', 'zzz_shaquille_o_neal', 'zzz_kobe_bryant', 'zzz_shaq', 'zzz_knick', 'zzz_los_angeles_laker', 'zzz_kobe']
[40-1-0.25086]: ['layoff', 'customer', 'employer', 'worker', 'manufacturing', 'supplier', 'retail', 'rent', 'retailer', 'shopper']
[41-1-0.22187]: ['acres', 'environmentalist', 'forest', 'environmental', 'land', 'germ', 'radioactive', 'timber', 'biological', 'wildlife']
[42-1-0.14201]: ['zzz_playstation', 'gamer', 'zzz_birthday', 'astrascope', 'zzz_news_america', 'zzz_touch_tone', 'brompton', 'zzz_clip_and_save', 'zzz_astrologer', 'zzz_xii']
[43-1-0.41952]: ['pointer', 'layup', 'jumper', 'rebound', 'outrebounded', 'halftime', 'fouled', 'foul', 'basket', 'buzzer']
[44-1-0.34321]: ['tablespoon', 'teaspoon', 'cup', 'pepper', 'chopped', 'saucepan', 'onion', 'garlic', 'oven', 'sauce']
[45-1-0.40768]: ['megabytes', 'user', 'download', 'modem', 'desktop', 'mp3', 'software', 'computer', 'digital', 'files']
[46-0.95-0.33261]: ['juror', 'execution', 'jury', 'murder', 'inmates', 'defendant', 'prosecutor', 'robbery', 'penalty', 'zzz_timothy_mcveigh']
[47-1-0.39516]: ['zzz_senate', 'zzz_house_republican', 'bill', 'zzz_mccain_feingold', 'zzz_d_wis', 'amendment', 'filibuster', 'zzz_r_ariz', 'unregulated', 'legislation']
[48-1-0.3287]: ['zzz_aid', 'zzz_hiv', 'infected', 'zzz_fda', 'genetically', 'epidemic', 'crop', 'medicines', 'zzz_world_health_organization', 'drug']
[49-1-0.35675]: ['student', 'teacher', 'curriculum', 'school', 'classroom', 'math', 'standardized', 'colleges', 'educator', 'faculty']
\end{Verbatim}

\subsection{Topic words on Wikitext-103:}
LDA Collapsed Gibbs sampling:
NPMI=0.289, TU=0.754
\begin{Verbatim}[breaklines=true, fontsize=\tiny]
[ 0 - 0.8 - 0.27197]:  ['design', 'model', 'vehicle', 'coin', 'engine', 'version', 'production', 'power', 'car', 'machine']
[ 1 - 0.80625 - 0.21883]:  ['specie', 'bird', 'ha', 'plant', 'brown', 'tree', 'white', 'nest', 'genus', 'fruit']
[ 2 - 0.78333 - 0.30133]:  ['film', 'role', 'award', 'production', 'movie', 'actor', 'million', 'director', 'scene', 'release']
[ 3 - 0.80333 - 0.25923]:  ['al', 'empire', 'city', 'emperor', 'army', 'roman', 'greek', 'byzantine', 'war', 'arab']
[ 4 - 0.85625 - 0.2574]:  ['star', 'planet', 'earth', 'sun', 'mass', 'space', 'moon', 'light', 'ha', 'surface']
[ 5 - 0.95 - 0.32639]:  ['storm', 'tropical', 'hurricane', 'wind', 'km', 'cyclone', 'damage', 'mph', 'day', 'depression']
[ 6 - 0.9 - 0.35437]:  ['child', 'family', 'life', 'woman', 'father', 'mother', 'friend', 'death', 'wife', 'son']
[ 7 - 0.85 - 0.3176]:  ['police', 'day', 'people', 'death', 'prison', 'murder', 'report', 'killed', 'trial', 'reported']
[ 8 - 0.82 - 0.28613]:  ['german', 'war', 'soviet', 'germany', 'russian', 'french', 'polish', 'poland', 'russia', 'france']
[ 9 - 0.78958 - 0.28276]:  ['god', 'church', 'christian', 'temple', 'religious', 'century', 'religion', 'text', 'ha', 'saint']
[ 10 - 0.68667 - 0.24509]:  ['american', 'state', 'war', 'york', 'washington', 'united', 'virginia', 'john', 'fort', 'general']
[ 11 - 0.8 - 0.29384]:  ['king', 'henry', 'england', 'john', 'royal', 'edward', 'william', 'english', 'son', 'scotland']
[ 12 - 0.68667 - 0.31251]:  ['match', 'championship', 'event', 'world', 'team', 'won', 'title', 'wrestling', 'champion', 'final']
[ 13 - 0.75 - 0.31763]:  ['island', 'ship', 'french', 'british', 'sea', 'navy', 'captain', 'port', 'fleet', 'coast']
[ 14 - 0.95 - 0.31526]:  ['chinese', 'china', 'japanese', 'japan', 'vietnam', 'singapore', 'kong', 'philippine', 'government', 'vietnamese']
[ 15 - 0.85625 - 0.17395]:  ['food', 'ice', 'harry', 'restaurant', 'ha', 'product', 'wine', 'meat', 'king', 'potter']
[ 16 - 0.8 - 0.36086]:  ['state', 'president', 'election', 'republican', 'campaign', 'vote', 'senate', 'governor', 'house', 'party']
[ 17 - 0.86667 - 0.25547]:  ['route', 'road', 'highway', 'state', 'county', 'north', 'ny', 'east', 'street', 'south']
[ 18 - 0.72 - 0.26666]:  ['ship', 'gun', 'fleet', 'mm', 'inch', 'war', 'german', 'class', 'navy', 'ton']
[ 19 - 0.95 - 0.35007]:  ['air', 'aircraft', 'flight', 'force', 'no.', 'squadron', 'fighter', 'pilot', 'operation', 'wing']
[ 20 - 0.77 - 0.24508]:  ['race', 'stage', 'team', 'lap', 'car', 'point', 'driver', 'lead', 'won', 'place']
[ 21 - 0.86667 - 0.25987]:  ['san', 'spanish', 'la', 'california', 'texas', 'mexico', 'state', 'el', 'american', 'francisco']
[ 22 - 0.475 - 0.39568]:  ['album', 'song', 'music', 'track', 'released', 'record', 'single', 'release', 'chart', 'number']
[ 23 - 0.62292 - 0.27391]:  ['century', 'castle', 'wall', 'building', 'built', 'church', 'stone', 'house', 'site', 'ha']
[ 24 - 0.73958 - 0.22785]:  ['element', 'nuclear', 'ha', 'energy', 'metal', 'number', 'form', 'gas', 'group', 'chemical']
[ 25 - 0.55667 - 0.42035]:  ['club', 'match', 'season', 'team', 'league', 'cup', 'goal', 'final', 'scored', 'player']
[ 26 - 0.9 - 0.44177]:  ['force', 'army', 'division', 'battle', 'battalion', 'attack', 'infantry', 'troop', 'brigade', 'regiment']
[ 27 - 0.75333 - 0.23014]:  ['british', 'london', 'australian', 'australia', 'war', 'wale', 'royal', 'victoria', 'world', 'britain']
[ 28 - 0.85625 - 0.22846]:  ['black', 'white', 'horse', 'red', 'flag', 'dog', 'blue', 'breed', 'green', 'ha']
[ 29 - 0.55667 - 0.31724]:  ['game', 'team', 'season', 'yard', 'point', 'player', 'play', 'coach', 'goal', 'football']
[ 30 - 0.70833 - 0.41145]:  ['band', 'song', 'rock', 'album', 'guitar', 'tour', 'music', 'record', 'group', 'recording']
[ 31 - 0.52125 - 0.24719]:  ['episode', 'series', 'season', 'character', 'ha', 'scene', 'television', 'viewer', 'michael', 'rating']
[ 32 - 0.78958 - 0.24362]:  ['ha', 'language', 'word', 'theory', 'social', 'world', 'term', 'human', 'form', 'idea']
[ 33 - 0.81667 - 0.34101]:  ['court', 'law', 'state', 'case', 'act', 'legal', 'justice', 'judge', 'decision', 'united']
[ 34 - 0.75625 - 0.19829]:  ['specie', 'animal', 'ha', 'female', 'male', 'shark', 'large', 'long', 'population', 'water']
[ 35 - 0.9 - 0.31353]:  ['book', 'work', 'published', 'story', 'art', 'writing', 'painting', 'writer', 'poem', 'magazine']
[ 36 - 0.71667 - 0.25432]:  ['building', 'park', 'city', 'street', 'house', 'museum', 'foot', 'room', 'hotel', 'center']
[ 37 - 0.95 - 0.34335]:  ['station', 'line', 'train', 'bridge', 'railway', 'service', 'passenger', 'construction', 'built', 'tunnel']
[ 38 - 0.85 - 0.30801]:  ['school', 'university', 'student', 'college', 'program', 'member', 'education', 'national', 'research', 'science']
[ 39 - 0.61667 - 0.29633]:  ['government', 'party', 'political', 'minister', 'member', 'national', 'country', 'leader', 'state', 'power']
[ 40 - 0.64333 - 0.28726]:  ['game', 'season', 'league', 'run', 'baseball', 'hit', 'home', 'team', 'series', 'major']
[ 41 - 0.60125 - 0.24256]:  ['character', 'series', 'story', 'man', 'bond', 'comic', 'ha', 'set', 'star', 'effect']
[ 42 - 0.775 - 0.33565]:  ['music', 'work', 'opera', 'musical', 'performance', 'play', 'composer', 'theatre', 'orchestra', 'piece']
[ 43 - 0.9 - 0.30473]:  ['company', 'million', 'business', 'market', 'bank', 'cost', 'sale', 'price', 'country', 'industry']
[ 44 - 0.57125 - 0.23648]:  ['episode', 'series', 'television', 'simpson', 'homer', 'season', 'ha', 'character', 'network', 'bart']
[ 45 - 0.75625 - 0.2758]:  ['river', 'water', 'area', 'lake', 'mountain', 'park', 'creek', 'ha', 'mile', 'valley']
[ 46 - 0.33458 - 0.24179]:  ['game', 'player', 'character', 'released', 'series', 'version', 'video', 'final', 'release', 'ha']
[ 47 - 0.75625 - 0.27016]:  ['cell', 'disease', 'ha', 'protein', 'treatment', 'risk', 'effect', 'blood', 'people', 'case']
[ 48 - 0.62292 - 0.19694]:  ['city', 'ha', 'town', 'area', 'population', 'local', 'school', 'india', 'century', 'district']
[ 49 - 0.59167 - 0.32599]:  ['song', 'video', 'number', 'single', 'chart', 'music', 'week', 'performance', 'madonna', 'performed']
\end{Verbatim}
Online LDA:
NPMI=0.282, TU=0.776
\begin{Verbatim}[breaklines=true, fontsize=\tiny]
[ 0 - 1 - 0.34845]:  ['chinese', 'japanese', 'china', 'japan', 'singapore', 'kong', 'hong', 'korean', 'malaysia', 'emperor']
[ 1 - 0.65667 - 0.36749]:  ['season', 'club', 'game', 'team', 'football', 'league', 'goal', 'yard', 'cup', 'match']
[ 2 - 0.86667 - 0.3889]:  ['music', 'work', 'opera', 'musical', 'performance', 'composer', 'orchestra', 'theatre', 'concert', 'piano']
[ 3 - 0.73333 - 0.43867]:  ['force', 'division', 'army', 'battalion', 'battle', 'war', 'brigade', 'attack', 'infantry', 'regiment']
[ 4 - 0.83056 - 0.26379]:  ['film', 'role', 'production', 'award', 'movie', 'actor', 'best', 'director', 'released', 'ha']
[ 5 - 0.91667 - 0.34322]:  ['german', 'soviet', 'war', 'germany', 'russian', 'polish', 's', 'hitler', 'jew', 'nazi']
[ 6 - 0.88333 - 0.23038]:  ['art', 'painting', 'work', 'oxford', 'artist', 'museum', 'cambridge', 'blue', 'london', 'van']
[ 7 - 0.80333 - 0.23849]:  ['australia', 'match', 'australian', 'test', 'run', 'england', 'wicket', 'cricket', 'team', 'inning']
[ 8 - 1 - 0.30542]:  ['company', 'million', 'business', 'bank', 'market', 'sale', 'sold', 'food', 'product', 'price']
[ 9 - 0.41556 - 0.24511]:  ['series', 'episode', 'character', 'scene', 'star', 'doctor', 'ha', 'television', 'set', 'season']
[ 10 - 0.82 - 0.24655]:  ['race', 'second', 'lap', 'team', 'car', 'stage', 'driver', 'point', 'lead', 'place']
[ 11 - 0.51556 - 0.23493]:  ['episode', 'season', 'series', 'television', 'character', 'ha', 'rating', 'homer', 'simpson', 'scene']
[ 12 - 0.50556 - 0.20267]:  ['country', 'world', 'state', 'government', 'ha', 'national', 'international', 'united', 'woman', 'people']
[ 13 - 0.45389 - 0.24179]:  ['game', 'player', 'character', 'released', 'series', 'version', 'video', 'ha', 'release', 'final']
[ 14 - 0.9 - 0.27909]:  ['la', 'el', 'latin', 'puerto', 'mexico', 'american', 'spanish', 'del', 'brazil', 'argentina']
[ 15 - 0.68889 - 0.21473]:  ['water', 'sea', 'shark', 'fish', 'ha', 'ft', 'island', 'area', 'whale', 'specie']
[ 16 - 0.74 - 0.27888]:  ['man', 'comic', 'story', 'issue', 'book', 'magazine', 'character', 'spider', 'series', 'harry']
[ 17 - 0.52333 - 0.31687]:  ['game', 'season', 'team', 'league', 'player', 'run', 'point', 'career', 'second', 'played']
[ 18 - 0.68889 - 0.29798]:  ['book', 'work', 'published', 'novel', 'ha', 'writing', 'wrote', 'life', 'story', 'poem']
[ 19 - 0.83056 - 0.28515]:  ['cell', 'disease', 'ha', 'virus', 'protein', 'cause', 'treatment', 'study', 'used', 'symptom']
[ 20 - 0.73889 - 0.29223]:  ['church', 'god', 'christian', 'century', 'king', 'bishop', 'religious', 'catholic', 'ha', 'death']
[ 21 - 0.95 - 0.3451]:  ['station', 'line', 'service', 'train', 'railway', 'bridge', 'construction', 'passenger', 'opened', 'built']
[ 22 - 0.85 - 0.24117]:  ['island', 'spanish', 'san', 'french', 'colony', 'dutch', 'bay', 'spain', 'francisco', 'colonial']
[ 23 - 0.86667 - 0.34357]:  ['ship', 'gun', 'fleet', 'navy', 'war', 'inch', 'mm', 'class', 'naval', 'battleship']
[ 24 - 0.85 - 0.24488]:  ['british', 'expedition', 'ship', 'royal', 'britain', 'captain', 'sir', 'london', 'ice', 'party']
[ 25 - 0.75833 - 0.45318]:  ['band', 'album', 'song', 'rock', 'record', 'music', 'guitar', 'released', 'recording', 'tour']
[ 26 - 0.78056 - 0.26289]:  ['used', 'energy', 'nuclear', 'metal', 'gas', 'element', 'water', 'ha', 'chemical', 'carbon']
[ 27 - 0.61556 - 0.21828]:  ['character', 'ha', 'storyline', 'series', 'season', 'relationship', 'tell', 'said', 'paul', 'dr.']
[ 28 - 0.73889 - 0.20052]:  ['animal', 'specie', 'fossil', 'known', 'bone', 'specimen', 'like', 'ha', 'genus', 'skull']
[ 29 - 0.95 - 0.22467]:  ['design', 'coin', 'model', 'version', 'dollar', 'structure', 'computer', 'window', 'mint', 'user']
[ 30 - 1 - 0.14881]:  ['manchester', 'bach', 'leigh', 'liverpool', 'wheeler', 'cantata', 'movement', 'naruto', 'christmas', 'shaw']
[ 31 - 0.93333 - 0.34847]:  ['air', 'aircraft', 'flight', 'squadron', 'no.', 'force', 'pilot', 'wing', 'fighter', 'mission']
[ 32 - 0.68056 - 0.2899]:  ['used', 'number', 'use', 'ha', 'example', 'using', 'set', 'section', 'different', 'case']
[ 33 - 0.73889 - 0.26738]:  ['building', 'century', 'house', 'castle', 'built', 'church', 'wall', 'ha', 'tower', 'st']
[ 34 - 0.75 - 0.30962]:  ['said', 'police', 'case', 'day', 'people', 'court', 'trial', 'report', 'right', 'murder']
[ 35 - 0.72222 - 0.23387]:  ['school', 'university', 'student', 'college', 'state', 'program', 'national', 'center', 'ha', 'city']
[ 36 - 0.80833 - 0.21909]:  ['horse', 'dog', 'breed', 'animal', 'parson', 'used', 'century', 'wolf', 'pony', 'sheep']
[ 37 - 0.75 - 0.34074]:  ['state', 'party', 'court', 'election', 'law', 'government', 'president', 'act', 'committee', 'vote']
[ 38 - 0.68333 - 0.24268]:  ['american', 'state', 'war', 'york', 'washington', 'united', 'virginia', 'john', 'white', 'fort']
[ 39 - 0.73889 - 0.17892]:  ['specie', 'ha', 'bird', 'male', 'female', 'white', 'tree', 'brown', 'population', 'genus']
[ 40 - 0.65833 - 0.38782]:  ['song', 'album', 'music', 'single', 'number', 'chart', 'video', 'track', 'released', 'week']
[ 41 - 0.7 - 0.27283]:  ['king', 'empire', 'battle', 'army', 'henry', 'son', 'war', 'roman', 'french', 'greek']
[ 42 - 0.75556 - 0.24232]:  ['river', 'area', 'city', 'park', 'ha', 'town', 'creek', 'mile', 'south', 'county']
[ 43 - 0.86667 - 0.25378]:  ['route', 'highway', 'road', 'u', 'state', 'ny', 'north', 'county', 'street', 'east']
[ 44 - 0.63333 - 0.25196]:  ['government', 'military', 'force', 'war', 'croatian', 'vietnam', 'croatia', 'vietnamese', 'army', 'state']
[ 45 - 0.85556 - 0.27502]:  ['star', 'planet', 'earth', 'sun', 'space', 'mass', 'ha', 'orbit', 'light', 'moon']
[ 46 - 1 - 0.26647]:  ['al', 'india', 'temple', 'indian', 'arab', 'muslim', 'tamil', 'ibn', 'egyptian', 'israeli']
[ 47 - 0.80333 - 0.32352]:  ['match', 'championship', 'team', 'event', 'world', 'won', 'wrestling', 'title', 'tournament', 'champion']
[ 48 - 0.95 - 0.32639]:  ['storm', 'tropical', 'hurricane', 'wind', 'km', 'cyclone', 'damage', 'mph', 'day', 'depression']
[ 49 - 0.9 - 0.34157]:  ['child', 'family', 'woman', 'life', 'father', 'mother', 'wife', 'friend', 'home', 'daughter']
\end{Verbatim}

ProdLDA:
NPMI=0.4, TU=0.624
\begin{Verbatim}[breaklines=true, fontsize=\tiny]
[0-0.85-0.43559]: ['legislature', 'gubernatorial', 'nomination', 'republican', 'statewide', 'governor', 'democrat', 'candidacy', 'senate', 'legislative']
[1-0.48333-0.35108]: ['game', 'player', 'metacritic', 'sequel', 'ign', 'gameplay', 'character', 'film', 'visuals', 'grossing']
[2-0.95-0.46624]: ['glacial', 'basalt', 'volcanic', 'glaciation', 'temperature', 'lava', 'pyroclastic', 'magma', 'sedimentary', 'sediment']
[3-0.75-0.45842]: ['uefa', 'cup', 'scored', 'midfielder', 'goalkeeper', 'victory', 'equaliser', 'wembley', 'fa', 'goalless']
[4-0.58667-0.25822]: ['specie', 'secretion', 'tissue', 'genus', 'vertebrate', 'taxonomy', 'phylogenetic', 'gland', 'symptom', 'habitat']
[5-0.43333-0.50367]: ['terminus', 'intersects', 'highway', 'intersection', 'interchange', 'concurrency', 'northeast', 'roadway', 'renumbering', 'junction']
[6-0.81667-0.45658]: ['touchdown', 'bcs', 'overtime', 'season', 'fumble', 'yard', 'playoff', 'fumbled', 'halftime', 'defensive']
[7-0.78333-0.47278]: ['aircraft', 'squadron', 'reconnaissance', 'sortie', 'raaf', 'bomber', 'avionics', 'operational', 'airfield', 'airframe']
[8-0.43333-0.51711]: ['highway', 'intersects', 'intersection', 'interchange', 'terminus', 'renumbering', 'concurrency', 'northeast', 'roadway', 'realigned']
music TV	
[9-0.40667-0.47917]: ['chart', 'peaked', 'billboard', 'mtv', 'debuted', 'song', 'video', 'album', 'riaa', 'cinquemani']
[10-0.48333-0.40831]: ['aircraft', 'squadron', 'sortie', 'mm', 'reconnaissance', 'aft', 'torpedo', 'knot', 'destroyer', 'armament']
[11-0.37-0.15898]: ['taxonomy', 'intersects', 'specie', 'whitish', 'phylogenetic', 'iucn', 'highway', 'genus', 'underpart', 'habitat']
[12-0.7-0.17913]: ['kÃ¶ppen', 'census', 'demography', 'campus', 'population', 'hectare', 'km2', 'constituency', 'enrollment', 'borough']
[13-0.52333-0.36319]: ['album', 'music', 'studio', 'lyric', 'allmusic', 'recording', 'song', 'musical', 'filmfare', 'bassist']
[14-0.43333-0.4715]: ['km', 'mph', 'tropical', 'westward', 'rainfall', 'flooding', 'convection', 'landfall', 'extratropical', 'storm']
[15-0.75-0.38372]: ['reign', 'ecclesiastical', 'archbishop', 'vassal', 'papacy', 'legate', 'ruler', 'papal', 'earldom', 'chronicler']
[16-0.51667-0.40183]: ['artillery', 'casualty', 'destroyer', 'battalion', 'squadron', 'reinforcement', 'troop', 'regiment', 'guadalcanal', 'convoy']
[17-0.93333-0.29541]: ['doctrine', 'parliament', 'hitler', 'socialism', 'philosopher', 'constitutional', 'theologian', 'critique', 'bucer', 'marxism']
[18-0.71667-0.49424]: ['championship', 'rematch', 'pinfall', 'shawn', 'disqualification', 'wwe', 'smackdown', 'backstage', 'referee', 'match']
[19-0.73667-0.28076]: ['temperature', 'diameter', 'density', 'oxidation', 'latitude', 'acidic', 'specie', 'dioxide', 'molecular', 'carbonate']
[20-0.6-0.33918]: ['championship', 'match', 'defeated', 'rematch', 'randy', 'referee', 'backstage', 'storyline', 'ign', 'summerslam']
[21-0.53333-0.42777]: ['game', 'player', 'sequel', 'metacritic', 'ign', 'gameplay', 'visuals', 'character', 'protagonist', 'gamespot']
[22-0.7-0.42088]: ['inning', 'batting', 'scored', 'unbeaten', 'batted', 'debut', 'scoring', 'wicket', 'bowled', 'opener']
[23-0.48333-0.4502]: ['mph', 'km', 'landfall', 'tropical', 'storm', 'hurricane', 'rainfall', 'flooding', 'extratropical', 'saffir']
[24-0.53333-0.38377]: ['episode', 'funny', 'decides', 'actor', 'nielsen', 'aired', 'filming', 'comedy', 'discovers', 'asks']
[25-0.58667-0.40961]: ['glee', 'chart', 'futterman', 'billboard', 'peaked', 'debuted', 'slezak', 'mtv', 'lyrically', 'song']
[26-0.7-0.11981]: ['demography', 'kÃ¶ppen', 'railway', 'stadium', 'infrastructure', 'census', 'constituency', 'campus', 'km2', 'stadion']
[27-0.58333-0.38244]: ['episode', 'actor', 'filming', 'script', 'comedy', 'funny', 'discovers', 'producer', 'sepinwall', 'film']
[28-0.68333-0.34977]: ['legislature', 'constitutional', 'governorship', 'appoint', 'election', 'legislative', 'treaty', 'diplomatic', 'elected', 'democrat']
[29-0.46667-0.41143]: ['season', 'playoff', 'league', 'nhl', 'game', 'rookie', 'touchdown', 'player', 'coach', 'goaltender']
[30-0.48333-0.49504]: ['mph', 'km', 'tropical', 'westward', 'landfall', 'flooding', 'northwestward', 'rainfall', 'northeastward', 'extratropical']
[31-0.65-0.44581]: ['amidships', 'conning', 'frigate', 'fleet', 'broadside', 'waterline', 'casemates', 'torpedo', 'mm', 'knot']
[32-0.40667-0.47498]: ['chart', 'peaked', 'billboard', 'album', 'video', 'debuted', 'song', 'riaa', 'mtv', 'phonographic']
[33-0.58333-0.52608]: ['interchange', 'terminus', 'intersects', 'highway', 'intersection', 'roadway', 'eastbound', 'westbound', 'freeway', 'route']
[34-0.48333-0.46306]: ['brigade', 'casualty', 'troop', 'infantry', 'artillery', 'flank', 'battalion', 'commanded', 'division', 'regiment']
[35-0.63333-0.35619]: ['episode', 'actor', 'filming', 'realizes', 'nielsen', 'discovers', 'asks', 'mulder', 'scully', 'viewer']
[36-0.52333-0.43397]: ['album', 'recording', 'allmusic', 'song', 'music', 'lyric', 'studio', 'musical', 'vocal', 'guitarist']
[37-0.75-0.43839]: ['bishopric', 'archbishop', 'ecclesiastical', 'clergy', 'consecrated', 'chronicler', 'papacy', 'lordship', 'archbishopric', 'papal']
[38-0.8-0.50362]: ['batting', 'inning', 'batted', 'hitter', 'batsman', 'fielder', 'nl', 'outfielder', 'unbeaten', 'rbi']
[39-0.6-0.44881]: ['mm', 'knot', 'torpedo', 'aft', 'amidships', 'boiler', 'conning', 'waterline', 'cruiser', 'horsepower']
[40-0.53667-0.29499]: ['specie', 'habitat', 'genus', 'iucn', 'taxonomy', 'vegetation', 'morphology', 'mammal', 'underpart', 'plumage']
[41-0.48333-0.41737]: ['infantry', 'casualty', 'troop', 'battalion', 'artillery', 'reinforcement', 'brigade', 'flank', 'division', 'army']
[42-0.51667-0.4231]: ['season', 'nhl', 'playoff', 'game', 'rookie', 'shutout', 'player', 'league', 'roster', 'goaltender']
[43-0.83333-0.31359]: ['treaty', 'mamluk', 'politburo', 'diplomatic', 'sovereignty', 'constitutional', 'militarily', 'abbasid', 'emir', 'gdp']
[44-0.71667-0.30233]: ['finite', 'soluble', 'integer', 'infinity', 'molecule', 'protein', 'infinite', 'molecular', 'computational', 'oxidation']
[45-0.56667-0.37782]: ['midfielder', 'cup', 'match', 'defeat', 'midfield', 'uefa', 'fa', 'defeated', 'championship', 'debut']
[46-0.71667-0.47156]: ['molecule', 'membrane', 'protein', 'eukaryote', 'oxidation', 'molecular', 'soluble', 'metabolism', 'metabolic', 'microscopy']
[47-0.85333-0.36357]: ['continuo', 'cantata', 'soundtrack', 'chorale', 'bwv', 'recitative', 'album', 'guitar', 'bach', 'music']
[48-0.58667-0.39756]: ['taxonomy', 'specie', 'genus', 'morphology', 'morphological', 'phylogenetic', 'clade', 'taxonomic', 'phylogeny', 'iucn']
[49-0.95-0.47451]: ['prognosis', 'diagnostic', 'behavioral', 'clinical', 'symptom', 'diagnosis', 'cognitive', 'abnormality', 'therapy', 'intravenous']
\end{Verbatim}

NTM-R:
NPMI=0.215, TU=0.912
\begin{Verbatim}[breaklines=true, fontsize=\tiny]
[0-0.85-0.13957]: ['m', 'enterprise', 'commander', 'bungie', 'generation', 'election', 'candidate', 'hd', 'roddenberry', 'society']
[1-0.95-0.18795]: ['liturgical', 'altarpiece', 'liturgy', 'fugue', 'cetacean', 'picts', 'anatomical', 'pictish', 'riata', 'grammatical']
[2-0.95-0.31937]: ['colfer', 'futterman', 'monteith', 'herodotus', 'slezak', 'karofsky', 'cheerleading', 'santana', 'xerxes', 'plutarch']
[3-0.7-0.15281]: ['cleveland', 'maryland', 'kentucky', 'iowa', 'harrison', 'mar', 'ford', 'pa', 'olivia', 'tech']
[4-0.9-0.15532]: ['sr', 'pembroke', 'mersey', 'plough', 'whitby', 'gateshead', 'humber', 'altrincham', 'peterborough', 'lichtenstein']
[5-0.73333-0.076084]: ['md', 'indonesian', 'svalbard', 'kepler', 'runway', 'm', 'jenna', 'ice', 'antarctic', 'widerÃ¸e']
[6-0.95-0.16751]: ['resonator', 'impedance', 'goebbels', 'bormann', 'jAzef', 'maunsell', 'heydrich', 'duAan', 'fAhrer', 'waveguide']
[7-1-0.26747]: ['sired', 'ranulf', 'anjou', 'blois', 'thessalonica', 'andronikos', 'rabi', 'nicaea', 'angevin', 'bohemond']
[8-0.66667-0.15309]: ['nelson', 'mexican', 'iowa', 'swift', 'lewis', 'jackson', 'moore', 'mar', 'texas', 'dog']
[9-0.8-0.23519]: ['leng', 'tgs', 'inglis', 'donaghy', 'beatle', 'overdubs', 'fey', 'snl', 'futterman', 'clapton']
[10-1-0.40066]: ['refuel', 'floatplane', 'grumman', 'refueling', 'sonar', 'transatlantic', 'rendezvoused', 'tf', 'leyte', 'tinian']
[11-0.71667-0.058102]: ['lichtenstein', 'etty', 'pa', 'md', 'nude', 'aftershock', 'jovanoviÄ\x87', 'eruptive', 'dreaming', 'weyden']
[12-0.95-0.18683]: ['sauk', 'brig.', 'galena', 'seminole', 'frankfort', 'kentuckian', 'hoosier', 'holliday', 'punted', 'maj.']
[13-0.95-0.15503]: ['widerÃ¸e', 'dupont', 'brest', 'tripoli', 'madras', 'guadalcanal', 'cherbourg', 'yorktown', 'hannibal', 'bombay']
[14-0.95-0.26828]: ['vijayanagara', 'ghat', 'batik', 'madurai', 'coimbatore', 'varanasi', 'cetacean', 'thanjavur', 'uttar', 'marathi']
[15-0.8-0.21604]: ['johnson', 'van', 'jackson', 'taylor', 'smith', 'dutch', 'martin', 'nelson', 'adam', 'lewis']
[16-1-0.18763]: ['canuck', 'nhl', 'tampa', 'mlb', 'canadiens', 'rbi', 'cantata', 'bermuda', 'sox', 'athletics']
[17-0.88333-0.079676]: ['banksia', 'hd', 'thrower', 'pam', 'halo', 'bowler', 'scoring', 'spike', 'mar', 'quadruple']
[18-0.95-0.14555]: ['reelected', 'accredited', 'reelection', 'senatorial', 'sorority', 'unionist', 'phi', 'bsa', 'appointee', 'briarcliff']
[19-0.88333-0.12676]: ['wheelchair', 'iowa', 'wsdot', 'ssh', 'plutonium', 'psh', 'paralympics', 'sr', 'freestyle', 'ub']
[20-0.83333-0.0696]: ['ny', 'md', 'jna', 'henriksen', 'veronica', 'labial', 'torv', 'zng', 'm1', 'lindelof']
[21-0.95-0.13199]: ['squad', 'jordan', 'hamilton', 'shark', 'johnson', 'teammate', 'kansa', 'rochester', 'ranger', 'hockey']
[22-0.9-0.17853]: ['theater', 'doctor', 'texas', 'orchestra', 'san', 'grand', 'theatre', 'disney', 'arthur', 'bar']
[23-1-0.18027]: ['mintage', 'mycena', 'cheilocystidia', 'cystidia', 'breen', 'spongebob', 'numismatic', 'capon', 'obverse', 'ellipsoid']
[24-1-0.71697]: ['duchovny', 'vitaris', 'spotnitz', 'mulder', 'gillian', 'paranormal', 'shearman', 'pileggi', 'scully', 'handlen']
[25-0.95-0.30171]: ['tardis', 'eastenders', 'gillan', 'torchwood', 'catesby', 'walford', 'luftwaffe', 'moffat', 'daleks', 'dalek']
[26-1-0.36124]: ['martyn', 'swartzwelder', 'mirkin', 'wiggum', 'kirkland', 'sauropod', 'smithers', 'jacobson', 'milhouse', 'theropod']
[27-1-0.11983]: ['cookery', 'hindenburg', 'povenmire', 'kratos', 'blamey', 'plankton', 'hillenburg', 'alamein', 'tulagi', 'rearguard']
[28-0.93333-0.47153]: ['stravinsky', 'clarinet', 'berlioz', 'debussy', 'oratorio', 'op.', 'liszt', 'opÃ©ra', 'elgar', 'orchestration']
[29-0.95-0.27002]: ['phylum', 'fumble', 'yardage', 'bcs', 'scrimmage', 'bivalve', 'sportswriter', 'fumbled', 'fiba', 'punted']
[30-1-0.20817]: ['rican', 'afanasieff', 'fatale', 'dupri', 'myrmecia', 'femme', 'wallonia', 'musicnotes.com', 'erotica', 'intercut']
[31-0.95-0.12071]: ['maunsell', 'navigable', 'naktong', 'sprinter', 'hauling', 'doncaster', 'bridgwater', 'rijeka', 'lswr', 'stretford']
[32-1-0.2357]: ['constitutionality', 'habeas', 'scalia', 'appellate', 'unreasonable', 'brownlee', 'harlan', 'sotomayor', 'newt', 'brahman']
[33-0.95-0.20657]: ['csx', 'stub', 'resurfaced', 'legislated', 'widen', 'rejoining', 'widens', 'pulaski', 'drawbridge', 'leng']
[34-0.85-0.13878]: ['harrison', 'jersey', 'summit', 'flag', 'disney', 'doggett', 'beatles', 'township', 'amusement', 'roller']
[35-0.9-0.15949]: ['dia>x87m', 'uematsu', 'petACn', 'naruto', 'nobuo', 'nhu', 'itza', 'sasuke', 'kenshin', 'texians']
[36-0.93333-0.12961]: ['pulp', 'sf', 'delaware', 'wasp', 'reprint', 'ant', 'cent', 'hergÃ©', 'tintin', 'pa']
[37-0.85-0.32081]: ['mi', 'oricon', 'rpgfan', 'nobuo', 'uematsu', 'enix', 'dengeki', 'maeda', 'hamauzu', 'ovum']
[38-1-0.13418]: ['astronomical', 'michigan', 'roof', 'coaster', 'window', 'saginaw', 'lansing', 'bl', 'usher', 'stadium']
[39-0.83333-0.17217]: ['highness', 'medici', 'dodo', 'palatine', 'weyden', 'cosimo', 'mascarene', 'huguenot', 'opÃ©ra', 'catesby']
[40-1-0.22306]: ['edda', 'fragmentary', 'thanhouser', 'loki', 'odin', 'cameraman', 'eline', 'heming', 'norse', 'ua']
[41-0.85-0.23907]: ['tgs', 'tornado', 'poehler', 'donaghy', 'pawnee', 'jenna', 'offerman', 'schur', 'tate', 'severe']
[42-0.95-0.30993]: ['eruptive', 'riparian', 'pyroclastic', 'glaciation', 'volcanism', 'tectonic', 'headwater', 'andes', 'drier', 'tropic']
[43-0.93333-0.19441]: ['ctw', 'muppets', 'filmography', 'muppet', 'repertory', 'cooney', 'heterosexual', 'opÃ©ra', 'goldwyn', 'professorship']
[44-0.95-0.44768]: ['gamesradar', 'unlockable', 'gametrailers', 'novelization', 'dengeki', 'famitsu', 'rpgs', 'ps3', 'cg', 'overworld']
[45-0.9-0.14062]: ['stanley', 'tiger', 'harvard', 'hudson', 'baltimore', 'maryland', 'kg', 'morrison', 'nba', 'lb']
[46-0.85-0.15527]: ['angelou', 'eurovision', 'zng', 'sao', 'miloÅ¡eviÄ\x87', 'svalbard', 'tuÄ\x91man', 'knin', 'bahraini', 'jna']
[47-0.95-0.17175]: ['atrium', 'pv', 'stucco', 'cornice', 'emu', 'pilaster', 'pediment', 'neoclassical', 'briarcliff', 'biomass']
[48-0.85-0.13668]: ['flag', 'vietnam', 'enterprise', 'singapore', 'slave', 'korean', 'philippine', 'stewart', 'zero', 'nba']
[49-1-0.40665]: ['harvick', 'hamlin', 'biffle', 'rAikkAnen', 'sauber', 'kenseth', 'trulli', 'heidfeld', 'verstappen', 'fisichella']
\end{Verbatim}

W-LDA:
NPMI=0.464, TU=0.998
\begin{Verbatim}[breaklines=true, fontsize=\tiny]
[0-0.95-0.51584]: ['jma', 'outage', 'gust', 'typhoon', 'landfall', 'floodwaters', 'jtwc', 'saffir', 'rainbands', 'overflowed'] 
[1-1-0.51968]: ['byzantine', 'caliphate', 'caliph', 'abbasid', 'ibn', 'byzantium', 'constantinople', 'nikephoros', 'emir', 'alexios']
[2-0.95-0.60175]: ['dissipating', 'tropical', 'dissipated', 'extratropical', 'cyclone', 'shear', 'northwestward', 'southwestward', 'saffir', 'convection']
[3-1-0.54757]: ['purana', 'vishnu', 'shiva', 'sanskrit', 'worshipped', 'hindu', 'deity', 'devotee', 'mahabharata', 'temple']
[4-1-0.49348]: ['beatle', 'beatles', 'leng', 'clapton', 'lennon', 'harrison', 'mccartney', 'overdubs', 'ringo', 'spector']
[5-1-0.46882]: ['torpedoed', 'grt', 'ub', 'destroyer', 'flotilla', 'convoy', 'escorting', 'refit', 'kriegsmarine', 'narvik']
[6-1-0.42421]: ['campus', 'enrollment', 'undergraduate', 'alumnus', 'faculty', 'accredited', 'student', 'semester', 'graduate', 'tuition']
[7-1-0.39366]: ['politburo', 'stalin', 'soviet', 'sejm', 'lithuania', 'ussr', 'lithuanian', 'polish', 'ssr', 'gorbachev']
[8-1-0.42948]: ['protein', 'receptor', 'prognosis', 'symptom', 'intravenous', 'mrna', 'medication', 'diagnosis', 'abnormality', 'nucleotide']
[9-1-0.50012]: ['fuselage', 'avionics', 'airframe', 'boeing', 'airline', 'lbf', 'takeoff', 'cockpit', 'undercarriage', 'mach']
[10-1-0.45672]: ['raaf', 'jagdgeschwader', 'bf', 'messerschmitt', 'staffel', 'luftwaffe', 'oberleutnant', 'no.', 'usaaf', 'squadron']
[11-1-0.46824]: ['constitutionality', 'statute', 'appellate', 'unconstitutional', 'defendant', 'amendment', 'judicial', 'court', 'plaintiff', 'statutory']
[12-1-0.72662]: ['lap', 'sauber', 'ferrari', 'rAikkAnen', 'rosberg', 'heidfeld', 'barrichello', 'vettel', 'trulli', 'massa']
[13-1-0.45447]: ['ny', 'renumbering', 'realigned', 'routing', 'cr', 'hamlet', 'truncated', 'intersects', 'unsigned', 'intersecting']
[14-1-0.50035]: ['beyoncÃ©', 'madonna', 'rihanna', 'cinquemani', 'carey', 'musicnotes.com', 'mariah', 'idolator', 'gaga', 'britney']
[15-1-0.31089]: ['gatehouse', 'castle', 'chancel', 'anglesey', 'stonework', 'nave', 'moat', 'antiquarian', 'earthwork', 'bastion']
[16-1-0.48429]: ['freeway', 'interchange', 'md', 'undivided', 'concurrency', 'cloverleaf', 'northbound', 'southbound', 'sr', 'highway']
[17-1-0.41763]: ['electrification', 'railway', 'locomotive', 'tramway', 'electrified', 'freight', 'intercity', 'train', 'nsb', 'footbridge']
[18-1-0.29094]: ['shakira', 'minogue', 'sugababes', 'airplay', 'chart', 'oricon', 'amor', 'salsa', 'stefani', 'tejano']
[19-1-0.55855]: ['ihp', 'conning', 'amidships', 'casemates', 'barbette', 'waterline', 'ironclad', 'krupp', 'hotchkiss', 'battlecruisers']
[20-1-0.68992]: ['wwe', 'smackdown', 'pinfall', 'tna', 'ringside', 'wrestlemania', 'heavyweight', 'wrestling', 'summerslam', 'wrestled']
[21-1-0.35687]: ['plumage', 'underpart', 'viviparous', 'pectoral', 'iucn', 'upperparts', 'nestling', 'passerine', 'copulation', 'gestation']
[22-1-0.61635]: ['hitter', 'mlb', 'baseman', 'rbi', 'nl', 'strikeout', 'outfielder', 'fastball', 'pitcher', 'slugging']
[23-1-0.49182]: ['nomura', 'manga', 'famitsu', 'anime', 'enix', 'shÅ\x8dnen', 'fantasy', 'rpgfan', 'dengeki', 'nobuo']
[24-1-0.3997]: ['ebert', 'film', 'imax', 'afi', 'disney', 'grossing', 'spielberg', 'grossed', 'pixar', 'screenplay']
[25-1-0.62758]: ['multiplayer', 'platforming', 'nintendo', 'gamepro', 'gamerankings', 'eurogamer', 'gamecube', 'gamespot', 'gamespy', 'gameplay']
[26-1-0.47722]: ['parsec', 'orbit', 'orbiting', 'astronomer', 'kepler', 'luminosity', 'planetary', 'brightest', 'constellation', 'brightness']
[27-1-0.38927]: ['wicket', 'batsman', 'bowled', 'bowler', 'wisden', 'selector', 'equalised', 'cricketer', 'unbeaten', 'midfielder']
[28-1-0.22046]: ['puritan', 'congregation', 'settler', 'colony', 'rabbi', 'synagogue', 'massachusetts', 'colonist', 'virginia', 'hampshire']
[29-1-0.62088]: ['volcano', 'lava', 'magma', 'volcanic', 'eruption', 'pyroclastic', 'eruptive', 'caldera', 'volcanism', 'basalt']
[30-1-0.23967]: ['tardis', 'eastenders', 'sayid', 'rhimes', 'soap', 'walford', 'moffat', 'lindelof', 'realises', 'torchwood']
[31-1-0.49605]: ['finite', 'equation', 'theorem', 'impedance', 'algebraic', 'integer', 'mathematical', 'computation', 'multiplication', 'inverse']
[32-1-0.43899]: ['pilaster', 'pediment', 'portico', 'facade', 'cornice', 'facsade', 'architectural', 'architect', 'gable', 'marble']
[33-1-0.65639]: ['cystidia', 'spored', 'cheilocystidia', 'edibility', 'basidium', 'mycologist', 'hypha', 'hyaline', 'hymenium', 'spore']
[34-1-0.32705]: ['frigate', 'brig', 'musket', 'indiaman', 'privateer', 'ticonderoga', 'loyalist', 'cadiz', 'texians', 'rigging']
[35-1-0.51317]: ['marge', 'homer', 'bart', 'swartzwelder', 'wiggum', 'stewie', 'scully', 'groening', 'milhouse', 'simpson']
[36-1-0.58321]: ['krasinski', 'liz', 'halpert', 'jenna', 'rainn', 'tgs', 'dunder', 'pam', 'schrute', 'carell']
[37-1-0.46533]: ['halide', 'isotope', 'oxidation', 'oxide', 'aqueous', 'lanthanide', 'h2o', 'chloride', 'hydride', 'hydroxide']
[38-1-0.41953]: ['thrash', 'kerrang', 'bassist', 'frontman', 'band', 'guitarist', 'album', 'christgau', 'riff', 'nirvana']
[39-1-0.46206]: ['battalion', 'brigade', 'infantry', 'platoon', 'bridgehead', 'regiment', 'panzer', 'rok', 'pusan', 'counterattack']
[40-1-0.64531]: ['touchdown', 'fumble', 'quarterback', 'kickoff', 'punt', 'yardage', 'cornerback', 'linebacker', 'rushing', 'preseason']
[41-1-0.58504]: ['nhl', 'goaltender', 'defenceman', 'canuck', 'ahl', 'blackhawks', 'whl', 'hockey', 'defencemen', 'canadiens']
[42-1-0.44545]: ['inflorescence', 'banksia', 'pollinator', 'pollination', 'seedling', 'nectar', 'pollen', 'follicle', 'flowering', 'thiele']
[43-1-0.40939]: ['gubernatorial', 'republican', 'democrat', 'reelection', 'candidacy', 'senate', 'mintage', 'caucus', 'congressman', 'democratic']
[44-1-0.23055]: ['alamo', 'cyclotron', 'implosion', 'metallurgical', 'physicist', 'laboratory', 'physic', 'reactor', 'oppenheimer', 'testified']
[45-1-0.36894]: ['poem', 'angelou', 'poetry', 'prose', 'literary', 'poet', 'narrator', 'wollstonecraft', 'poetic', 'preface']
[46-1-0.40576]: ['northumbria', 'mercia', 'archbishop', 'papacy', 'earldom', 'bishopric', 'mercian', 'overlordship', 'papal', 'kingship']
[47-1-0.37166]: ['menu', 'gb', 'burger', 'apps', 'software', 'iphone', 'processor', 'user', 'apple', 'app']
[48-1-0.18914]: ['dia>x87m', 'labour', 'ngA', 'mp', 'liberal', 'nhu', 'rhodesia', 'protester', 'alberta', 'saigon']
[49-1-0.48897]: ['cantata', 'recitative', 'concerto', 'bach', 'libretto', 'berlioz', 'soloist', 'chorale', 'oboe', 'symphony']
\end{Verbatim}

\subsection{AGnews}
Online LDA:
\begin{Verbatim}[breaklines=true, fontsize=\tiny]
npmi=0.21322384969796335
[ 0 - 0.68803 - 0.22231]:  ['microsoft', 'software', 'window', 'security', 'version', 'new', 'ha', 'server', 'company', 'application']
[ 1 - 0.65048 - 0.19395]:  ['season', 'los', 'angeles', 'player', 'holiday', 'new', 'team', 'sport', 'forward', 'wa']
[ 2 - 0.74088 - 0.13022]:  ['year', 'ago', 'wa', 'ha', 'family', 'focus', 'com', 'saddam', 'month', 'british']
[ 3 - 0.73333 - 0.22068]:  ['trade', 'tax', 'organization', 'world', 'fund', 'u', 'year', 'boeing', 'international', 'enron']
[ 4 - 0.83214 - 0.20167]:  ['east', 'middle', 'country', 'new', 'king', 'world', 'saudi', 'approach', 'annual', 'era']
[ 5 - 0.83333 - 0.28552]:  ['israeli', 'palestinian', 'drug', 'gaza', 'israel', 'minister', 'strip', 'west', 'bank', 'prime']
[ 6 - 0.85588 - 0.23562]:  ['search', 'google', 'site', 'web', 'internet', 'public', 'engine', 'ha', 'yahoo', 'offering']
[ 7 - 0.64636 - 0.19748]:  ['scientist', 'study', 'say', 'researcher', 'new', 'human', 'ha', 'ap', 'expert', 'science']
[ 8 - 0.72255 - 0.25]:  ['court', 'federal', 'case', 'charge', 'judge', 'trial', 'wa', 'said', 'law', 'ha']
[ 9 - 1 - 0.19845]:  ['japan', 'japanese', 'tokyo', 'texas', 'powerful', 'heavy', 'rain', 'indonesia', 'networking', 'typhoon']
[ 10 - 1 - 0.18393]:  ['aid', 'mark', 'worker', 'italian', 'italy', 'wake', 'relief', 'forest', 'doubt', 'option']
[ 11 - 0.66667 - 0.2254]:  ['iraq', 'hostage', 'said', 'iraqi', 'militant', 'french', 'group', 'release', 'islamic', 'wa']
[ 12 - 0.9 - 0.19566]:  ['state', 'united', 'press', 'canadian', 'canada', 'cp', 'toronto', 'nation', 'ottawa', 'martin']
[ 13 - 0.66833 - 0.19107]:  ['game', 'olympic', 'athens', 'point', 'coach', 'night', 'team', 'wa', 'football', 'gold']
[ 14 - 0.63755 - 0.25872]:  ['billion', 'million', 'company', 'said', 'deal', 'bid', 'ha', 'group', 'buy', 'agreed']
[ 15 - 0.56969 - 0.19212]:  ['company', 'executive', 'chief', 'said', 'new', 'york', 'amp', 'ha', 'financial', 'exchange']
[ 16 - 1 - 0.22169]:  ['according', 'report', 'released', 'university', 'school', 'book', 'published', 'student', 'survey', 'newspaper']
[ 17 - 0.95 - 0.16021]:  ['news', 'german', 'germany', 'nyse', 'nasdaq', 'gold', 'dutch', 'field', 'corporation', 'berlin']
[ 18 - 0.76548 - 0.13072]:  ['gt', 'lt', 'http', 'reuters', 'york', 'new', 'post', 'm', 'font', 'sans']
[ 19 - 0.84048 - 0.16701]:  ['house', 'white', 'new', 'national', 'ap', 'hong', 'kong', 'intelligence', 'republican', 'senate']
[ 20 - 0.80588 - 0.18459]:  ['ha', 'moon', 'earth', 'scientist', 'planet', 'mile', 'mar', 'titan', 'nasa', 'image']
[ 21 - 0.73803 - 0.20273]:  ['computer', 'world', 'pc', 'drive', 'personal', 'new', 'ibm', 'power', 'hard', 'ha']
[ 22 - 1 - 0.20013]:  ['free', 'agent', 'pick', 'pair', 'single', 'centre', 'sweep', 'choice', 'crowd', 'carter']
[ 23 - 0.76303 - 0.18581]:  ['music', 'online', 'digital', 'apple', 'store', 'ha', 'new', 'industry', 'player', 'ipod']
[ 24 - 0.59588 - 0.24328]:  ['president', 'minister', 'bush', 'prime', 'john', 'said', 'government', 'war', 'iraq', 'ha']
[ 25 - 0.95 - 0.17342]:  ['giant', 'oil', 'russian', 'gas', 'baseball', 'yukos', 'bond', 'major', 'moscow', 'auction']
[ 26 - 0.80667 - 0.24213]:  ['space', 'nasa', 'flight', 'station', 'said', 'plane', 'launch', 'international', 'airport', 'commercial']
[ 27 - 0.80667 - 0.26711]:  ['people', 'said', 'killed', 'attack', 'police', 'baghdad', 'city', 'force', 'iraqi', 'official']
[ 28 - 0.67255 - 0.22167]:  ['quot', 'wa', 'said', 'thing', 'want', 'better', 'know', 'say', 'ha', 'need']
[ 29 - 0.95 - 0.23242]:  ['england', 'champion', 'match', 'goal', 'stage', 'league', 'home', 'wednesday', 'trophy', 'captain']
[ 30 - 1 - 0.18207]:  ['european', 'hurricane', 'union', 'florida', 'ivan', 'eu', 'france', 'coast', 'storm', 'island']
[ 31 - 0.80667 - 0.23308]:  ['change', 'nuclear', 'iran', 'agency', 'said', 'program', 'global', 'weapon', 'nation', 'security']
[ 32 - 0.68803 - 0.23633]:  ['service', 'phone', 'technology', 'mobile', 'wireless', 'company', 'new', 'internet', 'ha', 'chip']
[ 33 - 0.83922 - 0.13841]:  ['ha', 'turning', 'heat', 'team', 'bar', 'managed', 'seattle', 'lewis', 'connecticut', 'allen']
[ 34 - 1 - 0.20403]:  ['san', 'francisco', 'johnson', 'diego', 'stewart', 'hotel', 'testing', 'living', 'room', 'jose']
[ 35 - 0.79 - 0.25845]:  ['job', 'cut', 'airline', 'said', 'plan', 'u', 'million', 'cost', 'air', 'bankruptcy']
[ 36 - 1 - 0.17114]:  ['victim', 'taiwan', 'blow', 'philippine', 'suffered', 'steve', 'singapore', 'overnight', 'delivered', 'gate']
[ 37 - 0.74714 - 0.18515]:  ['india', 'new', 'radio', 'pakistan', 'indian', 'satellite', 'minister', 'la', 'delhi', 'said']
[ 38 - 0.95 - 0.37642]:  ['election', 'presidential', 'president', 'party', 'vote', 'campaign', 'candidate', 'political', 'opposition', 'russia']
[ 39 - 0.85667 - 0.21321]:  ['china', 'south', 'north', 'korea', 'said', 'talk', 'chinese', 'beijing', 'africa', 'official']
[ 40 - 0.925 - 0.26416]:  ['cup', 'world', 'open', 'round', 'final', 'championship', 'win', 'race', 'second', 'grand']
[ 41 - 0.71548 - 0.26602]:  ['sunday', 'ap', 'game', 'touchdown', 'season', 'yard', 'quarterback', 'new', 'running', 'victory']
[ 42 - 0.81667 - 0.15514]:  ['research', 'quote', 'profile', 'black', 'wa', 'property', 'williams', 'heavyweight', 'said', 'accepted']
[ 43 - 0.64881 - 0.22719]:  ['price', 'oil', 'reuters', 'stock', 'new', 'u', 'york', 'rate', 'high', 'dollar']
[ 44 - 0.70714 - 0.25335]:  ['series', 'red', 'new', 'sox', 'york', 'game', 'night', 'boston', 'yankee', 'run']
[ 45 - 0.83088 - 0.19549]:  ['game', 'video', 'announcement', 'watch', 'paul', 'ha', 'nintendo', 'mass', 'lose', 'fact']
[ 46 - 0.765 - 0.26435]:  ['sale', 'percent', 'profit', 'said', 'reported', 'quarter', 'share', 'year', 'earnings', 'reuters']
[ 47 - 0.71588 - 0.17818]:  ['manager', 'club', 'ha', 'united', 'manchester', 'league', 'arsenal', 'old', 'wa', 'chelsea']
[ 48 - 0.76667 - 0.21224]:  ['australia', 'test', 'leader', 'arafat', 'australian', 'yasser', 'wa', 'palestinian', 'day', 'said']
[ 49 - 0.74088 - 0.23108]:  ['just', 'like', 'big', 'year', 'look', 'time', 'wa', 'good', 'little', 'ha']
uniqueness=0.802	
\end{Verbatim}

LDA Collapsed Gibbs sampling:
\begin{Verbatim}[breaklines=true, fontsize=\tiny]
npmi=0.23902729002814144
[ 0 - 0.81667 - 0.32929]:  ['palestinian', 'leader', 'israeli', 'gaza', 'west', 'israel', 'official', 'arafat', 'yasser', 'sunday']
[ 1 - 0.9 - 0.24206]:  ['space', 'nasa', 'international', 'station', 'scientist', 'launch', 'earth', 'mission', 'moon', 'star']
[ 2 - 0.82778 - 0.19331]:  ['chief', 'executive', 'company', 'bid', 'rival', 'oracle', 'board', 'ha', 'peoplesoft', 'offer']
[ 3 - 0.69167 - 0.23957]:  ['sunday', 'game', 'season', 'touchdown', 'sport', 'yard', 'running', 'quarterback', 'network', 'left']
[ 4 - 0.79 - 0.23012]:  ['dollar', 'reuters', 'rate', 'economic', 'growth', 'federal', 'economy', 'reserve', 'euro', 'tuesday']
[ 5 - 0.505 - 0.24069]:  ['china', 'news', 'japan', 'reuters', 'monday', 'thursday', 'wednesday', 'reported', 'tuesday', 'report']
[ 6 - 0.76944 - 0.16612]:  ['game', 'industry', 'ha', 'player', 'video', 'sun', 'today', 'latest', 'sony', 'movie']
[ 7 - 0.69444 - 0.20597]:  ['phone', 'ha', 'market', 'mobile', 'world', 'company', 'maker', 'electronics', 'device', 'cell']
[ 8 - 0.56944 - 0.22276]:  ['ha', 'year', 'world', 'past', 'today', 'number', 'grand', 'month', 'time', 'half']
[ 9 - 1 - 0.2544]:  ['drug', 'health', 'heart', 'food', 'study', 'risk', 'researcher', 'child', 'medical', 'died']
[ 10 - 0.83333 - 0.18319]:  ['iraq', 'group', 'british', 'french', 'hostage', 'held', 'worker', 'militant', 'release', 'american']
[ 11 - 0.78333 - 0.31725]:  ['billion', 'million', 'company', 'deal', 'group', 'buy', 'agreed', 'sell', 'cash', 'stake']
[ 12 - 0.73333 - 0.23883]:  ['government', 'country', 'region', 'nation', 'security', 'talk', 'peace', 'rebel', 'darfur', 'end']
[ 13 - 0.81111 - 0.21487]:  ['ha', 'make', 'big', 'making', 'television', 'question', 'doe', 'tv', 'work', 'set']
[ 14 - 0.70833 - 0.2839]:  ['point', 'coach', 'night', 'team', 'scored', 'game', 'university', 'football', 'season', 'victory']
[ 15 - 0.72333 - 0.22315]:  ['stock', 'share', 'york', 'street', 'investor', 'market', 'reuters', 'wall', 'higher', 'wednesday']
[ 16 - 0.85 - 0.27922]:  ['city', 'people', 'killed', 'iraq', 'iraqi', 'baghdad', 'force', 'bomb', 'attack', 'car']
[ 17 - 0.7 - 0.23109]:  ['san', 'hit', 'run', 'francisco', 'ap', 'night', 'home', 'victory', 'win', 'texas']
[ 18 - 0.85 - 0.26333]:  ['minister', 'prime', 'country', 'party', 'leader', 'pakistan', 'president', 'tony', 'afp', 'foreign']
[ 19 - 0.83333 - 0.30335]:  ['computer', 'technology', 'ibm', 'chip', 'intel', 'product', 'pc', 'announced', 'power', 'business']
[ 20 - 0.54278 - 0.16654]:  ['ha', 'press', 'change', 'ap', 'canadian', 'global', 'tuesday', 'thursday', 'international', 'year']
[ 21 - 0.88333 - 0.33114]:  ['software', 'microsoft', 'security', 'window', 'version', 'application', 'linux', 'operating', 'source', 'user']
[ 22 - 0.56333 - 0.15669]:  ['gt', 'lt', 'reuters', 'http', 'york', 'thursday', 'washington', 'tuesday', 'wednesday', 'post']
[ 23 - 1 - 0.23855]:  ['oil', 'price', 'high', 'record', 'crude', 'supply', 'barrel', 'concern', 'future', 'energy']
[ 24 - 0.83333 - 0.23538]:  ['plan', 'cut', 'airline', 'air', 'job', 'cost', 'line', 'bankruptcy', 'union', 'million']
[ 25 - 0.6 - 0.29362]:  ['service', 'network', 'wireless', 'company', 'internet', 'technology', 'business', 'communication', 'customer', 'announced']
[ 26 - 0.56111 - 0.15573]:  ['wa', 'ap', 'contract', 'ha', 'yesterday', 'left', 'list', 'monday', 'free', 'signed']
[ 27 - 0.95 - 0.38325]:  ['court', 'federal', 'case', 'judge', 'lawsuit', 'law', 'filed', 'legal', 'claim', 'trial']
[ 28 - 0.86667 - 0.27833]:  ['president', 'election', 'bush', 'john', 'presidential', 'ap', 'campaign', 'vote', 'kerry', 'house']
[ 29 - 0.68333 - 0.21267]:  ['world', 'lead', 'championship', 'cup', 'sunday', 'round', 'shot', 'saturday', 'title', 'tiger']
[ 30 - 0.80833 - 0.28389]:  ['red', 'series', 'boston', 'game', 'sox', 'league', 'york', 'yankee', 'baseball', 'houston']
[ 31 - 0.83333 - 0.19603]:  ['state', 'united', 'nation', 'nuclear', 'program', 'iran', 'secretary', 'weapon', 'washington', 'official']
[ 32 - 0.80833 - 0.26289]:  ['police', 'wa', 'attack', 'man', 'accused', 'war', 'charged', 'arrested', 'terrorist', 'yesterday']
[ 33 - 0.81667 - 0.21302]:  ['people', 'hurricane', 'thousand', 'home', 'coast', 'storm', 'florida', 'missing', 'official', 'powerful']
[ 34 - 0.725 - 0.24607]:  ['month', 'report', 'consumer', 'government', 'showed', 'september', 'job', 'august', 'week', 'october']
[ 35 - 0.70833 - 0.21241]:  ['research', 'group', 'firm', 'quote', 'bank', 'profile', 'company', 'business', 'monday', 'investment']
[ 36 - 0.95 - 0.21915]:  ['quot', 'thing', 'called', 'word', 'don', 'good', 'story', 'told', 'work', 'staff']
[ 37 - 0.72778 - 0.16637]:  ['ap', 'motor', 'ha', 'scientist', 'plant', 'general', 'human', 'long', 'great', 'remains']
[ 38 - 0.95 - 0.25207]:  ['percent', 'sale', 'profit', 'quarter', 'reported', 'earnings', 'store', 'loss', 'retailer', 'rose']
[ 39 - 0.75667 - 0.17671]:  ['russian', 'thursday', 'school', 'russia', 'los', 'angeles', 'ap', 'major', 'wednesday', 'california']
[ 40 - 0.53 - 0.24683]:  ['reuters', 'week', 'south', 'north', 'friday', 'tuesday', 'monday', 'wednesday', 'thursday', 'korea']
[ 41 - 0.525 - 0.22387]:  ['wa', 'year', 'time', 'ago', 'yesterday', 'day', 'week', 'earlier', 'long', 'history']
[ 42 - 0.64444 - 0.23892]:  ['million', 'security', 'company', 'public', 'ha', 'fund', 'pay', 'exchange', 'commission', 'regulator']
[ 43 - 0.625 - 0.1675]:  ['day', 'test', 'today', 'australia', 'india', 'australian', 'yesterday', 'england', 'saturday', 'team']
[ 44 - 0.6 - 0.24946]:  ['open', 'world', 'final', 'set', 'cup', 'champion', 'saturday', 'reach', 'round', 'win']
[ 45 - 0.85 - 0.23868]:  ['league', 'champion', 'club', 'goal', 'manager', 'england', 'real', 'manchester', 'madrid', 'arsenal']
[ 46 - 0.53333 - 0.27788]:  ['week', 'time', 'season', 'start', 'year', 'home', 'day', 'early', 'end', 'weekend']
[ 47 - 0.81667 - 0.24605]:  ['european', 'trade', 'union', 'german', 'tax', 'world', 'eu', 'germany', 'organization', 'commission']
[ 48 - 0.85 - 0.27251]:  ['online', 'search', 'web', 'google', 'internet', 'site', 'music', 'apple', 'user', 'service']
[ 49 - 0.86667 - 0.24667]:  ['olympic', 'athens', 'gold', 'medal', 'won', 'american', 'men', 'woman', 'world', 'olympics']
uniqueness=0.7559999999999999
\end{Verbatim}

ProdLDA:
\begin{Verbatim}[breaklines=true, fontsize=\tiny]
[ 0 - 0.78667 - 0.27803]:  ['directory', 'netscape', 'flaw', 'xp', 'itunes', 'server', 'midrange', 'user', 'gmail', 'fujitsu']
[ 1 - 0.17 - 0.28389]:  ['lt', 'gt', 'serif', 'arial', 'helvetica', 'verdana', 'font', 'sans', 'm', 'http']
[ 2 - 0.69167 - 0.20085]:  ['moon', 'lunar', 'spacecraft', 'saturn', 'rover', 'mar', 'lived', 'utah', 'parachute', 'shuttle']
[ 3 - 0.66167 - 0.2175]:  ['touchdown', 'yard', 'scored', 'dodger', 'inning', 'st', 'pujols', 'seahawks', 'slam', 'astros']
[ 4 - 0.93333 - 0.19771]:  ['trent', 'jumper', 'tennessee', 'overcame', 'keith', 'cub', 'touchdown', 'milwaukee', 'season', 'mvp']
[ 5 - 0.44167 - 0.24495]:  ['crude', 'barrel', 'oil', 'price', 'nikkei', 'opec', 'midsession', 'stock', 'heating', 'rose']
[ 6 - 0.75833 - 0.19283]:  ['allawi', 'iyad', 'abuja', 'nepal', 'yonhap', 'pervez', 'eta', 'militant', 'sudan', 'iraqi']
[ 7 - 0.17 - 0.28389]:  ['lt', 'gt', 'http', 'font', 'serif', 'arial', 'helvetica', 'verdana', 'sans', 'm']
[ 8 - 0.825 - 0.14285]:  ['cup', 'phelps', 'scored', 'qualifier', 'cardinal', 'homered', 'federer', 'colt', 'magic', 'roger']
[ 9 - 0.87 - 0.22113]:  ['sharapova', 'wimbledon', 'unbeaten', 'roddick', 'inning', 'champion', 'brett', 'postseason', 'homer', 'rivera']
[ 10 - 0.49167 - 0.33562]:  ['insurgent', 'stronghold', 'baghdad', 'killed', 'iraqi', 'gaza', 'raid', 'israeli', 'killing', 'palestinian']
[ 11 - 0.80333 - 0.23854]:  ['ipod', 'imac', 'desktop', 'xp', 'pt', 'embedded', 'apple', 'erp', 'com', 'window']
[ 12 - 0.9 - 0.19538]:  ['abuja', 'sudanese', 'hideout', 'kabul', 'jerusalem', 'karzai', 'ariel', 'captive', 'hamid', 'damascus']
[ 13 - 0.8 - 0.28904]:  ['msn', 'priority', 'server', 'hd', 'lan', 'infoworld', 'user', 'notebook', 'workstation', 'linux']
[ 14 - 0.44167 - 0.22566]:  ['oil', 'crude', 'nikkei', 'inventory', 'price', 'barrel', 'trader', 'output', 'greenspan', 'opec']
[ 15 - 0.81667 - 0.34766]:  ['telescope', 'spacecraft', 'relativity', 'earth', 'hubble', 'backwards', 'planet', 'circling', 'planetary', 'cassini']
[ 16 - 0.17 - 0.28389]:  ['lt', 'gt', 'http', 'serif', 'arial', 'helvetica', 'verdana', 'font', 'sans', 'm']
[ 17 - 0.87 - 0.20465]:  ['pitched', 'rutherford', 'piscataway', 'pedro', 'felix', 'shutout', 'pete', 'martinez', 'inning', 'kazmir']
[ 18 - 0.68667 - 0.28964]:  ['version', 'smart', 'msn', 'antivirus', 'window', 'browser', 'feature', 'malicious', 'compatible', 'xp']
[ 19 - 0.325 - 0.24661]:  ['crude', 'oil', 'barrel', 'heating', 'output', 'price', 'nikkei', 'opec', 'stock', 'inventory']
[ 20 - 0.81667 - 0.16414]:  ['docomo', 'conspiracy', 'atomic', 'tehran', 'unused', 'iran', 'nuclear', 'regulatory', 'ntt', 'protocol']
[ 21 - 0.78333 - 0.33018]:  ['java', 'server', 'kodak', 'cingular', 'software', 'microsystems', 'apps', 'microsoft', 'ibm', 'mobile']
[ 22 - 0.83333 - 0.21765]:  ['cia', 'musharraf', 'yushchenko', 'tehran', 'pervez', 'enrichment', 'iran', 'conciliatory', 'irna', 'blair']
[ 23 - 0.95 - 0.17406]:  ['pitcher', 'acc', 'premiership', 'curt', 'tampa', 'jim', 'supersonics', 'raucous', 'cal', 'oakland']
[ 24 - 0.71667 - 0.23467]:  ['capsule', 'soyuz', 'cosmonaut', 'solar', 'astronaut', 'titan', 'lore', 'atmosphere', 'mar', 'genesis']
[ 25 - 1 - 0.20041]:  ['safin', 'marat', 'busch', 'cincinnati', 'aaron', 'singled', 'sidelined', 'raptor', 'hamstring', 'guillermo']
[ 26 - 0.675 - 0.2152]:  ['nordegren', 'astronaut', 'space', 'earth', 'pitcairn', 'moon', 'orbit', 'elin', 'nasa', 'craft']
[ 27 - 0.44167 - 0.37017]:  ['gaza', 'baghdad', 'israeli', 'wounded', 'militant', 'palestinian', 'muqtada', 'wounding', 'insurgent', 'jabalya']
[ 28 - 0.17 - 0.28389]:  ['lt', 'gt', 'serif', 'arial', 'helvetica', 'verdana', 'font', 'http', 'sans', 'm']
[ 29 - 0.73667 - 0.26993]:  ['xp', 'nvidia', 'window', 'processor', 'msn', 'java', 'tool', 'chipset', 'stack', 'modeling']
[ 30 - 0.25 - 0.23847]:  ['lt', 'gt', 'http', 'serif', 'arial', 'helvetica', 'verdana', 'font', 'sans', 'quarterly']
[ 31 - 0.525 - 0.25439]:  ['mysterious', 'mar', 'solar', 'cassini', 'nasa', 'earth', 'fossil', 'saturn', 'soyuz', 'moon']
[ 32 - 0.51667 - 0.31844]:  ['baghdad', 'israeli', 'gaza', 'wounding', 'iraqi', 'insurgent', 'wounded', 'bomb', 'policeman', 'troop']
[ 33 - 0.17 - 0.28389]:  ['lt', 'gt', 'http', 'font', 'serif', 'arial', 'helvetica', 'verdana', 'sans', 'm']
[ 34 - 0.95 - 0.15975]:  ['liverpool', 'vaughan', 'nash', 'blackburn', 'gerrard', 'locker', 'notre', 'nba', 'lomana', 'lualua']
[ 35 - 0.85833 - 0.16806]:  ['knockout', 'scored', 'kicker', 'fc', 'timberwolves', 'ticker', 'defending', 'semifinal', 'rooney', 'astros']
[ 36 - 0.67 - 0.19242]:  ['homered', 'alcs', 'brave', 'yard', 'sox', 'schnyder', 'cup', 'victory', 'inning', 'finale']
[ 37 - 0.575 - 0.27208]:  ['ansari', 'prize', 'astronaut', 'spacecraft', 'pitcairn', 'spaceshipone', 'nasa', 'parachute', 'moon', 'atmosphere']
[ 38 - 0.81667 - 0.22488]:  ['nuclear', 'putin', 'censure', 'standoff', 'prime', 'minister', 'thabo', 'darfur', 'hostage', 'iran']
[ 39 - 0.60833 - 0.3156]:  ['gaza', 'moqtada', 'militant', 'hamas', 'wounding', 'killing', 'wounded', 'sharon', 'ariel', 'grenade']
[ 40 - 0.9 - 0.33229]:  ['interoperability', 'provider', 'sender', 'authentication', 'microsystems', 'subscriber', 'adobe', 'enterprise', 'software', 'ietf']
[ 41 - 0.73333 - 0.30305]:  ['militant', 'wounding', 'sunni', 'mosque', 'killed', 'shiite', 'strip', 'multan', 'palestinian', 'suicide']
[ 42 - 0.83333 - 0.22536]:  ['mcgahee', 'referee', 'linebacker', 'elbow', 'willis', 'dame', 'astros', 'notre', 'rib', 'martinez']
[ 43 - 1 - 0.11352]:  ['larkin', 'clubhouse', 'chelsea', 'defensive', 'dolphin', 'wei', 'owen', 'dunlop', 'league', 'coordinator']
[ 44 - 0.70333 - 0.3561]:  ['firefox', 'compatible', 'browser', 'mozilla', 'desktop', 'user', 'platform', 'worm', 'xp', 'edition']
[ 45 - 0.49167 - 0.26583]:  ['oil', 'crude', 'price', 'barrel', 'opec', 'inventory', 'eased', 'heating', 'gasoline', 'disruption']
[ 46 - 0.85833 - 0.248]:  ['preseason', 'pass', 'match', 'quarterback', 'ahman', 'nedbank', 'touchdown', 'valencia', 'jacksonville', 'scored']
[ 47 - 0.95 - 0.16433]:  ['championship', 'fitchburg', 'colby', 'oliver', 'celtic', 'endicott', 'playoff', 'coach', 'victory', 'pga']
[ 48 - 0.88333 - 0.25886]:  ['recep', 'tayyip', 'erdogan', 'bosnian', 'nuclear', 'equatorial', 'minister', 'thatcher', 'anwar', 'elbaradei']
[ 49 - 0.77 - 0.1631]:  ['wismilak', 'wta', 'yankee', 'sox', 'omega', 'oakland', 'gatlin', 'calf', 'sharapova', 'inning']
\end{Verbatim}

NTM-R:
\begin{Verbatim}[breaklines=true, fontsize=\tiny]
[0-0.5-0.17034]: ['eisner', 'zook', 'coaching', 'disney', 'walt', 'jaguar', 'willingham', 'notre', 'vacant', 'tyrone']
[1-0.65-0.2067]: ['lt', 'gt', 'http', 'font', 'serif', 'arial', 'helvetica', 'verdana', 'br', 'm']
[2-0.85-0.27743]: ['d', 'nintendo', 'cassini', 'saturn', 'playstation', 'console', 'sony', 'portable', 'andreas', 'moon']
[3-1-0.19087]: ['critic', 'treatment', 'committee', 'university', 'responsibility', 'fallen', 'item', 'public', 'medicine', 'undergo']
[4-0.54762-0.19074]: ['sox', 'pedro', 'saddam', 'kerry', 'martinez', 'hussein', 'red', 'george', 'fallujah', 'allawi']
[5-1-0.36219]: ['xp', 'browser', 'mozilla', 'firefox', 'beta', 'desktop', 'processor', 'window', 'msn', 'flaw']
[6-0.47-0.13705]: ['warming', 'vijay', 'arctic', 'climate', 'singh', 'radar', 'specie', 'pt', 'importance', 'bird']
[7-0.68667-0.31398]: ['telescope', 'orbiting', 'saturn', 'ansari', 'mojave', 'astronaut', 'antenna', 'hubble', 'cassini', 'shuttle']
[8-0.68333-0.24017]: ['chelsea', 'madrid', 'mutu', 'spanish', 'striker', 'camacho', 'banned', 'jol', 'cska', 'referee']
[9-0.47667-0.19242]: ['striker', 'mutu', 'ferguson', 'harry', 'trafford', 'rooney', 'manchester', 'arsene', 'hamid', 'karzai']
[10-0.68667-0.18856]: ['administration', 'crew', 'human', 'shuttle', 'atomic', 'food', 'flu', 'russia', 'hubble', 'soyuz']
[11-0.32417-0.15603]: ['greenspan', 'priority', 'ryder', 'alan', 'curt', 'schilling', 'pedro', 'martinez', 'sox', 'pt']
[12-0.52-0.08896]: ['upgrading', 'arctic', 'vijay', 'helen', 'zdnet', 'volcano', 'bird', 'simulator', 'mount', 'pt']
[13-0.5025-0.21493]: ['rooney', 'manchester', 'trafford', 'coaching', 'football', 'greenspan', 'wayne', 'auburn', 'blackburn', 'eriksson']
[14-0.875-0.13414]: ['blair', 'athlete', 'nasa', 'football', 'florida', 'tony', 'dangerous', 'watchdog', 'patriot', 'informed']
[15-0.5-0.2087]: ['willingham', 'tyrone', 'zook', 'ron', 'eisner', 'jeffrey', 'notre', 'dame', 'meyer', 'sirius']
[16-0.85833-0.16508]: ['motogp', 'nicholls', 'premiership', 'qualifying', 'newell', 'newcastle', 'pole', 'graeme', 'kieron', 'bannister']
[17-0.66012-0.26168]: ['challenger', 'greenspan', 'liberal', 'convention', 'kerry', 'campaign', 'hostile', 'candidate', 'democrat', 'poll']
[18-1-0.3008]: ['medal', 'gold', 'safin', 'marat', 'federer', 'lleyton', 'phelps', 'seed', 'athens', 'henman']
[19-0.88333-0.14774]: ['bernie', 'jaguar', 'ferrari', 'racing', 'prix', 'hopkins', 'ovitz', 'hoya', 'association', 'brazilian']
[20-0.78095-0.19085]: ['kerry', 'republican', 'appropriate', 'bush', 'greece', 'safe', 'columbia', 'saddam', 'hostage', 'regard']
[21-0.84762-0.13865]: ['celebration', 'simply', 'kerry', 'museum', 'represented', 'thanksgiving', 'korea', 'college', 'coast', 'mount']
[22-0.61167-0.19311]: ['shuttle', 'astronaut', 'nasa', 'endangered', 'capsule', 'moscow', 'soyuz', 'malaysia', 'warn', 'sean']
[23-0.44833-0.31216]: ['rooney', 'ferguson', 'blackburn', 'liverpool', 'arsenal', 'arsene', 'premiership', 'wenger', 'benitez', 'manchester']
[24-1-0.20159]: ['quarterly', 'earnings', 'profit', 'forecast', 'offset', 'nikkei', 'income', 'profile', 'higher', 'weighed']
[25-0.93333-0.20844]: ['corruption', 'genetic', 'handling', 'social', 'legislation', 'merck', 'dna', 'independent', 'cloning', 'vioxx']
[26-0.9-0.24207]: ['enrichment', 'uranium', 'tehran', 'iran', 'nuclear', 'suspend', 'sanction', 'freeze', 'atomic', 'negotiator']
[27-0.7125-0.15795]: ['mutu', 'hugo', 'greenspan', 'jailed', 'overturn', 'madrid', 'ottawa', 'chavez', 'conviction', 'spanish']
[28-0.56167-0.36871]: ['genesis', 'capsule', 'shuttle', 'space', 'soyuz', 'crew', 'nasa', 'spaceshipone', 'manned', 'astronaut']
[29-0.93333-0.16113]: ['kobe', 'eliot', 'attorney', 'bryant', 'guilty', 'ovitz', 'spitzer', 'milosevic', 'slobodan', 'enron']
[30-0.46167-0.11766]: ['obtaining', 'helen', 'erp', 'mount', 'priority', 'upgrading', 'radar', 'pyongyang', 'zdnet', 'pt']
[31-0.65-0.26263]: ['arial', 'verdana', 'helvetica', 'serif', 'font', 'sans', 'm', 'br', 'post', 'reg']
[32-0.49333-0.26054]: ['ferguson', 'trafford', 'manchester', 'alan', 'alex', 'newcastle', 'singh', 'tottenham', 'rooney', 'skipper']
[33-0.59583-0.25973]: ['republican', 'voter', 'convention', 'tax', 'congressional', 'poll', 'web', 'saddam', 'greenspan', 'social']
[34-0.95-0.20341]: ['oracle', 'peoplesoft', 'java', 'verizon', 'cingular', 'acquire', 'microsystems', 'hostile', 'takeover', 'conway']
[35-0.51429-0.19294]: ['martinez', 'sox', 'pedro', 'schilling', 'happen', 'curt', 'kerry', 'yankee', 'red', 'moon']
[36-0.95-0.22597]: ['ariel', 'sharon', 'manmohan', 'gaza', 'allawi', 'najaf', 'settler', 'aziz', 'iyad', 'kashmir']
[37-0.68667-0.29305]: ['climate', 'emission', 'kyoto', 'arctic', 'carbon', 'warming', 'dioxide', 'shuttle', 'hubble', 'scientific']
[38-0.44333-0.4275]: ['rooney', 'trafford', 'everton', 'ferguson', 'nistelrooy', 'arsene', 'striker', 'ruud', 'manchester', 'wenger']
[39-0.47333-0.18842]: ['meyer', 'trafford', 'tyrone', 'willingham', 'dame', 'notre', 'vogts', 'ferguson', 'berti', 'ron']
[40-0.50833-0.40095]: ['newcastle', 'premier', 'bolton', 'arsenal', 'premiership', 'chelsea', 'everton', 'blackburn', 'charlton', 'rooney']
[41-0.78333-0.21501]: ['putin', 'russian', 'chechen', 'beslan', 'vladimir', 'moscow', 'jakarta', 'spanish', 'canadian', 'kong']
[42-0.50417-0.17625]: ['importance', 'greenspan', 'priority', 'republican', 'legislative', 'poverty', 'alan', 'democratic', 'ryder', 'obtaining']
[43-1-0.30192]: ['homered', 'inning', 'homer', 'astros', 'touchdown', 'nl', 'peyton', 'pitched', 'clemens', 'yard']
[44-1-0.36627]: ['wounding', 'bomber', 'detonated', 'exploded', 'wounded', 'suicide', 'killing', 'injuring', 'mosque', 'bomb']
[45-0.35512-0.11672]: ['ryder', 'priority', 'pt', 'erp', 'vijay', 'obtaining', 'com', 'importance', 'greenspan', 'kerry']
[46-0.64345-0.20764]: ['assessment', 'academic', 'social', 'hong', 'kong', 'infrastructure', 'convention', 'kerry', 'greenspan', 'welfare']
[47-0.6-0.18425]: ['eisner', 'willingham', 'zook', 'tyrone', 'ovitz', 'spurrier', 'coordinator', 'chief', 'vice', 'walt']
[48-0.825-0.11548]: ['material', 'phone', 'biodegradable', 'hypersonic', 'asaravala', 'nasa', 'huygens', 'genesis', 'audiovox', 'iran']
[49-0.75833-0.15448]: ['hispano', 'madrid', 'barcelona', 'psv', 'charlton', 'kiev', 'premiership', 'russian', 'abbey', 'hartson']
\end{Verbatim}

W-LDA:
\begin{Verbatim}[breaklines=true, fontsize=\tiny]
[0-1-0.17838]: ['sale', 'quarter', 'retailer', 'idc', 'grew', 'slower', 'seasonally', 'unemployment', 'compared', 'july']
[1-1-0.50711]: ['najaf', 'baghdad', 'insurgent', 'shiite', 'fallujah', 'muqtada', 'mosul', 'iraqi', 'sadr', 'wounding']
[2-1-0.17183]: ['mae', 'fannie', 'vioxx', 'arthritis', 'enron', 'merck', 'accounting', 'celebrex', 'conrad', 'sanjay']
[3-1-0.3828]: ['arsene', 'wenger', 'arsenal', 'ferguson', 'premiership', 'nistelrooy', 'manchester', 'chelsea', 'striker', 'newcastle']
[4-1-0.2062]: ['bakar', 'arrested', 'hamza', 'suspect', 'jakarta', 'indonesian', 'bashir', 'murder', 'filmmaker', 'guantanamo']
[5-1-0.2292]: ['copyright', 'kazaa', 'copyrighted', 'piracy', 'movie', 'recording', 'lycos', 'liable', 'sharman', 'riaa']
[6-1-0.11278]: ['submarine', 'helen', 'kathmandu', 'volcano', 'maoist', 'earthquake', 'locust', 'mount', 'airliner', 'chicoutimi']
[7-1-0.51741]: ['prix', 'formula', 'schumacher', 'ecclestone', 'barrichello', 'rubens', 'ferrari', 'silverstone', 'jenson', 'bernie']
[8-1-0.29278]: ['enrichment', 'uranium', 'iran', 'tehran', 'atomic', 'nuclear', 'vienna', 'freeze', 'iaea', 'iranian']
[9-1-0.29095]: ['ipod', 'apple', 'nintendo', 'd', 'itunes', 'portable', 'music', 'obtaining', 'playstation', 'sony']
[10-1-0.3764]: ['saturn', 'spacecraft', 'cassini', 'moon', 'capsule', 'nasa', 'genesis', 'astronaut', 'space', 'orbit']
[11-0.18905-0.1813]: ['year', 'ha', 'say', 'time', 'new', 'make', 'world', 'ap', 'wa', 'state']
[12-1-0.18423]: ['slobodan', 'milosevic', 'augusto', 'pinochet', 'nobel', 'cloning', 'wangari', 'maathai', 'yugoslav', 'embryo']
[13-1-0.52732]: ['lleyton', 'federer', 'hewitt', 'mauresmo', 'wta', 'amelie', 'agassi', 'marat', 'sharapova', 'safin']
[14-1-0.19904]: ['equatorial', 'guinea', 'thatcher', 'norodom', 'pitcairn', 'coup', 'sihanouk', 'prince', 'throne', 'mercenary']
[15-1-0.45693]: ['speedway', 'nascar', 'dale', 'earnhardt', 'busch', 'talladega', 'kurt', 'raceway', 'breeder', 'nextel']
[16-1-0.13999]: ['martha', 'stewart', 'prison', 'kobe', 'sentence', 'quattrone', 'ghraib', 'lying', 'bryant', 'steroid']
[17-1-0.25499]: ['medal', 'athens', 'olympic', 'phelps', 'hamm', 'gymnastics', 'kenteris', 'sprinter', 'olympics', 'freestyle']
[18-1-0.30382]: ['manmohan', 'kashmir', 'shaukat', 'aziz', 'musharraf', 'pervez', 'jintao', 'kyoto', 'hu', 'erdogan']
[19-1-0.14067]: ['peoplesoft', 'eliot', 'mclennan', 'spitzer', 'oracle', 'marsh', 'cingular', 'tender', 'ipo', 'initial']
[20-1-0.21738]: ['ryder', 'wicket', 'pga', 'montgomerie', 'icc', 'langer', 'birdie', 'vijay', 'indie', 'jimenez']
[21-0.35571-0.15125]: ['say', 'year', 'ha', 'new', 'wa', 'make', 'outsourcing', 'time', 'quot', 'report']
[22-1-0.3434]: ['darfur', 'sudan', 'sudanese', 'khartoum', 'kofi', 'annan', 'congo', 'bin', 'osama', 'powell']
[23-1-0.20598]: ['eisner', 'ovitz', 'walt', 'disney', 'antitrust', 'microsystems', 'kodak', 'eastman', 'contentguard', 'java']
[24-1-0.2336]: ['willingham', 'tyrone', 'spurrier', 'notre', 'nhl', 'dame', 'zook', 'coaching', 'coach', 'mutu']
[25-1-0.1955]: ['profile', 'quote', 'research', 'yukos', 'lukoil', 'conocophillips', 'earnings', 'quarterly', 'gazprom', 'profit']
[26-0.22238-0.20984]: ['year', 'ha', 'time', 'say', 'new', 'check', 'wa', 'world', 'make', 'said']
[27-0.9-0.22791]: ['greenspan', 'alan', 'reserve', 'chairman', 'federal', 'social', 'budget', 'boom', 'economy', 'survey']
\end{Verbatim}

\subsection{DBPedia}
LDA Collapsed Gibbs sampling
\begin{Verbatim}[breaklines=true, fontsize=\tiny]
npmi=0.2569786099627621
[ 0 - 0.71667 - 0.24385]:  ['company', 'group', 'based', 'international', 'owned', 'founded', 'service', 'airline', 'largest', 'operates']
[ 1 - 0.85 - 0.26205]:  ['island', 'area', 'coast', 'small', 'bay', 'western', 'northern', 'long', 'water', 'pacific']
[ 2 - 0.80909 - 0.25008]:  ['wa', 'car', 'produced', 'model', 'motor', 'sport', 'engine', 'sold', 'production', 'vehicle']
[ 3 - 0.76667 - 0.25635]:  ['city', 'york', 'located', 'building', 'street', 'center', 'hotel', 'tower', 'park', 'hall']
[ 4 - 0.86667 - 0.28198]:  ['journal', 'hospital', 'research', 'medical', 'established', 'society', 'published', 'field', 'health', 'science']
[ 5 - 0.9 - 0.24606]:  ['south', 'north', 'america', 'east', 'central', 'africa', 'eastern', 'southern', 'europe', 'carolina']
[ 6 - 0.92 - 0.22337]:  ['state', 'united', 'washington', 'american', 'massachusetts', 'kingdom', 'jersey', 'oregon', 'maryland', 'boston']
[ 7 - 0.80909 - 0.29375]:  ['wa', 'november', 'october', 'march', 'august', 'september', 'december', 'april', 'june', 'july']
[ 8 - 0.8 - 0.14349]:  ['german', 'ha', 'germany', 'people', 'municipality', 'time', 'swedish', 'norwegian', 'village', 'norway']
[ 9 - 0.76667 - 0.29049]:  ['minister', 'president', 'served', 'born', 'general', 'politician', 'government', 'court', 'chief', 'office']
[ 10 - 0.725 - 0.19019]:  ['county', 'texas', 'ohio', 'district', 'city', 'florida', 'community', 'located', 'west', 'virginia']
[ 11 - 0.93333 - 0.23736]:  ['family', 'moth', 'white', 'black', 'mm', 'brown', 'red', 'green', 'adult', 'feed']
[ 12 - 0.77 - 0.27565]:  ['american', 'michael', 'david', 'john', 'smith', 'robert', 'james', 'scott', 'tom', 'mark']
[ 13 - 0.66667 - 0.28403]:  ['historic', 'house', 'national', 'built', 'place', 'register', 'building', 'listed', 'located', 'home']
[ 14 - 0.81667 - 0.22537]:  ['award', 'chinese', 'ha', 'china', 'international', 'hong', 'kong', 'received', 'traditional', 'academy']
[ 15 - 0.78333 - 0.30854]:  ['series', 'book', 'written', 'comic', 'child', 'story', 'published', 'set', 'character', 'manga']
[ 16 - 0.71667 - 0.33138]:  ['born', 'play', 'played', 'league', 'footballer', 'club', 'professional', 'football', 'player', 'major']
[ 17 - 0.80909 - 0.2279]:  ['wa', 'canadian', 'canada', 'british', 'ontario', 'columbia', 'quebec', 'son', 'toronto', 'september']
[ 18 - 0.91667 - 0.29944]:  ['church', 'england', 'st.', 'catholic', 'parish', 'st', 'christian', 'roman', 'located', 'saint']
[ 19 - 1 - 0.30692]:  ['california', 'san', 'la', 'spanish', 'mexico', 'brazil', 'los', 'angeles', 'francisco', 'el']
[ 20 - 0.8 - 0.34877]:  ['album', 'released', 'record', 'single', 'label', 'music', 'studio', 'hit', 'debut', 'country']
[ 21 - 0.70909 - 0.27038]:  ['wa', 'john', 'william', 'british', 'george', 'charles', 'james', 'thomas', 'robert', 'edward']
[ 22 - 0.85909 - 0.23405]:  ['wa', 'year', 'early', 'late', 'time', 'century', 'originally', 'bridge', 'period', 'date']
[ 23 - 0.86667 - 0.21604]:  ['mountain', 'range', 'located', 'hill', 'ft', 'peak', 'park', 'mount', 'metre', 'valley']
[ 24 - 0.86667 - 0.23436]:  ['school', 'high', 'public', 'student', 'located', 'secondary', 'grade', 'academy', 'middle', 'independent']
[ 25 - 0.83667 - 0.23066]:  ['work', 'art', 'museum', 'artist', 'american', 'history', 'painter', 'ha', 'modern', 'library']
[ 26 - 0.85 - 0.30557]:  ['born', 'world', 'won', 'summer', 'team', 'championship', 'event', 'medal', 'olympics', 'competed']
[ 27 - 0.69167 - 0.25058]:  ['member', 'politician', 'born', 'house', 'party', 'representative', 'served', 'elected', 'january', 'district']
[ 28 - 0.86667 - 0.31682]:  ['university', 'college', 'education', 'campus', 'institute', 'private', 'program', 'founded', 'institution', 'science']
[ 29 - 0.73667 - 0.27641]:  ['music', 'singer', 'born', 'musician', 'american', 'producer', 'jazz', 'blue', 'band', 'composer']
[ 30 - 1 - 0.23266]:  ['french', 'life', 'france', 'needed', 'young', 'le', 'woman', 'citation', 'man', 'paris']
[ 31 - 0.81667 - 0.29395]:  ['company', 'business', 'founded', 'service', 'product', 'inc.', 'firm', 'corporation', 'industry', 'headquartered']
[ 32 - 0.78333 - 0.29084]:  ['specie', 'family', 'genus', 'plant', 'snail', 'endemic', 'sea', 'marine', 'gastropod', 'mollusk']
[ 33 - 0.73409 - 0.13352]:  ['wa', 'republic', 'hockey', 'national', 'ice', 'turkey', 'czech', 'arabic', 'april', 'central']
[ 34 - 0.85 - 0.2474]:  ['river', 'lake', 'tributary', 'romania', 'flow', 'km', 'creek', 'mile', 'area', 'water']
[ 35 - 0.85 - 0.2667]:  ['specie', 'plant', 'habitat', 'native', 'forest', 'common', 'tree', 'tropical', 'endemic', 'natural']
[ 36 - 0.8 - 0.43845]:  ['album', 'released', 'band', 'rock', 'studio', 'live', 'song', 'recorded', 'track', 'release']
[ 37 - 0.71742 - 0.2448]:  ['navy', 'war', 'ship', 'world', 'royal', 'launched', 'wa', 'ii', 'named', 'built']
[ 38 - 0.70076 - 0.1717]:  ['india', 'indian', 'ha', 'wa', 'english', 'government', 'union', 'national', 'tamil', 'sri']
[ 39 - 0.90909 - 0.15204]:  ['wa', 'london', 'king', 'brother', 'irish', 'dutch', 'age', 'ireland', 'philippine', 'scottish']
[ 40 - 0.56167 - 0.19999]:  ['born', 'american', 'football', 'russian', 'national', 'played', 'player', 'professional', 'michigan', 'free']
[ 41 - 0.9 - 0.20425]:  ['japanese', 'italian', 'japan', 'game', 'television', 'video', 'based', 'production', 'medium', 'entertainment']
[ 42 - 0.76742 - 0.25129]:  ['wa', 'class', 'built', 'line', 'railway', 'locomotive', 'service', 'station', 'operated', 'unit']
[ 43 - 0.76742 - 0.25105]:  ['wa', 'aircraft', 'designed', 'built', 'design', 'world', 'air', 'force', 'light', 'construction']
[ 44 - 0.76667 - 0.32288]:  ['published', 'book', 'magazine', 'story', 'writer', 'newspaper', 'author', 'short', 'fiction', 'science']
[ 45 - 0.86667 - 0.16642]:  ['ha', 'australia', 'australian', 'zealand', 'store', 'wale', 'centre', 'south', 'chain', 'mall']
[ 46 - 0.81667 - 0.1886]:  ['ha', 'bank', 'small', 'form', 'crater', 'large', 'greek', 'named', 'called', 'meaning']
[ 47 - 0.9 - 0.35836]:  ['film', 'directed', 'starring', 'star', 'drama', 'comedy', 'role', 'produced', 'written', 'movie']
[ 48 - 0.74167 - 0.27242]:  ['mi', 'village', 'km', 'county', 'poland', 'approximately', 'district', 'kilometre', 'administrative', 'gmina']
[ 49 - 0.725 - 0.29974]:  ['district', 'village', 'province', 'county', 'population', 'census', 'rural', 'iran', 'persian', 'family']
uniqueness=0.8080000000000002
\end{Verbatim}

Online LDA:
\begin{Verbatim}[breaklines=true, fontsize=\tiny]
npmi=0.23031030285194948
[ 0 - 0.81845 - 0.24681]:  ['wa', 'son', 'john', 'born', 'william', 'george', 'father', 'died', 'henry', 'law']
[ 1 - 0.81667 - 0.26355]:  ['located', 'center', 'hotel', 'city', 'building', 'street', 'store', 'tower', 'centre', 'opened']
[ 2 - 1 - 0.15847]:  ['swedish', 'poet', 'republic', 'danish', 'sweden', 'nova', 'congo', 'nigeria', 'israel', 'kenya']
[ 3 - 0.72417 - 0.18097]:  ['wa', 'england', 'london', 'english', 'british', 'irish', 'ireland', 'county', 'cricketer', 'great']
[ 4 - 0.76845 - 0.22279]:  ['won', 'russian', 'born', 'summer', 'wa', 'world', 'olympics', 'medal', 'championship', 'competed']
[ 5 - 0.88333 - 0.26265]:  ['river', 'tributary', 'flow', 'mile', 'creek', 'km', 'water', 'bay', 'near', 'north']
[ 6 - 0.51583 - 0.22305]:  ['wa', 'historic', 'house', 'building', 'built', 'national', 'place', 'register', 'located', 'county']
[ 7 - 0.7625 - 0.24764]:  ['wa', 'aircraft', 'designed', 'built', 'design', 'engine', 'developed', 'produced', 'light', 'fighter']
[ 8 - 0.825 - 0.16935]:  ['class', 'railway', 'locomotive', 'municipality', 'line', 'service', 'bus', 'serbian', 'czech', 'built']
[ 9 - 0.85 - 0.25305]:  ['california', 'san', 'sea', 'snail', 'marine', 'family', 'gastropod', 'specie', 'mollusk', 'mexico']
[ 10 - 1 - 0.26365]:  ['italian', 'la', 'spanish', 'italy', 'spain', 'el', 'del', 'arabic', 'mexican', 'turkish']
[ 11 - 1 - 0.32138]:  ['chinese', 'china', 'hong', 'kong', 'traditional', 'pinyin', 'radio', 'taiwan', 'singapore', 'vietnam']
[ 12 - 0.8375 - 0.25622]:  ['journal', 'research', 'published', 'society', 'peer-reviewed', 'study', 'academic', 'established', 'wa', 'field']
[ 13 - 1 - 0.1873]:  ['le', 'hall', 'rose', 'albert', 'belgian', 'awarded', 'fame', 'jean', 'ray', 'philip']
[ 14 - 0.8375 - 0.22743]:  ['art', 'museum', 'wa', 'century', 'early', 'history', 'late', 'castle', 'work', 'known']
[ 15 - 1 - 0.16756]:  ['island', 'king', 'martin', 'scottish', 'scotland', 'prince', 'alabama', 'miller', 'rhode', 'isle']
[ 16 - 1 - 0.23044]:  ['bank', 'financial', 'puerto', 'branch', 'exchange', 'prison', 'stock', 'real', 'investment', 'rico']
[ 17 - 0.71429 - 0.34343]:  ['born', 'play', 'played', 'footballer', 'football', 'professional', 'club', 'player', 'currently', 'league']
[ 18 - 0.745 - 0.28877]:  ['mi', 'village', 'km', 'poland', 'kilometre', 'district', 'county', 'administrative', 'gmina', 'voivodeship']
[ 19 - 0.72917 - 0.24186]:  ['wa', 'navy', 'ship', 'built', 'royal', 'war', 'class', 'launched', 'named', 'commissioned']
[ 20 - 0.90417 - 0.15579]:  ['french', 'france', 'needed', 'citation', 'airline', 'wa', 'norwegian', 'paris', 'air', 'international']
[ 21 - 0.61845 - 0.25364]:  ['wa', 'born', 'politician', 'minister', 'president', 'party', 'served', 'member', 'national', 'government']
[ 22 - 0.7875 - 0.25469]:  ['magazine', 'published', 'wa', 'newspaper', 'comic', 'news', 'daily', 'medium', 'issue', 'weekly']
[ 23 - 0.41012 - 0.17541]:  ['member', 'house', 'district', 'wa', 'representative', 'born', 'politician', 'served', 'state', 'american']
[ 24 - 0.70417 - 0.13367]:  ['family', 'moth', 'genus', 'specie', 'described', 'mm', 'brown', 'wa', 'bulbophyllum', 'feed']
[ 25 - 0.53512 - 0.24432]:  ['american', 'played', 'league', 'wa', 'football', 'born', 'major', 'professional', 'baseball', 'season']
[ 26 - 0.77083 - 0.16115]:  ['church', 'hockey', 'parish', 'wa', 'st', 'ice', 'christian', 'located', 'cathedral', 'england']
[ 27 - 0.83333 - 0.21282]:  ['game', 'service', 'los', 'video', 'software', 'technology', 'angeles', 'network', 'based', 'medium']
[ 28 - 0.72083 - 0.26739]:  ['world', 'war', 'wa', 'ii', 'military', 'force', 'army', 'union', 'american', 'civil']
[ 29 - 0.93333 - 0.16956]:  ['crater', 'dutch', 'painter', 'far', 'moon', 'netherlands', 'ha', 'rim', 'wall', 'active']
[ 30 - 0.69917 - 0.29778]:  ['district', 'village', 'province', 'population', 'wa', 'county', 'census', 'rural', 'iran', 'persian']
[ 31 - 0.85 - 0.24835]:  ['lake', 'mountain', 'located', 'range', 'peak', 'hill', 'area', 'north', 'park', 'mount']
[ 32 - 1 - 0.17088]:  ['polish', 'golden', 'gordon', 'camp', 'hero', 'knight', 'gate', 'super', 'princess', 'blood']
[ 33 - 0.75 - 0.28135]:  ['specie', 'family', 'genus', 'plant', 'endemic', 'habitat', 'tropical', 'forest', 'natural', 'subtropical']
[ 34 - 0.7375 - 0.31567]:  ['book', 'novel', 'published', 'wa', 'story', 'author', 'written', 'series', 'writer', 'fiction']
[ 35 - 0.93333 - 0.17547]:  ['south', 'australia', 'australian', 'north', 'carolina', 'western', 'wale', 'africa', 'african', 'jersey']
[ 36 - 0.95 - 0.15799]:  ['new', 'zealand', 'hampshire', 'don', 'wave', 'stewart', 'brunswick', 'carter', 'barry', 'auckland']
[ 37 - 0.86667 - 0.21505]:  ['state', 'united', 'texas', 'kingdom', 'florida', 'georgia', 'oregon', 'ohio', 'virginia', 'american']
[ 38 - 0.77083 - 0.20659]:  ['company', 'wa', 'founded', 'group', 'based', 'owned', 'ha', 'corporation', 'product', 'business']
[ 39 - 0.7875 - 0.20195]:  ['japanese', 'wa', 'series', 'japan', 'car', 'manga', 'model', 'motor', 'produced', 'van']
[ 40 - 1 - 0.22157]:  ['german', 'germany', 'portuguese', 'wilson', 'berlin', 'von', 'austria', 'jewish', 'austrian', 'nelson']
[ 41 - 0.95 - 0.22885]:  ['india', 'canada', 'canadian', 'indian', 'ontario', 'columbia', 'quebec', 'british', 'toronto', 'tamil']
[ 42 - 0.81667 - 0.22995]:  ['new', 'york', 'city', 'connecticut', 'queen', 'manhattan', 'morris', 'american', 'sÃ£o', 'hudson']
[ 43 - 0.59762 - 0.20615]:  ['born', 'known', 'music', 'american', 'singer', 'best', 'ha', 'artist', 'musician', 'band']
[ 44 - 0.80417 - 0.37837]:  ['album', 'released', 'wa', 'record', 'band', 'studio', 'label', 'song', 'single', 'music']
[ 45 - 1 - 0.24352]:  ['st.', 'catholic', 'roman', 'philippine', 'saint', 'louis', 'paul', 'lady', 'mary', 'sister']
[ 46 - 0.73333 - 0.2347]:  ['specie', 'known', 'native', 'plant', 'common', 'leaf', 'tree', 'family', 'flower', 'grows']
[ 47 - 0.66583 - 0.18822]:  ['school', 'high', 'located', 'public', 'student', 'district', 'secondary', 'county', 'grade', 'wa']
[ 48 - 0.72083 - 0.28238]:  ['film', 'directed', 'wa', 'starring', 'star', 'written', 'drama', 'based', 'comedy', 'produced']
[ 49 - 0.82083 - 0.24591]:  ['university', 'college', 'education', 'located', 'hospital', 'institute', 'wa', 'science', 'campus', 'degree']
uniqueness=0.81
\end{Verbatim}

ProdLDA:
\begin{Verbatim}[breaklines=true, fontsize=\tiny]
[ 0 - 0.45 - 0.29022]:  ['football', 'league', 'played', 'born', 'hockey', 'nhl', 'player', 'draft', 'olympics', 'footballer']
[ 1 - 0.45 - 0.35073]:  ['politician', 'served', 'representative', 'elected', 'senate', 'constituency', 'assembly', 'election', 'minister', 'representing']
[ 2 - 0.44167 - 0.30271]:  ['leaf', 'grows', 'specie', 'plant', 'cm', 'mm', 'flowering', 'perennial', 'native', 'herb']
[ 3 - 0.29333 - 0.46587]:  ['album', 'released', 'chart', 'billboard', 'track', 'band', 'studio', 'release', 'compilation', 'label']
[ 4 - 0.38333 - 0.34704]:  ['league', 'born', 'football', 'played', 'hockey', 'professional', 'footballer', 'playing', 'nhl', 'player']
[ 5 - 0.44333 - 0.32381]:  ['film', 'directed', 'story', 'written', 'starring', 'fantasy', 'horror', 'fiction', 'manga', 'series']
[ 6 - 0.86667 - 0.26989]:  ['peer-reviewed', 'journal', 'editor-in-chief', 'scientific', 'springer', 'research', 'magazine', 'publication', 'aspect', 'review']
[ 7 - 0.56667 - 0.21863]:  ['tributary', 'river', 'flow', 'mountain', 'crater', 'lake', 'sawtooth', 'rim', 'permit', 'southwest']
[ 8 - 0.49333 - 0.31468]:  ['film', 'directed', 'starring', 'written', 'story', 'supporting', 'cannes', 'series', 'book', 'drama']
[ 9 - 0.64167 - 0.28355]:  ['album', 'released', 'manga', 'comic', 'edition', 'anime', 'volume', 'series', 'serialized', 'song']
[ 10 - 0.40833 - 0.33874]:  ['grows', 'leaf', 'flowering', 'specie', 'plant', 'tall', 'native', 'flower', 'shrub', 'erect']
[ 11 - 0.26667 - 0.24182]:  ['mi', 'kilometre', 'voivodeship', 'gmina', 'lie', 'administrative', 'km', 'approximately', 'village', 'poland']
[ 12 - 0.48333 - 0.36666]:  ['historic', 'register', 'building', 'built', 'added', 'dwelling', 'revival', 'roof', 'listed', 'gable']
[ 13 - 0.71667 - 0.31923]:  ['university', 'education', 'institution', 'peer-reviewed', 'undergraduate', 'college', 'affiliated', 'journal', 'graduate', 'academic']
[ 14 - 0.35 - 0.24971]:  ['mi', 'lie', 'km', 'voivodeship', 'gmina', 'kilometre', 'approximately', 'administrative', 'poland', 'regional']
[ 15 - 0.6 - 0.34841]:  ['navy', 'ship', 'commissioned', 'laid', 'launched', 'submarine', 'hm', 'bremen', 'twenty-four', 'naval']
[ 16 - 0.41667 - 0.28862]:  ['school', 'college', 'student', 'high', 'public', 'grade', 'university', 'republican', 'education', 'senate']
[ 17 - 0.53333 - 0.34759]:  ['historic', 'register', 'built', 'porch', 'revival', 'added', 'brick', 'church', 'dwelling', 'listed']
[ 18 - 0.81667 - 0.25104]:  ['peer-reviewed', 'journal', 'quarterly', 'indexed', 'topic', 'publishes', 'provides', 'technology', 'healthcare', 'privately']
[ 19 - 0.43333 - 0.35964]:  ['league', 'played', 'football', 'born', 'player', 'professional', 'season', 'fc', 'footballer', 'nba']
[ 20 - 0.29167 - 0.39506]:  ['district', 'census', 'romanized', 'population', 'iran', 'persian', 'rural', 'province', 'village', 'county']
[ 21 - 0.39333 - 0.48241]:  ['album', 'released', 'peaked', 'band', 'chart', 'release', 'ep', 'billboard', 'label', 'studio']
[ 22 - 0.29167 - 0.39506]:  ['district', 'romanized', 'census', 'population', 'iran', 'persian', 'rural', 'province', 'county', 'village']
[ 23 - 0.39333 - 0.43828]:  ['album', 'released', 'song', 'studio', 'band', 'release', 'chart', 'music', 'record', 'dvd']
[ 24 - 0.61667 - 0.22685]:  ['mountain', 'river', 'tributary', 'lake', 'divide', 'confluence', 'flow', 'lunar', 'km2', 'westward']
[ 25 - 0.45 - 0.3877]:  ['politician', 'served', 'assembly', 'minister', 'constituency', 'elected', 'legislative', 'election', 'deputy', 'republican']
[ 26 - 0.26667 - 0.24182]:  ['mi', 'lie', 'kilometre', 'gmina', 'voivodeship', 'km', 'administrative', 'approximately', 'village', 'poland']
[ 27 - 0.56667 - 0.21374]:  ['school', 'high', 'public', 'grade', 'located', 'student', 'unincorporated', 'co-educational', 'four-year', 'secondary']
[ 28 - 0.24167 - 0.25698]:  ['mi', 'village', 'district', 'voivodeship', 'gmina', 'lie', 'kilometre', 'county', 'population', 'administrative']
[ 29 - 0.29167 - 0.39506]:  ['district', 'romanized', 'census', 'population', 'iran', 'persian', 'rural', 'province', 'village', 'county']
[ 30 - 0.31833 - 0.47689]:  ['album', 'released', 'studio', 'song', 'band', 'billboard', 'release', 'chart', 'track', 'recorded']
[ 31 - 0.56 - 0.31184]:  ['film', 'directed', 'starring', 'story', 'written', 'silent', 'comedy', 'star', 'award', 'upcoming']
[ 32 - 0.43333 - 0.35178]:  ['league', 'played', 'football', 'born', 'player', 'won', 'season', 'professional', 'footballer', 'baseball']
[ 33 - 0.35833 - 0.45949]:  ['grows', 'leaf', 'stem', 'perennial', 'herb', 'centimeter', 'shrub', 'flowering', 'flower', 'plant']
[ 34 - 0.85 - 0.31775]:  ['aircraft', 'engine', 'kit', 'cc', 'conventional', 'convertible', 'car', 'kw', 'mid-size', 'configuration']
[ 35 - 0.5 - 0.38277]:  ['politician', 'elected', 'legislative', 'served', 'election', 'constituency', 'representative', 'cabinet', 'democratic', 'minister']
[ 36 - 0.41667 - 0.3537]:  ['habitat', 'specie', 'threatened', 'family', 'tropical', 'subtropical', 'moist', 'loss', 'endemic', 'natural']
[ 37 - 0.34167 - 0.35938]:  ['leaf', 'perennial', 'stem', 'flower', 'centimeter', 'plant', 'tall', 'grows', 'herb', 'specie']
[ 38 - 0.41667 - 0.37624]:  ['specie', 'habitat', 'tropical', 'subtropical', 'family', 'moist', 'threatened', 'endemic', 'lowland', 'loss']
[ 39 - 0.46 - 0.34544]:  ['film', 'directed', 'written', 'novel', 'starring', 'story', 'drama', 'novella', 'comedy', 'fantasy']
[ 40 - 0.95 - 0.27988]:  ['software', 'company', 'headquartered', 'investment', 'inc.', 'provider', 'operates', 'product', 'develops', 'privately']
[ 41 - 0.55 - 0.35021]:  ['navy', 'ship', 'warship', 'commissioned', 'destroyer', 'hm', 'laid', 'launched', 'lt.', 'war']
[ 42 - 0.56667 - 0.25608]:  ['school', 'grade', 'high', 'public', 'located', 'student', 'preparatory', 'caters', 'secondary', 'coeducational']
[ 43 - 0.34333 - 0.45168]:  ['album', 'released', 'chart', 'hit', 'song', 'record', 'band', 'billboard', 'studio', 'compilation']
[ 44 - 0.51667 - 0.23861]:  ['flow', 'lake', 'rim', 'elevation', 'river', 'crater', 'tributary', 'mountain', 'tidal', 'lunar']
[ 45 - 0.76 - 0.31916]:  ['film', 'directed', 'starring', 'hai', 'role', 'remake', 'hindi', 'lead', 'telugu', 'sen']
[ 46 - 0.51667 - 0.28064]:  ['specie', 'habitat', 'tropical', 'subtropical', 'family', 'moist', 'mollusk', 'threatened', 'gastropod', 'montane']
[ 47 - 0.58333 - 0.34982]:  ['historic', 'register', 'building', 'two-story', 'built', 'brick', 'doric', 'listed', 'roof', 'pile']
[ 48 - 0.55 - 0.33823]:  ['navy', 'laid', 'ship', 'commissioned', 'destroyer', 'sponsored', 'launched', 'mrs.', 'hm', 'command']
[ 49 - 0.85 - 0.37107]:  ['motor', 'vehicle', 'engine', 'bmw', 'manufactured', 'motorcycle', 'aircraft', 'hp', 'car', 'automaker']
\end{Verbatim}

NTM-R:
\begin{Verbatim}[breaklines=true, fontsize=\tiny]
[0-1-0.17993]: ['muricidae', 'murex', 'snail', 'gastropod', 'mollusk', 'thrash', 'melodic', 'mordella', 'superfamily', 'peaked']
[1-0.41644-0.13796]: ['taxonomy', 'algae', 'specifically', 'tephritid', 'tephritidae', 'ray-finned', 'fruit', 'bromeliad', 'coordinate', 'fly']
[2-0.62333-0.21805]: ['policy', 'suggest', 'obama', 'israeli', 'recognition', 'banking', 'firm', 'intelligence', 'african', 'advice']
[3-1-0.24395]: ['league', 'afl', 'football', 'batsman', 'right-handed', 'rugby', 'right-arm', 'vfl', 'premiership', 'midfielder']
[4-0.66012-0.11442]: ['baron', 'bates', 'ray-finned', 'pc', 'chacteau', 'gcmg', 'statesman', 'mcgill', 'cooke', 'mildred']
[5-0.43644-0.22398]: ['specifically', 'algae', 'ray-finned', 'taxonomy', 'suggest', 'seeking', 'reduce', 'increasing', 'aim', 'objective']
[6-0.29739-0.070157]: ['ray-finned', 'taxonomy', 'bates', 'tillandsia', 'algae', 'viscount', 'specifically', 'schaus', 'earl', 'pc']
[7-0.62958-0.20941]: ['algae', 'taxonomy', 'avoid', 'achieve', 'unique', 'balance', 'finding', 'laying', 'everyday', 'feel']
[8-0.78095-0.10182]: ['bates', 'peck', 'brendan', 'fraser', 'lillian', 'sylvia', 'archibald', 'tillandsia', 'carabidae', 'mabel']
[9-0.73125-0.19545]: ['algae', 'israeli', 'keeping', 'sort', 'meant', 'approach', 'arab', 'equivalent', 'dealing', 'south-western']
[10-1-0.14948]: ['faboideae', 'scotia', 'quebec', 'ftse', 'ferry', 'cruise', 'halifax', 'olsztyn.before', 'nova', 'm']
[11-0.555-0.2645]: ['economic', 'aim', 'policy', 'civil', 'responsibility', 'keeping', 'weapon', 'diplomatic', 'turning', 'possibility']
[12-0.67436-0.25298]: ['seeking', 'continuing', 'effort', 'diplomatic', 'specifically', 'maintain', 'culture', 'regarding', 'monitoring', 'cell']
[13-0.51429-0.10161]: ['deh', 'tillandsia', 'viscount', 'bates', 'meyrick', 'talbot', 'mildred', 'earl', 'archibald', 'eliza']
[14-0.68103-0.22395]: ['economic', 'improved', 'critical', 'lack', 'emphasis', 'specifically', 'preparing', 'taxonomy', 'protest', 'immigration']
[15-0.86429-0.16632]: ['bates', 'incomplete', 'smith', 'watson', 'mccarthy', 'johnston', 'perkins', 'gould', 'editor', 'mann']
[16-0.41644-0.21118]: ['algae', 'taxonomy', 'specifically', 'establishing', 'handling', 'increase', 'economic', 'keeping', 'difficult', 'ray-finned']
[17-1-0.088445]: ['eupithecia', 'geometridae', 'scopula', 'baluchestan', 'sistan', 'coleophora', 'coleophoridae', 'urdu', 'pterophoridae', 'arctiidae']
[18-0.74762-0.11147]: ['marquess', 'styled', 'bates', 'meyrick', 'viscount', 'nobleman', 'deh', 'engraver', 'pietro', 'bavaria']
[19-0.52061-0.223]: ['taxonomy', 'algae', 'specifically', 'unable', 'aim', 'funding', 'analysis', 'maintain', 'finding', 'priority']
[20-1-0.21449]: ['olympics', 'fencer', 'bulgarian', 'swimmer', 'competed', 'gymnast', 'eurovision', 'medalist', 'handball', 'budapest']
[21-1-0.31143]: ['senate', 'republican', 'constituency', 'representing', 'janata', 'attorney', 'election', 'legislative', 'delegate', 'caucus']
[22-1-0.22003]: ['clinical', 'healthcare', 'campus', 'peer-reviewed', 'undergraduate', 'theological', 'coeducational', 'publishes', 'adventist', 'preparatory']
[23-0.86-0.2594]: ['possibility', 'risk', 'counter', 'regime', 'need', 'profile', 'minimum', 'meant', 'mission', 'relevant']
[24-1-0.3246]: ['painting', 'sculpture', 'poem', 'drawing', 'museum', 'art', 'exhibition', 'illustrator', 'collection', 'poetry']
[25-0.785-0.25122]: ['tax', 'intelligence', 'controversial', 'possibility', 'reason', 'situation', 'security', 'credit', 'keeping', 'grass']
[26-0.39978-0.079958]: ['ray-finned', 'tephritidae', 'algae', 'tephritid', 'taxonomy', 'tillandsia', 'ulmus', 'elm', 'specifically', 'lago']
[27-0.59458-0.24169]: ['crisis', 'difficult', 'algae', 'iraq', 'driven', 'possibility', 'identification', 'instance', 'policy', 'change']
[28-0.74762-0.16082]: ['bates', 'firm', 'fowler', 'economist', 'nicholson', 'consulting', 'reynolds', 'banking', 'watkins', 'reid']
[29-0.63061-0.22814]: ['taxonomy', 'specifically', 'algae', 'contact', 'possibility', 'mind', 'prepare', 'robust', 'increasingly', 'significant']
[30-1-0.42495]: ['romania', 'tributary', 'valea', 'olt', 'river', 'mica83', 'pacracul', 'izvorul', 'racul', 'headwater']
[31-1-0.152]: ['bony', 'epoch', 'centimetre', 'grape', 'prehistoric', 'glacier', 'grevillea', 'volcanic', 'massif', 'hispanicized']
[32-0.57667-0.28435]: ['crisis', 'allow', 'possibility', 'increased', 'virtually', 'balance', 'belonging', 'difficult', 'protection', 'gain']
[33-0.65625-0.15701]: ['algae', 'castle', 'bringing', 'chacteau', 'energy', 'taxonomy', 'campaign', 'possibility', 'affected', 'assigned']
[34-1-0.35761]: ['hm', 'destroyer', 'minesweeper', 'sloop', 'navy', 'frigate', 'hmcs', 'patrol', 'admiral', 'clemson-class']
[35-0.715-0.27075]: ['committee', 'protection', 'planning', 'advisory', 'policy', 'virtually', 'movement', 'suggest', 'intervention', 'wroca82aw']
[36-0.44894-0.22927]: ['algae', 'taxonomy', 'suggest', 'virtually', 'balance', 'showing', 'specifically', 'ideal', 'purpose', 'build']
[37-0.26644-0.1139]: ['ray-finned', 'chacteau', 'bromeliad', 'algae', 'taxonomy', 'tephritidae', 'tillandsia', 'tephritid', 'pitcairnia', 'specifically']
[38-1-0.41582]: ['homebuilt', 'ultralight', 'trike', 'undercarriage', 'ready-to-fly-aircraft', 'low-wing', 'two-seat', 'single-engine', 'monoplane', 'single-seat']
[39-0.41728-0.19018]: ['taxonomy', 'polish', 'algae', 'specifically', 'striking', 'netherlands', 'suggest', 'finding', 'maintain', 'possibility']
[40-0.44208-0.059865]: ['tephritid', 'taxonomy', 'ray-finned', 'tephritidae', 'ulidiidae', 'algae', 'neoregelia', 'tillandsia', 'mantis', 'picture-winged']
[41-0.66833-0.27628]: ['virtually', 'sector', 'requires', 'showing', 'monitoring', 'emphasis', 'resulting', 'impact', 'possibility', 'concern']
[42-0.83333-0.15431]: ['incomplete', 'firm', 'jenkins', 'dixon', 'emma', 'nigel', 'watkins', 'consultant', 'investment', 'dc']
[43-1-0.49868]: ['threatened', 'ecuador.its', 'forests.it', 'habitat', 'arecaceae', 'loss', 'family.it', 'montane', 'moist', 'subtropical']
[44-0.635-0.28883]: ['effectively', 'difficult', 'emphasis', 'possibility', 'potential', 'diplomatic', 'concerned', 'illegal', 'emerging', 'crisis']
[45-1-0.37187]: ['horror', 'fantasy', 'thriller', 'drama', 'comedy', 'comedy-drama', 'starring', 'directed', 'anthology', 'sequel']
[46-1-0.15964]: ['pornographic', 'hop', 'clothing', 'hip', 'thoroughbred', 'retailer', 'arranger', 'dj', 'stand-up', 'store']
[47-0.59061-0.19672]: ['algae', 'allowing', 'taxonomy', 'improvement', 'charge', 'laying', 'invasion', 'policy', 'expensive', 'specifically']
[48-0.88333-0.10236]: ['bromeliad', 'olyÄ\x81', 'olya', 'bulbophyllum', 'poznaÅ\x84', 'poaceae', 'neoregelia', 'nowy', 'masovian', 'mazowiecki']
[49-1-0.12801]: ['herzegovina', 'bosnia', 'croatia', 'connected', 'estonia', 'municipality', 'kuyavian-pomeranian', 'northern-central', 'highway', 'kielce.the']
\end{Verbatim}

W-LDA:
\begin{Verbatim}[breaklines=true, fontsize=\tiny]
[0-1-0.3445]: ['tournament', 'championship', 'cup', 'tennis', 'career-high', 'ncaa', 'season', 'fifa', 'player', 'scoring']
[1-1-0.26173]: ['peer-reviewed', 'journal', 'publishes', 'wiley-blackwell', 'quarterly', 'opinion', 'editor-in-chief', 'topic', 'theoretical', 'biannual']
[2-1-0.28195]: ['snail', 'ally', 'fasciolariidae', 'gastropod', 'tulip', 'mollusk', 'spindle', 'circuit', 'muricidae', 'eulimidae']
[3-1-0.23924]: ['presenter', 'arranger', 'songwriter', 'multi-instrumentalist', 'performer', 'sitcom', 'actress', 'conductor', 'composer', 'comedian']
[4-1-0.40359]: ['pinyin', 'chinese', 'simplified', 'wade–giles', 'guangzhou', 'guangdong', 'yuan', 'jyutping', 'mandarin', 'taipei']
[5-1-0.23772]: ['shopping', 'mall', 'mixed-use', 'parking', 'm2', 'anchored', 'condominium', 'hotel', 'prison', 'high-rise']
[6-1-0.32526]: ['coleophora', 'coleophoridae', 'wingspan', 'august.the', 'elachista', 'elachistidae', 'larva', 'iberian', 'year.the', 'hindwings']
[7-1-0.29773]: ['solution', 'software', 'provider', 'multinational', 'telecommunication', 'nasdaq', 'investment', 'outsourcing', 'semiconductor', 'asset']
[8-1-0.39978]: ['inflorescence', 'erect', 'raceme', 'ovate', 'panicle', 'stem', 'leaflet', 'toothed', 'frond', 'lanceolate']
[9-1-0.37453]: ['made-for-tv', 'documentary', 'made-for-television', 'directed', 'screenplay', 'starring', 'comedy-drama', 'technicolor', 'sundance', 'film']
[10-1-0.22754]: ['translator', 'essayist', 'poet', 'novelist', 'literary', 'poetry', 'screenwriter', 'short-story', 'bridgeport', 'siedlce']
[11-1-0.24646]: ['summit', 'hiking', 'glacier', 'subrange', 'snowdonia', 'traversed', 'peak', 'glacial', 'pas', 'mountain']
[12-1-0.44789]: ['thrash', 'punk', 'metal', 'band', 'drummer', 'melodic', 'bassist', 'hardcore', 'demo', 'line-up']
[13-1-0.18155]: ['shortlisted', 'booker', 'newbery', 'young-adult', 'nobel', 'qal', 'marriage', 'prize', 'bestseller', 'autobiographical']
[14-1-0.44883]: ['kapoor', 'dharmendra', 'tamil-language', 'pivotal', 'bollywood', 'khanna', 'vinod', 'sinha', 'mithun', 'shetty']
[15-1-0.21445]: ['congressional', 'republican', 'iowa', 'arizona', 'kansa', 'missouri', 'diego', 'tempore', 'dodge', 'wyoming']
[16-1-0.24376]: ['fc', 'sergei', 'ssr', 'midfielder', 'división', 'russian', 'footballer', 'aleksandrovich', 'belarusian', 'vladimirovich']
[17-1-0.39128]: ['indonesia', 'lankan', 'indonesian', 'malaysia', 'java', 'jakarta', 'brunei', 'sri', 'lanka', 'sinhala']
[18-1-0.48659]: ['two-seat', 'fuselage', 'single-engine', 'monoplane', 'prototype', 'kw', 'airliner', 'single-engined', 'twin-engined', 'aircraft']
[19-1-0.40455]: ['wale', 'sydney', 'australian', 'brisbane', 'australia', 'queensland', 'melbourne', 'adelaide', 'nsw', 'perth']
[20-1-0.39939]: ['kerman', 'persian', 'jonubi', 'tehran', 'kermanshah', 'iran', 'isfahan', 'romanized', 'razavi', 'rural']
[21-1-0.27091]: ['rhode', 'oahu', 'hawaii', 'hawaiian', 'maui', 'honolulu', 'hawaii', 'mordella', 'massachusetts', 'tenebrionoidea']
[22-1-0.34612]: ['brandenburg', 'schleswig-holstein', 'und', 'saxony', 'germany', 'für', 'hamburg', 'mecklenburg-vorpommern', 'german', 'austria']
[23-1-0.3073]: ['register', 'historic', 'added', 'two-story', 'brick', 'massachusetts.the', 'armory', 'one-story', 'three-story', 'revival']
[24-1-0.32644]: ['fantasy', 'universe', 'paperback', 'hardcover', 'marvel', 'comic', 'role-playing', 'conan', 'sword', 'dungeon']
[25-1-0.16965]: ['railway', 'brewing', 'newspaper', 'brewery', 'ferry', 'tabloid', 'caledonian', 'daily', 'railroad', 'roster']
[26-1-0.2635]: ['french', 'du', 'la', 'château', 'france', 'playstation', 'renault', 'le', 'et', 'french-language']
[27-1-0.087707]: ['orchid', 'trance', 'dj', 'zanjan', 'techno', 'tappeh', 'orchidaceae', 'baden-württemberg', 'fabric', 'wasp']
[28-1-0.28215]: ['poland', 'administrative', 'voivodeship', 'north-west', 'gmina', 'mi', 'kielce', 'masovian', 'west-central', 'poznań']
[29-1-0.14504]: ['moth', 'geometridae', 'arctiidae', 'notodontidae', 'turridae', 'turrids', 'crambidae', 'eupithecia', 'raphitomidae', 'scopula']
[30-1-0.45268]: ['compilation', 'chart', 'billboard', 'hit', 'peaked', 'itunes', 'charted', 'riaa', 'remixes', 'airplay']
[31-1-0.36847]: ['leptodactylidae', 'eleutherodactylus', 'ecuador.its', 'forests.it', 'brazil.its', 'high-altitude', 'shrubland', 'subtropical', 'rivers.it', 'frog']
[32-1-0.46631]: ['vessel', 'patrol', 'navy', 'convoy', 'ship', 'anti-submarine', 'auxiliary', 'destroyer', 'escort', 'naval']
[33-1-0.36041]: ['undergraduate', 'postgraduate', 'doctoral', 'degree', 'faculty', 'bachelor', 'nursing', 'university', 'post-graduate', 'post-secondary']
[34-1-0.14203]: ['picture-winged', 'ulidiid', 'fly', 'tephritidae', 'firearm', 'tachinidae', 'ulidiidae', 'footwear', 'apparel', 'tephritid']
[35-1-0.15219]: ['prehistoric', 'bony', 'legume', 'faboideae', 'asteraceae', 'cephalopod', 'fabaceae', 'clam', 'daisy', 'bivalve']
[36-1-0.14339]: ['alberta', 'portland', 'oregon', 'columbia', 'vancouver', 'omaha', 'saskatchewan', 'davenport', 'hokkaidō', 'mysore']
[37-1-0.20091]: ['davidii', 'priory', 'dorset', 'exeter', 'surrey', 'buddleja', 'gloucestershire', 'deptford', 'wiltshire', 'edinburgh']
[38-1-0.39987]: ['church', 'diocese', 'parish', 'jesus', 'congregation', 'holy', 'christ', 'cathedral', 'deanery', 'roman']
[39-1-0.33474]: ['mascot', 'elementary', 'ib', 'kindergarten', 'enrollment', 'pre-kindergarten', 'school', 'secondary', 'preschool', 'high']
[40-1-0.19429]: ['pradesh', 'yugoslav', 'serbian', 'novi', 'andhra', 'india', 'cyrillic', 'mandal', 'maharashtra', 'kerala']
[41-1-0.10226]: ['bosnian', 'palm', 'turtle', 'thai', 'ready-to-fly-aircraft', 'supplied', 'lil', 'amateur', 'mixtape', 'rapper']
[42-1-0.42669]: ['sculpture', 'photography', 'gallery', 'painting', 'museum', 'exhibition', 'exhibited', 'curator', 'art', 'sculptor']
[43-1-0.38726]: ['tributary', 'pârâul', 'valea', 'romania', 'river', 'mureş', 'mic', 'transylvania', 'mică', 'olt']
[44-1-0.13235]: ['tillandsia', 'spider', 'salticidae', 'jumping', 'poaceae', 'praying', 'ant', 'neoregelia', 'mantis', 'neotropical']
[45-1-0.12438]: ['estonia', 'bistriţa', 'pärnu', 'ccm', 'michigan', 'estonian', 'tanzanian', 'lycaenidae', 'saare', 'tartu']
[46-1-0.38204]: ['santa', 'cruz', 'josé', 'luis', 'maría', 'mexican', 'carlos', 'cuba', 'juan', 'chilean']
[47-1-0.36108]: ['cabinet', 'minister', 'election', 'legislative', 'fáil', 'secretary', 'conservative', 'constituency', 'dála', 'teachta']
[48-1-0.40629]: ['italian', 'di', 'francesco', 'italy', 'baroque', 'giuseppe', 'lombardy', 'rome', 'carlo', 'luca']
[49-1-0.17494]: ['greek', 'greece', 'baluchestan', 'sistan', 'sixth', 'yorkshire', 'status', 'khash', 'chabahar', 'specialist']
\end{Verbatim}

\subsection{Yelp Review Polarity}

LDA Collapsed Gibbs sampling:
\begin{Verbatim}[breaklines=true, fontsize=\tiny]
npmi=0.23787181653390055
[ 0 - 0.85 - 0.25418]:  ['water', 'dirty', 'clean', 'smell', 'door', 'bathroom', 'wall', 'floor', 'hand', 'cleaning']
[ 1 - 0.59167 - 0.38849]:  ['steak', 'dish', 'restaurant', 'meal', 'dinner', 'cooked', 'potato', 'menu', 'lobster', 'dessert']
[ 2 - 0.58333 - 0.2649]:  ['walked', 'guy', 'asked', 'counter', 'lady', 'looked', 'girl', 'wanted', 'walk', 'door']
[ 3 - 0.52 - 0.27734]:  ['thing', 'make', 'ca', 'doe', 'kind', 'people', 'feel', 'wrong', 'stuff', 'big']
[ 4 - 0.67769 - 0.2165]:  ['burger', 'fry', 'cheese', 'onion', 'hot', 'ordered', 'good', 'mac', 'sweet', 'potato']
[ 5 - 0.58667 - 0.19482]:  ['wa', 'tasted', 'cold', 'dry', 'bland', 'ordered', 'taste', 'bad', 'looked', 'disappointed']
[ 6 - 0.61167 - 0.21671]:  ['club', 'people', 'night', 'music', 'girl', 'guy', 'party', 'friend', 'group', 'crowd']
[ 7 - 0.75333 - 0.21353]:  ['great', 'love', 'amazing', 'recommend', 'awesome', 'service', 'favorite', 'highly', 'loved', 'excellent']
[ 8 - 0.93333 - 0.26762]:  ['money', 'pay', 'extra', 'charge', 'dollar', 'paid', 'worth', 'free', 'cost', 'tip']
[ 9 - 0.78429 - 0.1659]:  ['vega', 'le', 'la', 'strip', 'trip', 'place', 'service', 'pour', 'montreal', 'san']
[ 10 - 0.68333 - 0.24412]:  ['car', 'work', 'guy', 'day', 'problem', 'needed', 'change', 'company', 'job', 'tire']
[ 11 - 0.66667 - 0.25441]:  ['phone', 'card', 'called', 'day', 'credit', 'company', 'told', 'number', 'business', 'month']
[ 12 - 0.58095 - 0.19303]:  ['staff', 'friendly', 'great', 'nice', 'coffee', 'super', 'clean', 'helpful', 'place', 'quick']
[ 13 - 0.6075 - 0.21777]:  ['service', 'bad', 'wa', 'time', 'experience', 'horrible', 'terrible', 'worst', 'slow', 'poor']
[ 14 - 0.69167 - 0.25876]:  ['drink', 'bar', 'night', 'happy', 'hour', 'friend', 'bartender', 'friday', 'saturday', 'cocktail']
[ 15 - 0.67333 - 0.19213]:  ['table', 'server', 'waitress', 'waiter', 'ordered', 'food', 'restaurant', 'seated', 'drink', 'water']
[ 16 - 0.71103 - 0.21877]:  ['pizza', 'sauce', 'cheese', 'wing', 'good', 'pasta', 'italian', 'slice', 'ordered', 'crust']
[ 17 - 0.76603 - 0.19741]:  ['breakfast', 'egg', 'wa', 'good', 'bacon', 'brunch', 'coffee', 'french', 'morning', 'pancake']
[ 18 - 0.56583 - 0.2383]:  ['line', 'time', 'people', 'hour', 'long', 'day', 'airport', 'late', 'wait', 'flight']
[ 19 - 0.88333 - 0.24362]:  ['room', 'hotel', 'stay', 'pool', 'casino', 'bed', 'stayed', 'night', 'strip', 'desk']
[ 20 - 0.67679 - 0.19974]:  ['place', 'love', 'super', 'dont', 'die', 'man', 'time', 'awesome', 'didnt', 'na']
[ 21 - 0.55417 - 0.20917]:  ['wa', 'hair', 'cut', 'time', 'wanted', 'short', 'groupon', 'left', 'long', 'looked']
[ 22 - 0.73667 - 0.23585]:  ['location', 'lot', 'parking', 'open', 'area', 'close', 'drive', 'street', 'ha', 'closed']
[ 23 - 0.95 - 0.27537]:  ['store', 'shop', 'item', 'buy', 'product', 'sale', 'bought', 'stuff', 'shopping', 'sell']
[ 24 - 0.625 - 0.23123]:  ['wa', 'husband', 'wife', 'friend', 'birthday', 'family', 'wanted', 'decided', 'mom', 'day']
[ 25 - 0.55269 - 0.23514]:  ['food', 'buffet', 'good', 'wa', 'crab', 'dinner', 'eat', 'seafood', 'shrimp', 'worth']
[ 26 - 0.72917 - 0.23019]:  ['dog', 'care', 'office', 'day', 'appointment', 'time', 'doctor', 'dr.', 'staff', 'patient']
[ 27 - 0.41603 - 0.25941]:  ['wa', 'good', 'pretty', 'nice', 'bit', 'thing', 'thought', 'kind', 'ok.', 'big']
[ 28 - 0.84103 - 0.25157]:  ['price', 'small', 'quality', 'high', 'portion', 'size', 'large', 'reasonable', 'worth', 'good']
[ 29 - 0.46198 - 0.25428]:  ['food', 'restaurant', 'good', 'eat', 'service', 'place', 'fast', 'eating', 'meal', 'average']
[ 30 - 0.68667 - 0.17554]:  ['ha', 'work', 'class', 'make', 'feel', 'gym', 'school', 'offer', 'member', 'doe']
[ 31 - 0.76103 - 0.25206]:  ['taco', 'chip', 'mexican', 'bean', 'food', 'salsa', 'good', 'burrito', 'bbq', 'sauce']
[ 32 - 0.50417 - 0.18603]:  ['wa', 'nail', 'time', 'day', 'massage', 'job', 'foot', 'work', 'experience', 'lady']
[ 33 - 0.53864 - 0.20588]:  ['sushi', 'roll', 'fish', 'good', 'fresh', 'place', 'menu', 'wa', 'chef', 'eat']
[ 34 - 0.86667 - 0.24305]:  ['review', 'star', 'yelp', 'experience', 'read', 'bad', 'reason', 'based', 'write', 'rating']
[ 35 - 0.58333 - 0.26748]:  ['wa', 'told', 'asked', 'manager', 'wanted', 'left', 'called', 'offered', 'gave', 'point']
[ 36 - 0.35364 - 0.22002]:  ['place', 'good', 'ha', 'pretty', 'people', 'friend', 'thing', 'lot', 'town', 'cheap']
[ 37 - 0.85 - 0.33754]:  ['cream', 'ice', 'chocolate', 'cake', 'tea', 'sweet', 'flavor', 'dessert', 'taste', 'delicious']
[ 38 - 0.75333 - 0.20119]:  ['local', 'phoenix', 'town', 'city', 'ha', 'live', 'street', 'area', 'downtown', 'valley']
[ 39 - 0.7825 - 0.23043]:  ['time', 'year', 'ha', 'visit', 'ago', 'week', 'couple', 'past', 'month', 'coming']
[ 40 - 0.75 - 0.19678]:  ['nice', 'area', 'decor', 'seating', 'inside', 'patio', 'feel', 'atmosphere', 'beautiful', 'bit']
[ 41 - 0.81667 - 0.27044]:  ['kid', 'game', 'watch', 'fun', 'big', 'play', 'tv', 'movie', 'lot', 'child']
[ 42 - 0.69 - 0.21958]:  ['customer', 'service', 'rude', 'business', 'owner', 'employee', 'attitude', 'care', 'people', 'manager']
[ 43 - 0.78333 - 0.41034]:  ['dish', 'chicken', 'rice', 'soup', 'fried', 'thai', 'noodle', 'sauce', 'beef', 'chinese']
[ 44 - 0.48936 - 0.23629]:  ['salad', 'chicken', 'wa', 'ordered', 'meal', 'food', 'soup', 'plate', 'dressing', 'good']
[ 45 - 0.63031 - 0.1956]:  ['beer', 'great', 'wine', 'selection', 'good', 'place', 'glass', 'menu', 'bar', 'list']
[ 46 - 0.78269 - 0.25272]:  ['sandwich', 'lunch', 'menu', 'option', 'bread', 'meat', 'fresh', 'special', 'good', 'choice']
[ 47 - 0.68333 - 0.27003]:  ['wa', 'boyfriend', 'thought', 'decided', 'felt', 'surprised', 'disappointed', 'impressed', 'excited', 'looked']
[ 48 - 0.9 - 0.18477]:  ['event', 'picture', 'seat', 'fun', 'art', 'ticket', 'photo', 'cool', 'music', 'stage']
[ 49 - 0.72917 - 0.23758]:  ['minute', 'order', 'wait', 'time', 'waiting', 'waited', 'long', 'hour', 'finally', 'min']
uniqueness=0.6839999999999999
\end{Verbatim}

Online LDA:
\begin{Verbatim}[breaklines=true, fontsize=\tiny]
npmi=0.23341299435543492
[ 0 - 0.42909 - 0.22503]:  ['customer', 'service', 'time', 'rude', 'people', 'place', 'employee', 'just', 'like', 'staff']
[ 1 - 0.93333 - 0.21528]:  ['happy', 'hour', 'shrimp', 'crab', 'seafood', 'pita', 'oyster', 'gyro', 'greek', 'hummus']
[ 2 - 0.77917 - 0.21341]:  ['airport', 'flight', 'ride', 'driver', 'cab', 'san', 'bus', 'u', 'hour', 'time']
[ 3 - 0.95 - 0.37322]:  ['cake', 'chocolate', 'dessert', 'cupcake', 'sweet', 'butter', 'pie', 'bakery', 'cream', 'cheesecake']
[ 4 - 0.95 - 0.18921]:  ['year', 'kid', 'old', 'ha', 'ago', 'family', 'used', 'daughter', 'son', 'child']
[ 5 - 0.37076 - 0.22117]:  ['wa', 'u', 'minute', 'order', 'table', 'food', 'did', 'came', 'time', 'asked']
[ 6 - 0.90909 - 0.23602]:  ['dirty', 'smell', 'clean', 'place', 'bathroom', 'sick', 'floor', 'smoke', 'hand', 'disgusting']
[ 7 - 0.80833 - 0.26455]:  ['thai', 'bbq', 'pork', 'curry', 'rib', 'meat', 'spicy', 'indian', 'pad', 'chicken']
[ 8 - 0.78333 - 0.22653]:  ['sandwich', 'bread', 'pho', 'meat', 'turkey', 'sub', 'wrap', 'beef', 'lunch', 'deli']
[ 9 - 1 - 0.2992]:  ['die', 'im', 'und', 'da', 'der', 'man', 'ich', 'war', 'ist', 'nicht']
[ 10 - 0.5875 - 0.36542]:  ['soup', 'rice', 'noodle', 'chinese', 'dish', 'chicken', 'bowl', 'fried', 'food', 'beef']
[ 11 - 0.95 - 0.24428]:  ['breakfast', 'egg', 'sunday', 'brunch', 'pancake', 'bacon', 'toast', 'waffle', 'french', 'morning']
[ 12 - 0.62159 - 0.33782]:  ['like', 'just', 'know', 'place', 'make', 'want', 'say', 'thing', 'look', 'people']
[ 13 - 0.32735 - 0.25599]:  ['wa', 'place', 'good', 'really', 'just', 'like', 'review', 'pretty', 'did', 'star']
[ 14 - 1 - 0.19794]:  ['dog', 'park', 'bagel', 'hot', 'course', 'pet', 'animal', 'cat', 'vet', 'golf']
[ 15 - 0.79242 - 0.23551]:  ['steak', 'wa', 'lobster', 'rib', 'cooked', 'potato', 'meat', 'prime', 'medium', 'filet']
[ 16 - 0.71833 - 0.24819]:  ['pizza', 'italian', 'crust', 'sauce', 'cheese', 'slice', 'order', 'pasta', 'good', 'topping']
[ 17 - 1 - 0.17949]:  ['free', 'machine', 'soda', 'photo', 'crepe', 'coke', 'gluten', 'christmas', 'diet', 'picture']
[ 18 - 0.9 - 0.22138]:  ['company', 'dress', 'shoe', 'work', 'house', 'home', 'shirt', 'new', 'apartment', 'wear']
[ 19 - 0.25159 - 0.25091]:  ['food', 'place', 'good', 'time', 'great', 'love', 'lunch', 'service', 'like', 'really']
[ 20 - 0.8375 - 0.26798]:  ['taco', 'chip', 'mexican', 'salsa', 'burrito', 'bean', 'food', 'margarita', 'tortilla', 'cheese']
[ 21 - 0.6875 - 0.23791]:  ['coffee', 'ice', 'cream', 'flavor', 'drink', 'like', 'cup', 'fruit', 'starbucks', 'yogurt']
[ 22 - 1 - 0.15334]:  ['wing', 'blue', 'buffalo', 'ranch', 'draft', 'philly', 'wild', 'sam', 'pint', 'diamond']
[ 23 - 0.675 - 0.21682]:  ['review', 'experience', 'ha', 'visit', 'make', 'star', 'quite', 'quality', 'staff', 'high']
[ 24 - 0.85 - 0.162]:  ['vega', 'la', 'strip', 'best', 'massage', 'trip', 'spa', 'casino', 'mall', 'cirque']
[ 25 - 0.93333 - 0.22988]:  ['night', 'friday', 'groupon', 'monday', 'truck', 'tuesday', 'thursday', 'deal', 'wednesday', 'flower']
[ 26 - 0.55159 - 0.24595]:  ['wa', 'restaurant', 'wine', 'dinner', 'menu', 'food', 'dish', 'meal', 'good', 'service']
[ 27 - 1 - 0.12768]:  ['queen', 'karaoke', 'frank', 'hockey', 'buzz', 'dairy', 'sing', 'jennifer', 'ave.', 'europe']
[ 28 - 0.72909 - 0.25806]:  ['wa', 'told', 'said', 'called', 'did', 'day', 'phone', 'asked', 'manager', 'card']
[ 29 - 0.8375 - 0.19815]:  ['music', 'ticket', 'movie', 'seat', 'fun', 'play', 'theater', 'time', 'playing', 'great']
[ 30 - 0.45159 - 0.20431]:  ['great', 'friendly', 'service', 'place', 'staff', 'food', 'love', 'amazing', 'recommend', 'good']
[ 31 - 0.75909 - 0.19737]:  ['nice', 'area', 'outside', 'table', 'inside', 'seating', 'patio', 'bar', 'place', 'view']
[ 32 - 0.35659 - 0.22838]:  ['wa', 'chicken', 'sauce', 'good', 'flavor', 'fried', 'ordered', 'like', 'little', 'just']
[ 33 - 0.9 - 0.26674]:  ['sushi', 'roll', 'fish', 'tuna', 'fresh', 'chef', 'salmon', 'japanese', 'rice', 'sashimi']
[ 34 - 0.875 - 0.19026]:  ['water', 'tea', 'glass', 'cup', 'drink', 'bottle', 'refill', 'iced', 'green', 'boba']
[ 35 - 0.50985 - 0.18241]:  ['club', 'night', 'drink', 'wa', 'people', 'girl', 'party', 'place', 'friend', 'line']
[ 36 - 0.7225 - 0.27453]:  ['price', 'buffet', 'worth', 'food', 'money', 'pay', 'better', 'quality', 'good', 'cost']
[ 37 - 0.32318 - 0.21363]:  ['wa', 'food', 'like', 'place', 'service', 'bad', 'ordered', 'tasted', 'good', 'just']
[ 38 - 0.44159 - 0.27947]:  ['wa', 'did', 'time', 'went', 'got', 'u', 'friend', 'came', 'just', 'day']
[ 39 - 0.7375 - 0.21797]:  ['class', 'office', 'care', 'time', 'doctor', 'dr.', 'appointment', 'gym', 'work', 'staff']
[ 40 - 0.65 - 0.32226]:  ['salad', 'cheese', 'bread', 'tomato', 'soup', 'dressing', 'mac', 'chicken', 'fresh', 'menu']
[ 41 - 0.65167 - 0.21533]:  ['burger', 'fry', 'cheese', 'onion', 'bun', 'ring', 'good', 'ordered', 'order', 'bacon']
[ 42 - 0.69159 - 0.19701]:  ['wa', 'car', 'hair', 'nail', 'did', 'time', 'salon', 'cut', 'job', 'tire']
[ 43 - 0.83333 - 0.25519]:  ['parking', 'car', 'line', 'door', 'lot', 'open', 'drive', 'closed', 'hour', 'sign']
[ 44 - 0.95 - 0.29985]:  ['le', 'et', 'la', 'pour', 'pa', 'que', 'est', 'en', 'une', 'je']
[ 45 - 0.8125 - 0.2528]:  ['store', 'shop', 'buy', 'item', 'sale', 'product', 'selection', 'price', 'shopping', 'like']
[ 46 - 0.56909 - 0.20493]:  ['bar', 'beer', 'drink', 'game', 'bartender', 'place', 'good', 'tv', 'selection', 'great']
[ 47 - 0.95 - 0.16546]:  ['box', 'package', 'post', 'jack', 'express', 'chris', 'hookah', 'office', 'ups', 'ship']
[ 48 - 0.75909 - 0.18445]:  ['location', 'place', 'phoenix', 'local', 'best', 'town', 'scottsdale', 'new', 'downtown', 'area']
[ 49 - 0.79242 - 0.21996]:  ['room', 'hotel', 'wa', 'stay', 'pool', 'bed', 'night', 'stayed', 'casino', 'desk']
uniqueness=0.738
\end{Verbatim}

ProdLDA:
\begin{Verbatim}[breaklines=true, fontsize=\tiny]
[0-0.40944-0.24083]: ['rib', 'brisket', 'bbq', 'fish', 'taco', 'mexican', 'catfish', 'cajun', 'salsa', 'okra']
[1-0.55111-0.15347]: ['greek', 'gyro', 'bland', 'atmosphere', 'tasteless', 'filthy', 'greasy', 'shish', 'mold', 'souvlaki']
[2-0.28444-0.18454]: ['catfish', 'bbq', 'hush', 'corn', 'rib', 'mac', 'taco', 'cajun', 'brisket', 'texas']
[3-0.49278-0.18204]: ['bland', 'tasteless', 'overpriced', 'disgusting', 'flavorless', 'edible', 'food', 'overrated', 'atmosphere', 'mediocre']
[4-1-0.17619]: ['airline', 'theater', 'airport', 'terminal', 'trail', 'stadium', 'exhibit', 'flight', 'airway', 'museum']
[5-1-0.13442]: ['buffet', 'chinese', 'crab', 'leg', 'bacchanal', 'dim', 'mein', 'wicked', 'seafood', 'carving']
[6-0.44167-0.14681]: ['pizza', 'wedding', 'italian', 'gluten', 'coordinator', 'delicious', 'crust', 'amazing', 'florist', 'birthday']
[7-0.71944-0.37625]: ['asada', 'carne', 'salsa', 'taco', 'burrito', 'thai', 'mexican', 'enchilada', 'tortilla', 'refried']
[8-1-0.48323]: ['est', 'tr\\u00e8s', 'retournerai', 'sont', 'endroit', 'peu', 'une', 'vraiment', 'oeufs', 'qui']
[9-0.38111-0.16431]: ['mac', 'rib', 'taco', 'chowder', 'love', 'yummy', 'salsa', 'brisket', 'chip', 'texas']
[10-0.37929-0.24311]: ['hash', 'burger', 'egg', 'breakfast', 'benedict', 'biscuit', 'toast', 'pancake', 'scrambled', 'corned']
[11-0.95-0.21124]: ['warranty', 'insurance', 'repair', 'contract', 'car', 'vehicle', 'bbb', 'cancel', 'rental', 'email']
[12-0.26262-0.25198]: ['breakfast', 'hash', 'egg', 'benedict', 'burger', 'toast', 'biscuit', 'brunch', 'omelet', 'pancake']
[13-1-0.22042]: ['suite', 'shower', 'hotel', 'elevator', 'pool', 'housekeeping', 'jacuzzi', 'bed', 'tub', 'amenity']
[14-1-0.2429]: ['foie', 'filet', 'gras', 'scallop', 'mignon', 'risotto', 'lobster', 'amuse', 'wine', 'creamed']
[15-0.47083-0.15043]: ['ceremony', 'chapel', 'pizza', 'wedding', 'minister', 'gluten', 'florist', 'bouquet', 'bianco', 'photographer']
[16-1-0.24788]: ['beer', 'pub', 'brewery', 'ale', 'brew', 'ipa', 'craft', 'bartender', 'game', 'draft']
[17-0.36444-0.17536]: ['taco', 'delicious', 'crawfish', 'margarita', 'cajun', 'bbq', 'mac', 'amazing', 'corn', 'fun']
[18-0.49-0.17937]: ['disgusting', 'filthy', 'tasteless', 'dirty', 'inedible', 'bland', 'dry', 'mediocre', 'gyro', 'gross']
[19-0.20206-0.22668]: ['indian', 'italian', 'naan', 'masala', 'pasta', 'tikka', 'atmosphere', 'pizza', 'food', 'india']
[20-0.65333-0.20453]: ['wash', 'wash.', 'vacuuming', 'rag', 'wiped', 'filthy', 'wipe', 'vacuum', 'vacuumed', 'car']
[21-0.27611-0.15756]: ['catfish', 'bbq', 'brisket', 'rib', 'cob', 'corn', 'margarita', 'mac', 'taco', 'hush']
[22-0.6625-0.19396]: ['pizza', 'crust', 'pepperoni', 'burger', 'wing', 'domino', 'fry', 'dog', 'topping', 'soggy']
[23-0.39179-0.36182]: ['indian', 'naan', 'italian', 'masala', 'tandoori', 'tikka', 'india', 'lassi', 'paneer', 'dosa']
[24-0.40762-0.16713]: ['indian', 'naan', 'bland', 'masala', 'tikka', 'underwhelming', 'uninspired', 'mediocre', 'ambiance', 'overpriced']
[25-0.36944-0.11527]: ['taco', 'margarita', 'mac', 'salsa', 'yummy', 'chip', 'shake', 'catfish', 'carne', 'potatoe']
[26-0.875-0.20124]: ['pita', 'hummus', 'cardio', 'gym', 'falafel', 'gyro', 'workout', 'greek', 'sandwich', 'produce']
[27-0.41778-0.21467]: ['mac', 'taco', 'rib', 'cob', 'juicy', 'burger', 'bbq', 'carne', 'bomb.com', 'delish']
[28-0.95-0.20403]: ['clothing', 'thrift', 'dress', 'jewelry', 'store', 'clearance', 'accessory', 'merchandise', 'cupcake', 'alteration']
[29-0.61944-0.12706]: ['crawfish', 'margarita', 'yummy', 'yum', 'sundae', 'nacho', 'delish', 'trifecta', 'love', 'taco']
[30-0.27607-0.29397]: ['indian', 'naan', 'tikka', 'masala', 'paneer', 'italian', 'food', 'india', 'korma', 'breakfast']
[31-0.49778-0.20012]: ['wash', 'atmosphere', 'dirty', 'filthy', 'wipe', 'wash.', 'rag', 'cleanliness', 'latte', 'cleaning']
[32-1-0.21306]: ['dr.', 'vet', 'doctor', 'dentist', 'instructor', 'dental', 'yoga', 'exam', 'nurse', 'grooming']
[33-1-0.41195]: ['sushi', 'yellowtail', 'nigiri', 'sashimi', 'tempura', 'miso', 'ayce', 'ramen', 'eel', 'tuna']
[34-0.37373-0.18235]: ['breakfast', 'benedict', 'excellent', 'toast', 'atmosphere', 'hash', 'highly', 'delicious', 'egg', 'brunch']
[35-1-0.20086]: ['community', 'institution', 'consistently', 'unmatched', 'management', 'culture', 'monopoly', 'estate', 'authentic', 'property']
[36-1-0.25813]: ['dance', 'bouncer', 'promoter', 'dj', 'x', 'dancing', 'club', 'dancer', 'dancefloor', 'guestlist']
[37-0.3704-0.1806]: ['indian', 'italian', 'pizza', 'atmosphere', 'pasta', 'naan', 'food', 'italy', 'ambiance', 'romantic']
[38-1-0.20674]: ['massage', 'manicure', 'pedicure', 'nail', 'salon', 'gel', 'stylist', 'pedi', 'cuticle', 'mani']
[39-1-0.20876]: ['manager', 'hostess', 'flagged', 'waited', 'seated', 'apology', 'acknowledged', 'rude', 'apologized', 'acknowledge']
[40-0.48333-0.1382]: ['wedding', 'chapel', 'ceremony', 'pizza', 'italian', 'gluten', 'photographer', 'minister', 'married', 'planner']
[41-0.62873-0.17883]: ['atmosphere', 'ambience', 'decor', 'food', 'indian', 'lawrenceville', 'cozy', 'ambiance', 'quaint', 'outdoor']
[42-0.33762-0.2659]: ['hash', 'breakfast', 'burger', 'benedict', 'egg', 'pancake', 'omelet', 'omelette', 'biscuit', 'brunch']
[43-0.41944-0.19644]: ['pizza', 'bianco', 'wedding', 'crust', 'italian', 'atmosphere', 'delicious', 'pepperoni', 'pizzeria', 'cibo']
[44-0.49-0.18319]: ['filthy', 'dirty', 'cleaner', 'bland', 'tasteless', 'mushy', 'disgusting', 'uneatable', 'gyro', 'rag']
[45-0.73929-0.23052]: ['frosting', 'cupcake', 'latte', 'bagel', 'coffee', 'barista', 'boba', 'pancake', 'donut', 'breakfast']
[46-1-0.26929]: ['cirque', 'acrobatics', 'soleil', 'performer', 'audience', 'stage', 'storyline', 'acrobatic', 'acrobat', 'tire']
[47-0.49595-0.19841]: ['burger', 'breakfast', 'hash', 'ronin', 'fry', 'shake', 'steak', 'bacon', 'toast', 'benedict']
[48-0.86111-0.082801]: ['edinburgh', 'atmosphere', 'cosy', 'acoustic', 'montreal', 'newington', 'landscaping', 'gameworks', 'ambience', 'pittsburgh']
[49-0.30429-0.33129]: ['indian', 'naan', 'italian', 'masala', 'tikka', 'paneer', 'tandoori', 'india', 'saag', 'pizza']
\end{Verbatim}

NTM-R:
\begin{Verbatim}[breaklines=true, fontsize=\tiny]
[0-0.1909-0.26952]: ['lincoln', 'proclaimed', 'proclaiming', 'rally', 'defended', 'civil', 'marching', 'marched', 'campaign', 'boycott']
[1-0.22741-0.22154]: ['independence', 'unsuccessfully', 'monument', 'proclaiming', 'marching', 'supported', 'challenged', 'tennessee', 'defended', 'emerged']
[2-0.14614-0.25689]: ['campaign', 'independence', 'defended', 'proclaiming', 'marched', 'missouri', 'drawn', 'marching', 'supported', 'proclaimed']
[3-0.21257-0.21085]: ['supported', 'fought', 'alabama', 'campaign', 'proclaiming', 'defended', 'marching', 'enthusiastically', 'mao', 'missouri']
[4-0.31407-0.17423]: ['proclaiming', 'campaign', 'nelson', 'marching', 'indiana', 'carolina', 'gay', 'unsuccessful', 'missouri', 'catholic']
[5-0.2586-0.25677]: ['defended', 'campaign', 'independence', 'marching', 'strongest', 'supported', 'proclaiming', 'sponsored', 'rally', 'leadership']
[6-0.11352-0.24402]: ['declaring', 'proclaiming', 'marched', 'marching', 'arkansas', 'defended', 'strongest', 'missouri', 'campaign', 'proclaimed']
[7-1-0.22502]: ['dance', 'dancing', 'bouncer', 'dj', 'danced', 'song', 'dancer', 'ipa', 'bartender', 'promoter']
[8-0.19602-0.24114]: ['proclaiming', 'road', 'supported', 'marched', 'campaign', 'marching', 'fought', 'capitol', 'lincoln', 'defended']
[9-0.092567-0.23741]: ['campaign', 'independence', 'proclaiming', 'proclaimed', 'lincoln', 'catholic', 'tennessee', 'supported', 'marching', 'marched']
[10-0.15347-0.20531]: ['proclaiming', 'defended', 'marching', 'capitol', 'alabama', 'marched', 'mustang', 'campaign', 'missouri', 'unsuccessfully']
[11-0.11936-0.23289]: ['banner', 'campaign', 'missouri', 'defended', 'marching', 'supported', 'proclaiming', 'marched', 'alabama', 'emerged']
[12-1-0.18502]: ['refund', 'voicemail', 'refused', 'unprofessional', 'supervisor', 'cox', 'ontrac', 'reschedule', 'apology', 'rudely']
[13-0.12697-0.21198]: ['proclaiming', 'supported', 'marching', 'defended', 'missouri', 'renamed', 'sponsored', 'marched', 'indiana', 'campaign']
[14-0.27399-0.26035]: ['campaign', 'proclaimed', 'proclaiming', 'fought', 'national', 'emerged', 'marched', 'marching', 'declaring', 'predecessor']
[15-1-0.29774]: ['dentist', 'dental', 'suis', 'je', 'sont', 'choix', 'est', 'peu', 'qui', 'fait']
[16-1-0.18789]: ['mocha', 'dunkin', 'latte', 'bagel', 'croissant', 'tire', 'einstein', 'cone', 'maple', 'scone']
[17-0.32514-0.22424]: ['campaign', 'defended', 'marched', 'marching', 'unsuccessfully', 'dame', 'proclaiming', 'ralph', 'federal', 'army']
[18-0.25847-0.21798]: ['adopted', 'defended', 'proclaiming', 'marched', 'nelson', 'vietnamese', 'lincoln', 'campaign', 'unsuccessfully', 'marching']
[19-0.25763-0.22588]: ['independence', 'supported', 'campaign', 'defended', 'marching', 'unsuccessfully', 'enthusiastically', 'presidential', 'nelson', 'mississippi']
[20-0.31388-0.19339]: ['proclaiming', 'marching', 'marched', 'boldly', 'unsuccessfully', 'maroon', 'supported', 'proclaim', 'arkansas', 'verdun']
[21-1-0.14517]: ['bellagio', 'tower', 'suite', 'shuttle', 'elevator', 'paris', 'monorail', 'continental', 'ami', 'hilton']
[22-0.19364-0.22081]: ['missouri', 'supported', 'proclaiming', 'marching', 'defended', 'campaign', 'battle', 'marched', 'indiana', 'puerto']
[23-0.25503-0.23828]: ['challenged', 'defended', 'marching', 'proclaiming', 'declaring', 'campaign', 'fought', 'unsuccessfully', 'monroe', 'kentucky']
[24-0.24245-0.21527]: ['lincoln', 'banner', 'campaign', 'proclaiming', 'marching', 'declaring', 'football', 'roosevelt', 'marched', 'supported']
[25-0.16281-0.26664]: ['marching', 'proclaiming', 'proclaimed', 'defended', 'independence', 'campaign', 'supported', 'civil', 'marched', 'mormon']
[26-0.95-0.53788]: ['ayce', 'goyemon', 'nigiri', 'sushi', 'sashimi', 'teharu', 'amaebi', 'oyshi', 'sakana', 'auswahl']
[27-0.35443-0.23179]: ['exception', 'campaign', 'defended', 'marching', 'claimed', 'revolution', 'boldly', 'marched', 'proclaiming', 'arkansas']
[28-0.29895-0.21784]: ['marching', 'emerged', 'boldly', 'declaring', 'marched', 'civil', 'notre', 'waterloo', 'defended', 'proclaiming']
[29-0.95-0.13247]: ['circus', 'sum', 'imitation', 'dim', 'para', 'carnival', 'lo', 'nigiri', 'bacchanal', 'boba']
[30-0.27428-0.20607]: ['campaign', 'community', 'proclaiming', 'thrilled', 'marching', 'proclaimed', 'unsuccessful', 'defended', 'supported', 'arkansas']
[31-0.31752-0.20652]: ['schedule', 'proclaiming', 'campaign', 'missouri', 'marched', 'revived', 'largely', 'marching', 'arkansas', 'unsuccessful']
[32-0.37102-0.22941]: ['proclaiming', 'defended', 'supported', 'campaign', 'mississippi', 'marching', 'marched', 'pancho', 'declared', 'illinois']
[33-1-0.16294]: ['mani', 'manicure', 'gel', 'pedicure', 'pedi', 'cuticle', 'asada', 'carne', 'waxing', 'eyebrow']
[34-0.22772-0.2491]: ['fought', 'voted', 'defended', 'marching', 'rally', 'campaign', 'proclaiming', 'independence', 'roosevelt', 'lincoln']
[35-1-0.34016]: ['paneer', 'der', 'und', 'zu', 'auch', 'nicht', 'ich', 'aber', 'essen', 'kann']
[36-0.17936-0.2285]: ['campaign', 'defended', 'convention', 'marching', 'nelson', 'proclaiming', 'lincoln', 'supported', 'catholic', 'marched']
[37-0.29847-0.19011]: ['lincoln', 'campaign', 'economy', 'indiana', 'proclaiming', 'marching', 'arkansas', 'avenue', 'dame', 'marched']
[38-1-0.15268]: ['mahi', 'mashed', 'undercooked', 'broccoli', 'wonton', 'chowder', 'overcooked', 'soggy', 'katsu', 'breading']
[39-0.25617-0.2291]: ['independence', 'campaign', 'defended', 'marching', 'civil', 'lincoln', 'proclaiming', 'popularity', 'marched', 'maryland']
[40-0.319-0.18811]: ['campaign', 'marching', 'begun', 'unsuccessfully', 'supported', 'mustang', 'alabama', 'proclaiming', 'tennessee', 'leaning']
[41-0.23145-0.21547]: ['indiana', 'chinese', 'fought', 'marched', 'marching', 'september', 'proclaimed', 'proclaiming', 'catholic', 'independence']
[42-0.24117-0.20143]: ['defended', 'colorado', 'marching', 'missouri', 'campaign', 'proclaiming', 'independence', 'marched', 'unsuccessfully', 'skyline']
[43-0.18681-0.22528]: ['campaign', 'independence', 'marching', 'proclaiming', 'rowdy', 'lincoln', 'defended', 'renamed', 'proclaimed', 'declaring']
[44-0.24283-0.20284]: ['chinese', 'defended', 'marched', 'proclaiming', 'independence', 'marching', 'universal', 'alabama', 'campaign', 'ralph']
[45-0.10685-0.22909]: ['marched', 'lincoln', 'proclaiming', 'unsuccessfully', 'marching', 'campaign', 'indiana', 'defended', 'proclaimed', 'revived']
[46-0.13688-0.2036]: ['campaign', 'marching', 'marched', 'emerged', 'indiana', 'puerto', 'proclaiming', 'tennessee', 'independence', 'missouri']
[47-0.2716-0.19729]: ['renamed', 'noodle', 'campaign', 'missouri', 'lincoln', 'defended', 'proclaiming', 'marched', 'resisted', 'proclaimed']
[48-0.35085-0.18812]: ['proclaiming', 'marching', 'campaign', 'boldly', 'marched', 'anti', 'arkansas', 'alamo', 'proclaim', 'kentucky']
[49-1-0.16444]: ['dog', 'cardio', 'grooming', 'vet', 'petsmart', 'gym', 'animal', 'membership', 'harkins', 'trainer']
\end{Verbatim}

W-LDA:
\begin{Verbatim}[breaklines=true, fontsize=\tiny]
[0-1-0.10334]: ['buffet', 'leg', 'wicked', 'crab', 'prime', 'station', 'bacchanal', 'wynn', 'carving', 'seafood']
[1-0.78333-0.19376]: ['register', 'cashier', 'employee', 'counter', 'starbucks', 'customer', 'barista', 'standing', 'store', 'stood']
[2-0.73333-0.23916]: ['music', 'dj', 'dance', 'band', 'chill', 'crowd', 'bar', 'fun', 'lounge', 'drink']
[3-0.65833-0.1593]: ['hostess', 'table', 'seated', 'u', 'minute', 'waited', 'server', 'sat', 'waitress', 'acknowledged']
[4-0.80833-0.23538]: ['cold', 'salad', 'lettuce', 'slow', 'sandwich', 'horrible', 'dressing', 'terrible', 'medium', 'steak']
[5-0.78333-0.42375]: ['starbucks', 'coffee', 'latte', 'espresso', 'baristas', 'barista', 'caffeine', 'mocha', 'iced', 'chai']
[6-0.65-0.33339]: ['asada', 'carne', 'burrito', 'taco', 'salsa', 'pastor', 'tortilla', 'mexican', 'pico', 'enchilada']
[7-0.58333-0.25355]: ['hash', 'pancake', 'breakfast', 'egg', 'toast', 'scrambled', 'omelet', 'biscuit', 'benedict', 'bagel']
[8-0.93333-0.31394]: ['tire', 'brake', 'mechanic', 'car', 'repair', 'dealership', 'engine', 'vehicle', 'warranty', 'leak']
[9-0.5-0.18936]: ['car', 'sandwich', 'breakfast', 'coffee', 'wash', 'burger', 'latte', 'fry', 'friendly', 'awesome']
[10-0.78333-0.16338]: ['pho', 'excellent', 'delicious', 'authentic', 'indian', 'amazing', 'best', 'chinese', 'outstanding', 'favorite']
[11-0.83333-0.23699]: ['filthy', 'dirty', 'disgusting', 'worst', 'health', 'waste', 'suck', 'horrible', 'gross', 'nasty']
[12-0.9-0.31227]: ['roasted', 'vinaigrette', 'creamy', 'tomato', 'goat', 'chocolate', 'rich', 'caramelized', 'squash', 'topped']
[13-0.76667-0.28784]: ['tortilla', 'enchilada', 'salsa', 'bean', 'chip', 'taco', 'fish', 'canned', 'tasted', 'refried']
[14-1-0.37871]: ['et', 'est', 'une', 'je', 'mais', 'qui', 'und', 'que', 'avec', 'dans']
[15-0.69167-0.19405]: ['reservation', 'table', 'wine', 'waiter', 'hostess', 'restaurant', 'seated', 'dining', 'party', 'arrived']
[16-0.9-0.20611]: ['nail', 'manicure', 'pedicure', 'gel', 'cuticle', 'polish', 'salon', 'pedi', 'toe', 'acrylic']
[17-0.9-0.18233]: ['dance', 'club', 'bouncer', 'x', 'promoter', 'vip', 'tao', 'dj', 'marquee', 'dancing']
[18-0.93333-0.3198]: ['nigiri', 'sushi', 'roll', 'sashimi', 'yellowtail', 'ayce', 'tempura', 'eel', 'tuna', 'uni']
[19-0.83333-0.34693]: ['ramen', 'noodle', 'broth', 'pho', 'vietnamese', 'curry', 'tofu', 'dumpling', 'bo', 'vermicelli']
[20-0.475-0.15906]: ['sushi', 'margarita', 'happy', 'hour', 'seated', 'table', 'reservation', 'drink', 'salsa', 'wine']
[21-1-0.29525]: ['brisket', 'bbq', 'rib', 'pulled', 'mac', 'pork', 'slaw', 'coleslaw', 'cole', 'meat']
[22-0.65333-0.17083]: ['sushi', 'consistently', 'happy', 'mexican', 'quality', 'consistent', 'location', 'pizza', 'ha', 'great']
[23-0.95-0.24093]: ['flight', 'airline', 'shuttle', 'cab', 'airport', 'driver', 'plane', 'delayed', 'airway', 'rental']
[24-0.95-0.29522]: ['steak', 'filet', 'steakhouse', 'ribeye', 'bone-in', 'mignon', 'rare', 'creamed', 'lobster', 'gras']
[25-0.78667-0.16426]: ['attentive', 'calamari', 'pleasantly', 'appetizer', 'happy', 'pizza', 'wa', 'great', 'enjoyed', 'loved']
[26-0.95-0.25875]: ['beer', 'tap', 'brewery', 'brew', 'pub', 'sport', 'craft', 'ale', 'draft', 'ipa']
[27-0.70833-0.21733]: ['waitress', 'came', 'asked', 'ordered', 'server', 'u', 'brought', 'table', 'drink', 'said']
[28-0.51667-0.25582]: ['breakfast', 'pancake', 'bagel', 'brunch', 'toast', 'egg', 'benedict', 'omelet', 'coffee', 'hash']
[29-0.83333-0.20873]: ['great', 'staff', 'friendly', 'helpful', 'atmosphere', 'service', 'excellent', 'knowledgeable', 'environment', 'clean']
[30-0.92-0.32171]: ['pizza', 'crust', 'pepperoni', 'slice', 'topping', 'dough', 'pizzeria', 'oven', 'ny', 'mozzarella']
[31-0.875-0.29161]: ['burger', 'bun', 'in-n-out', 'patty', 'shake', 'fry', 'milkshake', 'dog', 'smashburger', 'cheeseburger']
[32-1-0.24596]: ['cirque', 'soleil', 'acrobatics', 'audience', 'performer', 'stage', 'exhibit', 'performance', 'museum', 'theater']
[33-0.775-0.17923]: ['pad', 'thai', 'gyro', 'sandwich', 'curry', 'sub', 'pita', 'panang', 'chicken', 'tom']
[34-0.85-0.26535]: ['salon', 'massage', 'stylist', 'hair', 'facial', 'haircut', 'waxing', 'pedicure', 'spa', 'barber']
[35-0.74167-0.16379]: ['mexican', 'burger', 'food', 'wing', 'average', 'taco', 'overpriced', 'asada', 'bad', 'mediocre']
[36-0.93333-0.20984]: ['produce', 'grocery', 'market', 'trader', 'farmer', 'organic', 'bulk', 'park', 'store', 'supermarket']
[37-0.9-0.20996]: ['room', 'bed', 'shower', 'housekeeping', 'motel', 'hotel', 'stain', 'sheet', 'carpet', 'pillow']
[38-0.95-0.29802]: ['cupcake', 'frosting', 'cake', 'chocolate', 'cream', 'ice', 'yogurt', 'velvet', 'vanilla', 'boba']
[39-1-0.24957]: ['dr.', 'dentist', 'doctor', 'vet', 'dr', 'dental', 'patient', 'office', 'exam', 'clinic']
[40-0.8-0.16245]: ['bartender', 'game', 'bar', 'beer', 'band', 'dive', 'karaoke', 'football', 'song', 'jukebox']
[41-0.85-0.23321]: ['hotel', 'suite', 'spa', 'pool', 'casino', 'room', 'amenity', 'jacuzzi', 'stayed', 'spacious']
[42-1-0.16035]: ['view', 'fountain', 'bellagio', 'romantic', 'gabi', 'anniversary', 'ami', 'impeccable', 'pairing', 'mon']
[43-0.87-0.22016]: ['delivery', 'order', 'deliver', 'called', 'hung', 'pizza', 'phone', 'driver', 'delivered', 'answered']
[44-0.6-0.17233]: ['pho', 'closed', 'rude', 'bartender', 'customer', 'worst', 'suck', 'business', 'horrible', 'car']
[45-1-0.16971]: ['gym', 'contract', 'membership', 'cox', 'lease', 'fitness', 'apartment', 'account', 'tenant', 'trainer']
[46-0.88333-0.16832]: ['wine', 'bruschetta', 'tapa', 'cocktail', 'goat', 'date', 'martini', 'sangria', 'cozy', 'list']
[47-0.93333-0.23392]: ['clothing', 'clothes', 'shoe', 'accessory', 'store', 'dress', 'clearance', 'jewelry', 'pair', 'thrift']
[48-0.62-0.2303]: ['healthy', 'love', 'pizza', 'sandwich', 'favorite', 'hummus', 'gyro', 'fresh', 'pita', 'burger']
[49-0.9-0.20869]: ['chinese', 'mein', 'panda', 'bland', 'chow', 'rice', 'noodle', 'express', 'wonton', 'tasteless']
\end{Verbatim}

\end{document}